%% file: main.tex
\documentclass{article}

\usepackage{appendix}
\usepackage[english]{babel}
\usepackage{multirow}

\usepackage[hyphens]{url}

\usepackage[utf8]{inputenc}
\usepackage[T1]{fontenc}
\usepackage{longtable}
\usepackage{booktabs}
\usepackage{array}
\usepackage{pdflscape}

\usepackage[letterpaper,top=2cm,bottom=2cm,left=3cm,right=3cm,marginparwidth=1.75cm]{geometry}
\usepackage{enumitem}
\usepackage{amsmath}
\usepackage[colorlinks=true, allcolors=blue]{hyperref}
\usepackage{apacite}
\usepackage{lipsum}
\usepackage{threeparttable}
\usepackage{graphicx} % à mettre dans le préambule si pas déjà fait
\usepackage{placeins}
\usepackage{graphicx}
\usepackage{hyperref}
\usepackage{ragged2e}

\usepackage[justification=centering]{caption}
\newcommand{\tablenote}[2]{%
  \vspace{0.5ex}%
  {\noindent \footnotesize \justifying \textit{Note: #1 Reading: #2}\par}%
}

\newcommand{\tablereading}[1]{%
  \vspace{0.5ex}%
  {\noindent \footnotesize \justifying \textit{Reading: #1}\par}%
}

\newcommand{\tablenoteonly}[1]{%
  \vspace{0.5ex}%
  {\noindent \footnotesize \justifying \textit{Note: #1}\par}%
}

\title{Quotas for scholarship recipients: an efficient race-neutral alternative to affirmative action?\footnote{I am grateful to Marc Gurgand for supervising this work and providing invaluable feedback, and to Julien Grenet for serving as referee. I thank Camille Terrier, Renke Schmacker, and Rustamdjan Hakimov for their valuable advice and assistance with data access. I acknowledge the Parcoursup team, particularly Sonia Bonnafé and Serge Richard, for their assistance in helping me understand the data. I also want to thank Jonathan Roth for answering my technical questions. Finally, I would like to thank Camille Hémet, Martin Mugnier, and the participants of the Labor and Public Economics seminar for their constructive feedback. Access to some of the data used in this work was carried out in secure environments provided by CASD - Centre d'Accès Sécurisé aux Données (Ref. 10.3.34724/CASD).}}

\author{Louis Gleyo\\  \textit{Paris School of Economics}}

\begin{document}
\maketitle

\begin{abstract}
\noindent \textit{Since 2018, France's centralized higher education platform, Parcoursup, has implemented quotas for scholarship recipients, with program-specific thresholds based on the applicants' composition. Using difference-in-differences methods, I find that these quotas enabled scholarship students to access more selective programs, although the intention-to-treat effects remain modest. Matching methods reveal that the policy improved the scholarship students' waiting list positions relative to those of comparable non-scholarship peers, and simulations suggest that the modest effect could be attributed to the low intensity of the treatment. However, I detect no robust or lasting effects on the extensive margin of higher education access. Despite high policy salience, quotas did not affect the application behavior or pre-college investment of scholarship students, even among high achievers. These findings align with research on affirmative action, suggesting that such policies primarily benefit disadvantaged students who access selective institutions, rather than expanding total enrollment. Nevertheless, scholarship quotas demonstrate that race-neutral alternatives can effectively promote socioeconomic diversity in prestigious programs.}\\


\end{abstract}
\newpage 

\setcounter{tocdepth}{2}

\tableofcontents % <-- Ceci génère le sommaire

\newpage

\section{Introduction}

Access to French higher education, particularly elite institutions, remains marked by strong social inequalities \cite{bonneau2021quelle,bechichi2021segregation,benveniste2023like,schneider2023orientation}. Children from the lowest income decile have a threefold lower probability of accessing higher education than those from the highest decile \cite{bonneau2022unequal}. This stratification intensifies for selective programs: at equivalent academic levels, scholarship recipients formulate less ambitious preferences and enter less prestigious programs than non-scholarship students \cite{terrier2023confidence}. These disparities perpetuate social reproduction and question equal opportunity mechanisms in higher education.\\

French secondary education culminates in the baccalauréat exam, awarded in three tracks: general (academic preparation for higher education), technological (specialized programs, less selective than the general track), and professional (vocational training). Post-secondary programs fall into two main categories: selective programs, including two-year technical degrees (DUT and BTS) mainly for technological and vocational tracks, preparatory classes for Grandes Écoles (CPGE), and Grandes Écoles (prestigious specialized institutions in engineering, business, or public administration accessible through competitive admission); and non-selective programs, primarily three-year bachelor's degrees (Licence) at universities with open admission in most of these. This institutional architecture creates a clear hierarchy where Grandes Écoles represent the most prestigious pathway, accessible either through competitive preparatory classes or direct competitive admission. At the same time, universities often serve as the default option for students who are not admitted to selective programs.\\

France has used a centralized higher education admission system, known as APB (Admission Post-Bac) from 2007 to 2017, and then Parcoursup since 2018, through which all students must apply to post-secondary programs. In 2018, Parcoursup introduced a major institutional innovation: quotas for scholarship recipients in admission processes. This policy ensures that a minimum proportion of scholarship applicants receive admission offers in each program, calculated as "percentage of scholarship applicants + 2pp", with a 5\% floor. In France, scholarships are exclusively need-based, allocated according to strict income and family criteria, unlike merit-based scholarships common in other countries \cite{remigereau2024impact}. Until 2024, they're not automatic, and eligible parents should submit a request to receive them. This measure thus constitutes France's first national, large-scale 'race-neutral alternative' to affirmative action, contrasting with American or Brazilian approaches that combine social and ethnic criteria.\\

While numerous studies show affirmative action policies enable minorities to access more elite institutions \cite{bagde2016, otero2021affirmative, Bertrand2010} - and conversely that cessation of such policies degrades minority students' college quality \cite{card2005would,  Hinrichs2012, Backes2012,bleemer2022,yagan2016supply} - these studies find little or no evidence of effect on the extensive margin \cite{Hinrichs2012, Backes2012}. How can I explain these college quality effects? Several studies suggest that they result from changes in admissions committee behavior rather than changes in student application behavior \cite{card2005would,antonovics2013were,yagan2016supply}. However, some evidence indicates minority students\footnote{Throughout this article, I will use “minority student” as a synonym for “underrepresented minority student”, meaning categories of individuals who are underrepresented in higher education, particularly in terms of quality, based on ethnicity or economic and social status.} do make more ambitious applications under affirmative action \cite{akhtari2024,dickson2006does}. Studies also find null or positive effects of AA on pre-college investment \cite{antonovics2014effect, Francis2012, estevan2019redistribution, strifezzi2023, akhtari2024}. Race-neutral alternatives, such as x-percent rules requiring minimum proportions of top students from each high school to access prestigious colleges, show weak or null effects on minority students' college quality \cite{BLEEMER2023104839, Daugherty2014}. Conversely, institutional reports on Parcoursup quotas suggest effects contrary to the literature: minimal college quality impact \cite{courdescomptes_ore_2020}, but strong positive effects on application behavior \cite{courdescomptes_ore_2020,cesp2021} and the extensive margin \cite{cesp2021}. Therefore, this work addresses three research questions: the impact on students' final enrollment (extensive margin and college quality), application behaviors, and pre-college investment.\\

I use data from French higher education admission platforms: APB (2014-2017) and Parcoursup (2018-2020). I construct prestige scores for each program based on the average baccalaureate scores of admitted students. Using doubly robust difference-in-differences methods on cross-sectional data \cite{sant2020doubly}, I compare scholarship students with non-scholarship students having similar characteristics. I evaluate effects for each high school track (general, technical, professional), including Holm's sequential Bonferroni procedure for multiple hypothesis testing \cite{abdi2010holm} and honest difference-in-differences robustness tests \cite{rambachan2023more}. I apply this method to several outcomes: final enrollment (access to higher education, prestige of the final enrollment, type of program where the individual enrolls), application behavior (likelihood of applying in some types of programs and the average/maximum prestige of the applications), and pre-college investment, measured through final exam grades.\\

For the general track, I find a weak but robust and lasting effect on program prestige. I also find increased probability of admission to selective short-term DUT programs, but these effects are not robust. For the vocational track, I observed a decrease of approximately two percentage points in 2018 and 2019 for Bachelor's (not robust), offset by a 1 percentage point increase in the short-term BTS program (not significant). Regarding the extensive margin (enrollment in any program), pre-college investment (grades in the final high school exam, called "baccalauréat"), or application behavior, I find no robust or strong effects.\\

I therefore see that if I measure an effect on admission, it is concentrated on the intensive margin, and is not explained by student behavior, as explained by \citeA{card2005would}. These results are very opposed to the conclusions of institutional reports published on the subject \cite{courdescomptes_ore_2020,cesp2021}. The effects remain nonetheless weak, which is also explained by the fact that quotas are defined endogenously for each program: the lower the proportion of scholarship applicants a program has, the lower the quota for applying. This policy differs drastically from other affirmative action policies that exist outside France. My work nevertheless suffers from some limitations: apart from the fact that long-term outcomes are not studied here, I may fear a compositional change effect \cite{santanna2025differenceindifferencescompositionalchanges}, as the number of scholarship students increases each year.\\

In addition to these main estimations, I conduct complementary analyses to investigate the potential mechanisms that explain the obtained results. Using matching methods and before-and-after analysis, I demonstrate that scholarship students were slightly disadvantaged on waiting lists compared to non-scholarship students with similar profiles. I demonstrate that the introduction of quotas enabled scholarship students to substantially increase their chances of being admitted compared to non-scholarship students with similar characteristics. Using static simulation methods applied to the 2016 session on APB, I demonstrate that the observed effect on prestige is at least partially explained by changes in the type of program students attend. I also demonstrate that the 2pp bonus has a substantial impact on treatment intensity, i.e., the number of scholarship students who change programs due to quotas, suggesting that the weak effect is due to an initial low intensity of treatment (as a simple +2pp increase is sufficient to almost double the intensity of treatment).\\

The study allows several important contributions to the scientific literature on affirmative action. First, it demonstrates that it is possible to construct race-neutral alternatives that have a significant impact on the prestige of college admissions of disadvantaged students. Second, it helps strengthen the evidence on the weakness of a behavioral effect by exploiting not the ban of affirmative action but the creation of affirmative action. Moreover, behavioral effects are null regardless of whether we are "vague" about the amount of quotas (i.e., we inform students that they exist without specifying the amount, as in 2018 and 2019) or "transparent" (i.e., we provide the amount of quota for each program, as in 2020). Finally, the paper demonstrates that even when we set quotas calculated endogenously, there is a positive effect on admission. In contrast, some authors, such as \cite{couto2021parcoursup}, feared an increase in school segregation and social inequalities, a fear that this study refutes. New research will enable us to understand the long-term effects and enhance the robustness of our analyses.\\

The paper is structured as follows. In Section \ref{context}, I present the evaluated intervention, as well as the French educational context and the literature review on affirmative action. In Section \ref{hypothesis}, I present the hypotheses that will be tested. In Section \ref{data}, I present the data used and the data processing carried out to operationalize the outcomes. In Section \ref{strategy}, I present the empirical strategy employed, and the results are then presented in Section \ref{results}. In Section \ref{mechanisms}, I present complementary data analyses that will help us understand the potential mechanisms explaining the results. In Section \ref{discussion}, I discuss the interpretation of the results, limitations, and new research perspectives. Section \ref{conclusion} concludes by summarizing the results obtained.

\section{Context and intervention\label{context}}

\subsection{Higher Education in France}

\subsubsection{High-School Tracks}

In France, high school lasts three years (classes of \textit{seconde}, \textit{première}, and \textit{terminale}) and is organized into three major tracks: the General Baccalaureate, the Technological Baccalaureate, and the Vocational Baccalaureate. At the end of their curriculum, they prepare a diploma, the baccalaureate. The tracks determine the type of higher education program you could get admitted to.

\paragraph{General track} In the general track, until 2021, students in the general track were divided into three series: the "Scientific" specialization (STEM, natural sciences), the "Economics and Social" specialization (Economics, Social Sciences), and the "Literary" track (Humanities, Languages, Literature). Since 2021, a reform has altered the functioning of the track, and students must choose two electives from approximately ten options (Math, Physics, Economics and Social Sciences, Literature, etc.). Before and after 2021, all students had core courses common to the general track. The general track is widely considered the most prestigious of the three tracks, providing access to long-term studies and highly selective programs, such as CPGE and \textit{Grandes Écoles} (see below).

\paragraph{Technological track} For the technological track, students have a common core. In contrast to the general track, they have stronger and more specialized areas of focus; for example, the STI2D specialization prepares individuals for industrial careers, and the STMG specialization prepares them for business careers. It should be noted that the objective of this track is to prepare students for higher education studies, albeit in a shorter format (BTS, DUT), with the possibility for the very best students to enter a CPGE or a \textit{Grande École} via an adapted pathway.

\paragraph{Vocational track} The vocational track offers, through several dozen possible specializations, to train high school students specifically for specific jobs or families of jobs. Vocational baccalaureate holders can either start working immediately after completing their program or pursue short-term studies (BTS), typically for the most academically talented students.\\

\noindent Finally, it is worth noting that some high school students may be eligible to apply for a secondary education scholarship. This scholarship was granted upon request from parents (not automatically), with eligibility conditions based on financial resources (depending on income and the number of siblings). The operation and eligibility requirements for the awarding of these scholarships are detailed by \citeA{remigereau2024impact}. Until 2024, the scholarship was not automatically awarded, and parents had to apply for it manually. I have not found any official statistics on non-take-up; however, some official communications refer to a significant problem of non-take-up of scholarships \cite{blanquer_dussopt_2022_bourses}.

\subsubsection{Higher Education Programs}

French higher education is organized around several types of programs, which I will refer to as "Program Types" for simplicity. Each type of program has its characteristics, and some are specifically designed for certain high school tracks. Some programs are selective (can reject applicants even before putting them on the call list), while others are non-selective (cannot exclude a priori people from the call list). A description of the different types of training is provided in \autoref{principal:type_of_program}.\\

\begin{table}[ht]
\centering
\caption{Types of Higher Education Programs in France}
\label{principal:type_of_program}
\footnotesize
\begin{tabular}{|p{2cm}|p{5cm}|p{3cm}|p{1.5cm}|p{1.5cm}|}
\hline
\textbf{Program} & \textbf{Description} & \textbf{Target Population} & \textbf{Present on APB} & \textbf{Present on Parcoursup} \\
\hline
\textbf{Licence (Bachelor)} & 3-year program, equivalent to a Bachelor's degree. Most fields are non-selective. Allows students to pursue Master's studies. Also includes medical studies, known to be highly selective. & Primarily General, but also admits Technological and Vocational students within available spots. & Yes & Yes \\
\hline
\textbf{BTS} & Short, professional 2-year program, with limited possibility of continuing studies. & Primarily Vocational and Technological, also General. & Yes & Yes \\
\hline
\textbf{DUT} & 2-year university technological program, replaced by BUT (Bachelor Universitaire de Technologie) - 3-year program since 2021. Continuing studies are possible but rare. The program is often recognized as more advanced than BTS. & Primarily Technological and General. & Yes & Yes \\
\hline
\textbf{CPGE} & Classes Préparatoires aux Grandes Écoles, intensive, prestigious 2-year programs preparing for entrance exams to elite schools (engineering schools, ENS, business schools). & Primarily General, with some specific tracks for former technological high school students (TSI, TB, TPC) and Vocational (rare). & Yes & Yes \\
\hline
\textbf{Grandes Écoles} & Prestigious higher education institutions primarily training future senior executives. Includes engineering schools, business schools, and other specialized institutions (such as Sciences Po). Recruits are typically either directly after high school (from post-bac schools), after completing CPGE (after 2 years of study), or after earning a bachelor's degree. & Primarily General track, Technological (STI2D, STL) for some engineering schools, (STMG) for some business schools. & Partial & Yes (post-bac schools) \\
\hline
\end{tabular}
\end{table}

\noindent Before 2018, certain non-selective programs - some bachelor's degrees but not all - suffered from a mismatch between the number of applicants and the number of available spots. To address this problem, a lottery system was implemented, followed by selection processes; however, as \citeA{bechichi2021segregation} shows, most bachelor's programs remain minimally selective. Also, the type of program is not necessarily linked to the field of study: there are BTS programs in computer science, as well as DUTs and Bachelor's degrees. The most theoretical fields are reserved for Bachelor's degrees and CPGE. However, the type of program is closely linked to its selectivity and prestige. Grandes Écoles are elite higher education institutions that primarily recruit through competitive examinations following two to three years of intensive preparatory classes (Classes Préparatoires aux Grandes Écoles - CPGE). These preparatory classes, accessible after the baccalauréat, prepare students for entrance examinations to engineering, business, and other specialized schools. This system operates in parallel to the university system and is considered more selective and prestigious.\\

Other programs, such as medical and social training programs, are accessible through Parcoursup; however, they are not included in the analysis because they were not available on Admission Post-Bac (the higher education admission platform) before 2018.

\subsubsection{From \textit{Admission Post-Bac} to \textit{Parcoursup}}

Admission Post-Bac (APB, 2009-2017) was the central admission platform for French higher education. Applicants ranked their preferences in order, while programs established their call lists. A Gale-Shapley algorithm \cite{gale_shapley} was applied to produce stable assignments in three rounds. However, the system had a major dysfunction: the algorithm prioritized the rank of preferences as a criterion for competitive bachelor's degrees, allowing programs to access applicants' preferences and thus violating the theoretical conditions of the algorithm. Furthermore, bachelor's degrees, theoretically "non-selective" (supposed to admit all high school applicants located in the territory), began to lack places and resorted to random draws.
The consequences of these dysfunctions were dramatic during APB's last year in 2017: at the end of the main phase, only 541,204 applicants had received an admission offer and 92 bachelor's programs (some of which were labeled as "green dot", meaning without place shortage problems) finally had to resort to random draws among priority applicants \cite{CourDesComptes2017_APB}.\\

Parcoursup, launched in 2018, replaces APB with significant modifications: the elimination of random draws (replaced by selection in bachelor's programs with available places), the integration of more programs (including political science institutes, medical-social schools, and certain selective bachelor's degrees), and a change in the choice process. Applicants no longer rank their preferences in advance but instead decide "as they go" by accepting or rejecting received offers, which significantly slows down the assignment process. Scholarship quotas are also implemented in that same year to promote social diversity in selective programs. According to \citeA{frouillou2020have}, the introduction of the option for non-selective programs to rank applicants in order of preference has encouraged the adoption of recruitment practices similar to those of selective programs, notably by ranking students according to academic level.

\subsubsection{Inequalities in access to higher education in France}

Despite the quasi-free tuition fees in French public higher education, access to this level of education remains marked by profound social inequalities. Especially, the analysis of access to Grandes Écoles, considered the most prestigious institutions in the French educational system, reveals particularly marked inequalities. Among students whose parents belong to the lowest income decile, only 33\% access higher education, whereas this proportion reaches 90\% for students from the highest income decile. These disparities persist and intensify at higher education levels: approximately 35\% of students from the highest income decile attain a Master's level, compared to only 5\% for the lowest decile \cite{bonneau2022unequal}. Similar proportions are observed for access to elite programs (CPGE and Grandes Écoles).\\

The \textit{Institut des Politiques Publiques} report \cite{bonneau2021quelle} highlights a triple segmentation by gender, social origin, and geographical origin. Boys, students from highly privileged backgrounds, and those from the Île-de-France region (Paris region) benefit from privileged access to Grandes Écoles.
In 2016-2017, highly privileged students represented 64\% of Grandes Écoles enrollment, while they constitute only 23\% of the 20-24 age population. Conversely, students from disadvantaged backgrounds represent only 9\% of these institutions' enrollment, whereas they constitute 36\% of young people in this age group in the general population. It is worth noting that these orientation differences are only partially explained by academic performance gaps, suggesting the existence of differentiated behavioral factors or possible discrimination in recruitment processes.
This overrepresentation of privileged students reaches extreme levels in the most prestigious institutions. According to \citeA{benveniste2023like}, access to the most elite schools presents an even more unequal character: for students from ten of the most prestigious Grandes Écoles in France, children of ex-graduates have 80 times more chances of being admitted to these institutions.\\

Academic guidance inequalities largely precede entry into higher education. \citeA{schneider2023orientation} reveals that, for the generation of students who entered first grade in 2011, 92.9\% of executives' children request orientation to general and technological tracks in high school compared to only 53.1\% of unskilled workers' children. This track constitutes the privileged pathway to long and prestigious studies, as opposed to alternatives represented by vocational training or ``CAP". This early stratification of pathways partially explains the persistence of inequalities observed at higher education entry.\\

Even when restricting the analysis to high school students who have reached their final year, college choice inequalities persist in the transition to higher education. \citeA{terrier2023confidence} demonstrates that in 2021, girls and students from disadvantaged social backgrounds formulated less ambitious wishes than their male counterparts from privileged backgrounds, even at equivalent academic levels.

\subsection{Quotas in \textit{Parcoursup}: a large-scale affirmative action policy}
\subsubsection{How quotas work}

The quota system prevents scholarship recipients from being concentrated at the bottom of admission waiting lists, ensuring they receive fair consideration and equal opportunities. Originally designed for public institutions, quotas have been progressively extended to private institutions through agreements with the Ministry of Higher Education. This system operates exclusively during Parcoursup's main phase, not the supplementary phase.\\

The process unfolds in two stages. First, institutions determine which students will appear on the call list (this step is omitted for so-called "non-selective" Bachelor's programs). Second, institutions rank students according to their preferences. An automated algorithm systematically verifies quota compliance at each position $n$ in the call list by checking whether sufficient scholarship recipients appear between positions one and $n$. When the constraint is violated, the algorithm identifies the first scholarship recipient positioned after the applicant $n$ and promotes them to the position $n$, consequently demoting the original applicant in this position by one rank.\\

Consider a 25\% quota requirement. The algorithm first verifies that among the top 1 applicant, at least 25\% are scholarship recipients (i.e., the first applicant must be a scholarship recipient). If not, it places a scholarship recipient at the top of the call list. Next, it examines the top 5 applicants to ensure that at least 25\% are scholarship recipients (i.e., a minimum of two scholarship applicants among the first five applicants). If this condition fails, it promotes the first scholarship recipient ranked below the 5th position (e.g., 7th position) to the 5th position. This iterative process continues until either the list ends or no scholarship recipients remain available for promotion.\\

Developed by computer scientists Hugo Gimbert and Claire Mathieu, this open-source algorithm is documented (only in French) in official Ministry publications \cite{MESR2020Parcoursup}.

\begin{quote}
Let us consider a ranking group and denote by $q_B\%$ the minimum rate of scholarship holders in the group ($q_B$ ranging from 0 to 100). The algorithm guarantees a minimum proportion of scholarship holders among the applicants $C_1, \dots, C_n$ listed in the calling order: for every $k$, either at least $q_B\%$ of the applicants $C_1, \dots, C_k$ are scholarship holders; or none of the applicants $C_{k+1}, \dots, C_n$ are.

While the algorithm guarantees a minimum rate of scholarship holders being called, it cannot guarantee a minimum rate of them being admitted, since each applicant—including scholarship holders—is free to decline any admission offer.

The algorithm builds the calling list, starting from the academic ranking and promoting certain scholarship holders within it.

\textbf{Example.} Assume a minimum rate of 25\% of scholarship holders and an academic ranking $C_1, C_2, B_1, C_3,~~ C_4, C_5, B_2, C_6, ~~C_7, C_8, C_9, C_{10},~~ C_{11}, C_{12}, C_{13}, C_{14},~~ B_3, C_{15}, C_{16}, C_{17}$. The computed calling order becomes:\\ $B_1, C_1, C_2, C_3,~~B_2, C_4, C_5, C_6,~~ B_3, C_7, C_8, C_9, ~~C_{10},C_{11}, C_{12}, C_{13},~~ C_{14}, C_{15}, C_{16}, C_{17}$.

\hfill --- \textit{French Ministry of Higher Education and Research, 2020}
\end{quote}

The algorithm exhibits several notable properties. Scholarship recipients can only maintain or improve their ranking; they can never lose positions. Institutions retain incentives to rank applicants based on genuine preference, despite being aware of quotas. While institutions might strategically exclude scholarship recipients from waiting lists to prevent their promotion, they lack access to scholarship status information, rendering such strategies inoperable.\\

Crucially, these quotas are automatic (institutions do not need to calculate quotas themselves), mandatory (at least for public institutions), and apply specifically to call lists and not to final admission. If scholarship recipients offered admission through the quota system ultimately choose other institutions, the final admitted cohort may contain no scholarship recipients despite the presence of scholarship applicants on the call list and quotas.\\

The scholarship quota system coexists with other quota mechanisms based on baccalaureate tracks. BTS programs reserve spots for vocational baccalaureate holders, while DUT programs reserve places for technological baccalaureate holders. Unlike scholarship quotas, these constitute reserved places or "hard quotas". Originally in APB, these quotas were strengthened under Parcoursup. Non-selective undergraduate programs implement geographic quotas: when capacity is insufficient, students from the same academy as the target university receive priority. Quotas are set by regional rectors, with exceptions for boarding students, areas where certain bachelor's degrees are lacking (e.g., overseas territories - DROM-COM), and Île-de-France (which spans three academies). These quota systems fall outside the scope of this analysis.

\subsubsection{Quota calculation for each program}

Each program is assigned a different quota. In 2018, these quotas were set at the discretion of the academic rector for each concerned program. Starting in 2019, a calculation rule was imposed on all public programs (excluding apprenticeship programs) and many private programs, based on the rate of scholarship recipients among applicants in the program, with an additional two percentage points added. For example, if in a program $i$, 10\% of the applicants are scholarship recipients, then the quota is 12\%. Furthermore, there is a "floor" quota of 5\%:

\begin{equation}
    q_i = \max(\{\text{p}_{\text{scholarship recipients}, i} + 0.02; 0.05\}) \times 100
\end{equation}

\noindent The annual report of the Comité Ethique et Scientifique de Parcoursup \cite{CESP_Parcoursup_Rapport2_2020}, an independent committee composed of inspectors and researchers responsible for evaluating Parcoursup, indicates that this rule was well followed from 2019 onwards and allowed for real increases in quotas:

\begin{quote}
\textit{To be able to compare with 2018, let us consider programs with an admission capacity of at least 20. The result is clear: – in 2018, 51\% of programs had applied a scholarship recipient quota lower than the proportion of scholarship recipients among applicants. – In 2019, only 2\% of the programs applied a scholarship recipient quota lower than the proportion of scholarship recipients among applicants. It should be added that 86\% of the programs followed the ministry's recommendation by going beyond (+2 points). Moreover, except for a few private institutions, all programs adopted a quota of at least 5\%.}

\hfill --- \textit{Extract from the annual report of CESP, 2020}
\end{quote}

\noindent This quota system, to my knowledge, is unique. First, this means that quotas are calculated endogenously: the more a program attracts scholarship recipient applicants, the higher the scholarship recipient quota will be. Corollarily, some programs, which are already heavily attended by scholarship recipients (BTS), have a higher quota amount than the elite programs (CPGE). Ultimately, the main effect of quotas is to ensure that admitted scholarship recipients are not underrepresented \emph{compared to the proportion of scholarship recipient applicants}.

\subsubsection{Salience and display of quotas}

In 2018 and 2019, scholarship quotas existed, but their values for each program were not publicly disclosed. Instead, on each course implementing quotas, students could see a paragraph stating that scholarship quotas were in place, without knowing the exact value. Starting in 2020, the previous year's scholarship quota was displayed on each program page, making it visible to students. The current year's quota was not displayed because it is calculated based on the number of scholarship applicants, information that is not available until the end of the application phase.\\

\noindent Screenshots of quotas announcement on each Parcoursup program pages before 2019 ("opaque” condition) and from 2020 onwards ("transparent" condition), and their English translations, are presented in Appendix \ref{appendix:parcoursup_screenshots}.

\subsubsection{The effects of Parcoursup quotas}

Two institutional reports have attempted to assess the effects of Parcoursup quotas on scholarship students. \citeA{courdescomptes_ore_2020} concludes that there is a positive psychological effect for scholarship students, but no effect on student admissions.\\

In contrast, the report by the Comité Ethique et Scientifique de Parcoursup \citeA{cesp2021} concludes that there is a very strong effect on the extensive margin (large effect on the higher education continuation rate). For various reasons, I find that the statistical methods employed by the two institutions are inadequate and do not provide a reliable or unbiased causal estimate. The methodology of both reports is discussed in Appendix \ref{appendix:methodology}.\\

It is worth noting that a working paper by \citeA{remigereau2024impact} also examines the impact of scholarships on applications, revealing a positive effect on applications in selective programs for high-achieving males who are scholarship recipients. However, this refers to the effect of scholarships before the introduction of quotas.

\subsection{Previous evaluations of quotas}

Countries worldwide have implemented affirmative action (AA) policies to increase access to higher education. This review examines various policy designs and their effects, providing context for understanding quota-based systems, such as Parcoursup.

\subsubsection{Policy Design Variations}

\paragraph{Target Groups and Criteria} 
Affirmative action policies target different groups using various criteria. AA are traditionally \textbf{race-based programs} that use ethnic identity as the primary criterion, as seen in pre-2023 US university admissions and India's caste-based reservations. \textbf{Race-neutral alternatives} rely on socioeconomic factors like family income, geographic location, or school type. After the ban on race-based preferences in Texas \cite{hopwood1996}, local governments implemented the Top 10\% Rule, guaranteeing university admission to top-performing students from each high school \cite{Long2004, Cullen2013}, thereby including the best students from disadvantaged high schools. France's Parcoursup quotas are race-neutral, focusing solely on financial need through scholarship eligibility. Some countries combine approaches: Brazil's \citeA{brasil2012} uses both racial and socioeconomic criteria, while India now reserves places for both caste groups \cite{Bertrand2010} and economically disadvantaged students since 2019.

\paragraph{Implementation Methods}
Policy mechanisms vary significantly across countries. The US Supreme Court banned explicit quotas in 1978 \cite{regents1978}, and predetermined bonus points in 2003 \cite{gratz2003}, leading universities to adopt "holistic review" processes that considered race among many factors until a ban in 2023 of every kind of racial preferences \cite{students2023}. In contrast, India and Brazil use explicit quotas: India reserves 49.5\% of university places for specific groups (with some local state variations), while Brazil reserves 50\% for quota beneficiaries. France uses a unique "call quota" system. In this system, a minimum percentage of offers must be made to scholarship students. Still, it doesn't guarantee final enrollment outcomes (their quota is filled even if all scholarship students refuse enrollment).

\paragraph{Transparency and Communication}
Policy transparency affects student behavior and outcomes. US holistic review processes were deliberately opaque, while Brazilian and Indian systems explicitly display quota information on application platforms, since students must explicitly fill out questions about their ethnicity and are redirected to a special procedure on the application platform when they're eligible \cite{sisu-duvidas,nta_jee_main_2024}. France's approach evolved from vague mentions in 2018-2019 (the existence of quotas is mentioned on every program page, but not the specific amounts) to displaying the specific quota amounts for each program from 2020 onward.

\subsubsection{Effects on Minority Students}

\paragraph{Enrollment Outcomes} Affirmative action significantly affects minority enrollment patterns. Many studies show that AA state-level bans decrease the percentage of minority students in elite universities \cite{card2005would, Hinrichs2012, Backes2012,bleemer2022, yagan2016supply}. However, \citeA{Backes2012} and \citeA{Hinrichs2012} insist on the fact that there is no or very little evidence of effect on the total enrollment of minority students: there is a clear intensive margin effect, but only limited evidence for an extensive margin effect. In reaction to these bans, universities use non-racial information that is predictive of race to apply indirect affirmative action. Still, these systems are poorly targeting "good" minority students and compensate only partially for the fall in admission rates in elite schools for minority students \cite{antonovics2014effect_quality, Long2008}. Race-neutral alternatives, such as percentage plans, don't fully compensate for the effects of traditional AA \cite{Long2004}. \citeA{Daugherty2014} finds that there is no effect of the Texas Top-Percent Rule on the quality of the college attended by minority students. \citeA{BLEEMER2023104839} concludes that traditional quota-based AAs have a much greater effect on the admission of minority students than "holistic review", which is itself much more effective than Percent Plans. Finally, \citeA{Cullen2013} proves that this policy implies compositional change effects, with some good strategic students (or parents) entering (less competitive) disadvantaged schools to benefit from the Percent Rule. AA policies outside the US (and not using bans as an exogenous shock) also show a positive effect on enrollment in selective universities \cite{bagde2016,otero2021affirmative, Bertrand2010}.

\paragraph{Application Behavior} The evidence on how minority students' application behavior reacts to AA is very mixed. Some studies on AA bans suggest that the decline in admission rates of minority students is driven by college behavior, and that the application behavior of minority students remains stable before and after the ban \cite{card2005would} or negative effects concentrated on students unlikely to get admitted after the ban \cite{yagan2016supply,antonovics2013were}. \citeA{dickson2006does} found little negative effect on the extensive margin (applying to a college) for minority students. However, bans are a very specific type of exogenous shock, and one can imagine asymmetric effects, where the behavioral reaction is different when an AA is implemented rather than when it is removed (ratchet effect). Using the reinstatement of AA in some states, \citeA{akhtari2024} demonstrates a positive effect of AA reinstatement on the application of minority students, especially in the top half of the test score distribution.

\paragraph{Pre-College Investment} The theoretical effects of AA on minority students' motivation to invest in their skills before college (getting good grades, etc.) are ambiguous. However, empirical literature finds no clear negative effects. \citeA{antonovics2014effect} finds little evidence of a negative effect on the academic gap between majority and minority students at the pre-college level following the AA ban, thus implying a positive motivational effect of AA. In Brazil, \citeA{Francis2012} finds no negative effect on pre-college preparation of minority students. \citeA{strifezzi2023}, \citeA{estevan2019redistribution} and \citeA{akhtari2024} find a positive effect on pre-college human capital.

\paragraph{Long-term Outcomes} A much-discussed theory in the economics of education is the "mismatch hypothesis", which argues that underrepresented minority students admitted through AA are too weak to benefit from overly selective universities, and that they will fail in higher education, or at least not experience long-term benefits. However, a fairly recent literature falsifies this hypothesis. \citeA{bleemer2022} and \citeA{Antman2024} both conclude that AA bans lead to a decrease in educational attainment and labor market outcomes for underrepresented minority students, with some heterogeneity, supporting the idea that AA has positive long-term outcomes. Outside the US, \citeA{otero2021affirmative} and \citeA{Bertrand2010} both find long-term positive effects on income for minority students, while \citeA{bagde2016} finds no evidence of a "mismatch". In contrast, the effects on overall students (minority students who gain a new admission and majority students who potentially lost their admission) are less clear, with \citeA{bleemer2022} finding that positive effects on minority students exceed the negative effects on majority students, \citeA{otero2021affirmative} finding no effect on overall efficiency, and \citeA{Bertrand2010} finding a negative effect on majority students displaced due to AA, that largely offsets the positive effect on minority students.

\subsubsection{Gap in the literature}

\paragraph{Race-Neutral Alternative} I have observed that research on race-neutral alternatives primarily focuses on x-percent rule policies. The French context offers unique insights, as quotas based on socioeconomic status represent a large-scale natural experiment in race-neutral affirmative action implemented as a primary policy rather than a substitute. However, there is a limited rigorous evaluation of Parcoursup's effectiveness in promoting diversity compared to race-based AA.

\paragraph{Salience and complexity} The role of information and salience in AA effectiveness remains underexplored, particularly regarding how policy introductions create different informational dynamics than policy eliminations. To the best of my knowledge, no research has examined how different communication strategies, transparency levels, or the complexity of application processes affect student behavior during policy implementation. The contrast between transparent quota systems (Brazil, India) and opaque holistic review (pre-2023 US) suggests that policy design characteristics beyond the substantive rules may significantly influence behavioral outcomes. The French context again offers a unique opportunity to compare the "salient" (explicit quotas, Brazil/India) and "opaque" (holistic review, USA) conditions, as the quota system was initially opaque in the early years 2018-2019 (existing quotas but non-public value) before being displayed in a more salient way (public quotas) in 2020.

\paragraph{A unique kind of quotas}

A critical gap exists in understanding "call quotas" (\textit{quotas sur les appels}) versus final enrollment quotas, particularly in systems like Parcoursup, where offers must be made to target populations without guaranteeing final enrollment. This distinction has important implications for policy effectiveness, as the relationship between offers and final enrollment may vary systematically across different groups. \citeA{bechichi2021segregation} provides initial evidence on the introduction of Parcoursup in general, showing that the platform didn't reduce segregation in the first years of implementation. However, no empirical analysis focuses specifically on the "call quotas". Additionally, there is limited research on the endogenous determination of quotas. To the best of my knowledge, my work is the first study to evaluate an affirmative action policy where quotas are directly calculated endogenously from the proportion of scholarship applications in the program. \citeA{couto2021parcoursup} assumes that such quotas reinforce segregation by reproducing initial inequalities in distribution. I could, for example, hypothesize that these quotas send the wrong signal to scholarship students by encouraging them to apply to places where there are the most quotas, that is, initially, the places where there have historically been the most scholarship students, then reinforcing segregation.

\section{Hypothesis\label{hypothesis}}

The institutional reports about Parcoursup \cite{courdescomptes_ore_2020,cesp2021}, as well as evaluations of other affirmative action programs, enable us to formulate several research hypotheses regarding the impact of quotas on scholarship students. Overall, I can make hypotheses about the admission of scholarship students, their application behavior, and their academic performance.

\subsection{Effect on final enrollment of scholarship applicants}

\noindent \label{h1a} \textbf{H1a. Extensive margin – Access to higher education} The introduction of quotas affects the rate at which scholarship applicants pursue higher education.\\

\noindent \label{h1b} \textbf{H1b. Intensive margin - Prestige of the program in which the individual is finally admitted} The introduction of quotas affects the prestige of the program in which scholarship recipients are admitted, conditional on having been admitted to a program.\\

\noindent  \label{h1c} \textbf{H1c. Exploratory - Type of program in which the individual is finally admitted} The introduction of quotas affects the type of program in which the scholarship recipient is admitted. This hypothesis is exploratory, and I have no precise prediction about the direction of the effect or which type of program would be most affected.\\

\subsection{Effects on the behavior of scholarship applicants}

\noindent  \label{h2a}  \textbf{H2a. Extensive margin - Participation in Parcoursup} The introduction of quotas affects the probability of being an “active” applicant on Parcoursup, i.e., having made at least one application.\\

\noindent  \label{h2b} \textbf{H2b. Intensive margin - Prestige of the programs to which the student applies} The introduction of quotas affects the level of prestige of the programs to which scholarship recipients apply, conditional on having made at least one application. In other words, the introduction of quotas motivates individuals to apply to more prestigious programs.\\

\noindent  \label{h2c} \textbf{H2c. Exploratory - Type of program in which the individual applies} The introduction of quotas affects the type of program in which scholarship recipients apply. This hypothesis is exploratory, and I still have no precise prediction about the direction of the effect or which type of program would be most affected.

\subsection{Effects on pre-college academic performance of scholarship applicants}

\noindent  \label{h3a} \textbf{H3a. Pre-college academic performance} The introduction of quotas has an effect on the academic level at the end of the school year for scholarship applicants. 

\section{Data\label{data}}

\subsection{Data sources}

To test the research hypotheses, I utilize data from the APB (until 2017) and Parcoursup (from 2018) platforms. I combine raw data, anonymized by the Ministry of Higher Education but taken directly from the application \cite{casd_apb_2014_remontees,casd_apb_2015_remontees,casd_apb_2016_remontees,casd_apb_2017_remontees,casd_parcoursup_2018_remontees,casd_parcoursup_2019_remontees,casd_parcoursup_2020_remontees}, and data that have been pre-treated by the Ministry of Higher Education \cite{casd_apb_2014_bases,casd_apb_2015_bases,casd_apb_2016_bases,casd_apb_2017_bases,casd_parcoursup_2018_bases,casd_parcoursup_2019_bases,casd_parcoursup_2020_bases}. Data are provided by the Centre d'Accès Sécurisé aux Données (CASD), a French secure server for sensitive data. These data ensure a certain degree of comparability. As the Parcoursup information system uses the same architecture as APB, the tables used are similar. However, some fields may have been renamed, and some nomenclatures have changed, which required standardization work.\footnote{For example, the PCS nomenclature (used to categorize the socio-professional category of the parents) required standardization work, as the version used in APB was the 2003 nomenclature, and the version used in Parcoursup is the 2020 nomenclature.} Methodological details are specified in Appendix \ref{appendix:methodological_details}.

\subsection{Operationalization of outcomes}

\noindent The overall idea is to apply analysis to different outcomes, each serving to answer a different hypothesis. Here, I provide the operational definition of each of the outcomes used.

\subsubsection{Binary outcomes and academic performance}

\begin{description}[style=nextline]

    \item[\textbf{Admitted via Parcoursup (H1a)}]%
    A binary variable indicating whether an applicant received and accepted a final admission offer (i.e., has been enrolled) to a program via Parcoursup, either in the main or complementary phase. I do not verify whether the student was registered at the start of the school year, but only whether they have definitively accepted a program on Parcoursup. I want to emphasize that for this variable to be 1, the individual must have received an offer of admission \emph{and} accepted it. Additionally, if the individual did not pass the baccalaureate, they are automatically withdrawn from all their choices, so the variable is set to 0.

    \item[\textbf{Admitted in... X (H1c)}]%
    Indicates whether a Parcoursup applicant was finally enrolled in a program of type \textit{X}. Tested program types include: Bachelor's (Licence), DUT, BTS, CPGE, and Engineering Schools. Business schools, political science institutes, and social/paramedical schools are excluded as they were not on the previous APB platform. Therefore, multiple binary outcomes are tested for the hypothesis.

    \item[\textbf{Submitted an application (H2a)}]%
    Indicates whether an applicant submitted at least one application (wish) on Parcoursup.

    \item[\textbf{Submitted an application in... X (H2c)}]%
    Indicates whether a Parcoursup applicant submitted at least one wish in a program of type \textit{X}. The program types included are the same as defined above: Bachelor's, DUT, BTS, CPGE, and Engineering Schools. Again, multiple binary outcomes are tested for a single hypothesis.

    \item[\textbf{Baccalauréat Grade (H3a)}]%
    Score obtained in the standardized French high school final exam, used as an objective measure of academic achievement. Only the main session score is used, even in the case of resits. Although not used in the main admission phase, it may reflect motivational effects from earlier performance. It is preferred over the annual GPA (grades obtained in high school during the school year) due to its greater standardization.

\end{description}

\subsubsection{Quantifying academic prestige}

The hypotheses \textbf{H1b} and \textbf{H2b} investigate academic program prestige based on the characteristics of admitted students. Following \citeA{terrier2023confidence}, prestige is measured using the average baccalaureate scores of admitted students. I first compute each program's mean percentile rank of admitted students, then normalize these scores uniformly between 0 and 100, weighted by program size. A score of 99 indicates a top 1\% program in terms of selectivity. More formally, for each program $i \in \{1, 2, \ldots, I\}$, I first compute the mean percentile rank of finally enrolled students:

\begin{equation}
\bar{P}_i = \frac{1}{n_i} \sum_{j=1}^{n_i} p_{ij}
\label{eq:mean_percentile}
\end{equation}

\noindent I then normalize these mean percentiles across all programs to obtain the prestige score:

\begin{equation}
\text{Prestige}_i = \left( \frac{\sum\limits_{j:\ \bar{P}_j \leq \bar{P}_i} N_j}{\sum\limits_{k=1}^{I} N_k} \right) \times 100
\label{eq:prestige_pondere}
\end{equation}

where:
\begin{align*}
p_{ij} &= \text{baccalaureate percentile rank of student } j \text{ admitted to program } i \\
N_i &= \text{number of students finally enrolled in program } i \\
\bar{P}_i &= \text{mean baccalaureate percentile of students admitted to program } i \\
I &= \text{total number of programs in the sample}
\end{align*}
The idea is that your prestige score is the proportion of programs that have a lower mean percentile rank among admitted students. If you have all the best students in your program, then all the other programs have a lower mean percentile rank of students than your program ($\forall j\neq i, \bar{P_j}<\bar{P_i})$, then your Prestige is 100.\\

At this stage, each program is assigned a prestige score. From this, I construct the outcome variables for hypotheses \textbf{H1b} and \textbf{H2b}.

\begin{description}[style=nextline]
    \item[\textbf{Prestige of Final Admission (H1b)}]%
    The prestige score of the program in which the applicant was ultimately admitted and enrolled.
    
    \item[\textbf{Average Prestige of Applications (H1b)}]%
    The mean prestige score across all programs to which the student applied reflects the overall selectivity level of their application portfolio.
    
    \item[\textbf{Maximum Prestige of Applications (H1b)}]%
    The prestige score of the most selective program to which the individual applied, capturing their most "ambitious" application.
\end{description}

\noindent Note that hypothesis \textbf{H1b} is tested through two distinct outcomes (average and maximum prestige of applications), each subject to separate estimation. Note that in 2020, COVID-19 led to modified high school graduation average calculations, incorporating continuous assessment grades rather than relying solely on final exams. This methodological change will not affect our measurement if the programs admitting students with the highest continuous assessment grades also admit those with the highest overall graduation grades, which appears unlikely. Indeed, from one year to the next, the prestige of a given program remains very stable, without being perfectly identical. The correlation coefficient (weighted by the number of enrolments in the programs) is consistently around 0.85 (even for 2019 v. 2020, despite the small change in the high school graduation method).

\subsection{Descriptive statistics}

This analysis focuses on differences between scholarship recipients and non-recipients within the same academic track, rather than differences across tracks. Descriptive statistics (on outcomes and socio-demographic characteristics) are detailed in Appendix \ref{descriptive:general} (general track), Appendix \ref{descriptive:techno} (technological track), and Appendix \ref{descriptive: vocational} (vocational track). In each case, they are broken down by non-scholarship holders versus scholarship holders, and two different tables are produced for outcomes and socio-demographic characteristics.

\subsubsection{Revenue and eligibility}

\begin{figure}[htbp]
  \centering
  \caption{Proportion of scholarship recipients by income quintile (self-declared)}

  \includegraphics[width=1\textwidth, trim=0 0 0 0, clip]{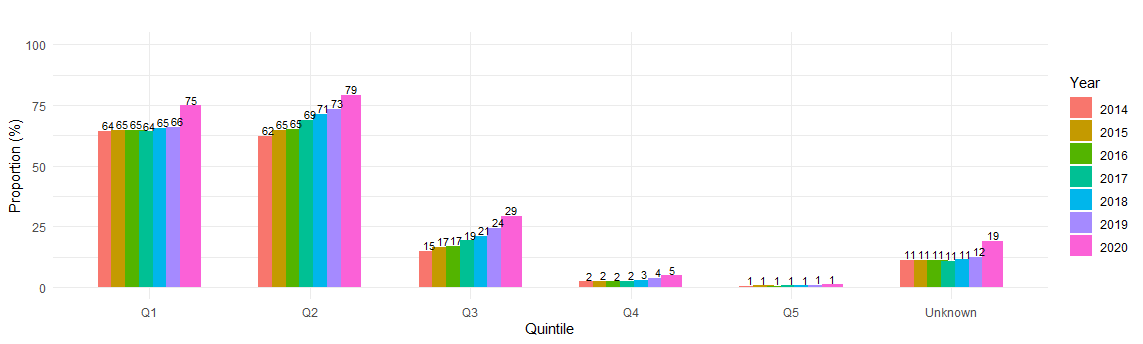}
  \tablereading{In 2018, 71\% of individuals in Quintile 2 are scholarship recipients}

  \label{proportion_revenu}
\end{figure}

Individuals have the option of providing information about their parents' taxable income and the number of siblings they have. However, this data is self-reported and should therefore be treated with caution, and cannot be used to perform RDD. \autoref{proportion_revenu} shows that individuals in Quintile 1 and Quintile 2 are the most likely to receive scholarships, and that the probability of receiving a scholarship increases each year (particularly in 2020) for these quintiles. I observe that the proportion of scholarship recipients is far from 100\% in these years, typically ranging from 60\% to 70\%, even though in the vast majority of cases, Quintiles 1 and 2 of income are supposed to be eligible for scholarships (regardless of the number of siblings). These statistics suggest that there is a “common support” of people who are supposed to be eligible but who do not apply for the scholarship. Furthermore, we observe a low proportion of people in Quintile 3 who are scholarship recipients, primarily because large families often exceed the eligibility threshold for scholarships.

\subsubsection{Socio-demographic Characteristics}

Scholarship recipients differ substantially from non-recipients across multiple dimensions. Recipients come from more disadvantaged backgrounds: in the general track, 17.9\% have a blue-collar worker as their primary legal guardian, compared to 10.5\% for non-recipients (2017). Additionally, 18\% have an inactive guardian, compared to 2.3\% for non-recipients. Single-parent households are more prevalent among recipients (25\% vs 7.2\% in general tracks). Recipients are also more likely to be female (62\% vs 55\% in general tracks) and foreign-born (12\% vs 4\% in 2017). Non-recipients attend socially advantaged schools more often than recipients.

\subsubsection{Academic Behavior and Characteristics}

\paragraph{General Track}
While recipients and non-recipients have similar admission rates to at least one program ($\approx$80\%), recipients exhibit weaker academic performance. They apply to less selective programs and are admitted to less selective institutions. Only 15\% of the recipients applied to CPGE, compared to 23\% of the non-recipients, with final admission rates of 5.9\% versus 11\%, respectively. Non-recipients are more likely to pursue the prestigious scientific specialization.
\paragraph{Technological Track}
Recipients are overrepresented in business programs (54\% of recipients go into STMG vs 47\% of non-recipients in 2017) but underrepresented in industrial programs (STI2D: 20\% vs 26\%). Despite having slightly better academic results than non-recipients, recipients make less ambitious applications. Non-recipients achieve slightly more selective admissions when admitted (e.g., 11\% reach DUT programs vs 9\% for recipients in 2017).
\paragraph{Vocational Track}
Recipients have a higher pursuit rate in higher education (47\% vs 37\% in 2017) but obtain slightly less selective placements upon admission. This difference stems partly from selective attrition: recipients are more likely to be admitted to less selective university programs (17\% vs 7.8\% in 2017), while BTS admission rates remain similar (29\% for both groups).

\subsubsection{Summary}

Scholarship recipients consistently come from more disadvantaged backgrounds. In the general track, they demonstrate lower academic performance, less ambitious application behavior, and receive less selective admissions — a pattern that persists even when controlling for academic results \citeA{terrier2023confidence}. However, in technological and vocational tracks, recipients do not show substantially worse outcomes than non-recipients.

\section{Empirical Strategy\label{strategy}}

\subsection{Periods and Study Populations}

\subsubsection{Periods}

The empirical estimation requires distinct pre-treatment and post-treatment periods. For pre-treatment periods, I use APB data from 2014, 2015, 2016, and 2017. Post-treatment periods include 2018, 2019, and 2020. The 2020 academic year presents specific considerations: baccalaureate exams were canceled due to the COVID-19 crisis, with grades calculated based on continuous assessment averages. Consequently, hypotheses related to baccalaureate performance are not tested for this year. However, the entire application phase was not directly affected by the COVID-19 crisis: the lockdown was announced only on the evening the applications closed, and I can therefore assume that the shock did not influence the applicants' behavior regarding their applications\footnote{As is the case every year, students still had the option of not confirming their choice in the following months (i.e., during lockdown), but in general, the vast majority of choices are confirmed.}.\\

I exclude the years 2021-2023 from the post-treatment analysis, despite data availability, due to the implementation of a major baccalaureate reform in 2021. This reform transformed the general track structure, as students previously distributed across three series now select from multiple specialized subjects. The reform substantially impacted student pathways, notably increasing gender inequalities in academic choices \citeA{theconversation2024reforme}. However, the reform's differentiated impact across socioeconomic gradients remains undocumented. Given the reform's massive and exogenous nature, it potentially affected scholarship recipients and the counterfactual group differently, violating the parallel trends assumption essential to our identification strategy.

\subsubsection{Study Populations}

I examine hypotheses across three distinct populations: final-year high school students in general, those in technological, and those in vocational tracks receiving scholarships. This separation reflects the tracks' divergent post-secondary pathways: vocational track graduates primarily pursue BTS programs, technological graduates target DUT programs, while CPGE students predominantly come from general tracks. Descriptive statistics reveal substantial behavioral differences across tracks, varying pre-treatment outcomes, and potentially heterogeneous reform effects. For the counterfactual group, I include only final-year students, excluding applicants seeking re-enrollment or returning to studies after interruption.\\

The effects on the prestige of the program to which the individual was admitted are only estimated on the population of students who were admitted to a program (regardless of which one) at the end of the procedure. Methodological details about the inclusion criteria are specified in Appendix \ref{appendix:methodological_details}

\subsection{Estimator}

I estimate treatment effects using an event-study design with 2017 as the reference period (the last year before treatment introduction). Pre-treatment periods compared to the reference include 2014, 2015, and 2016, while post-treatment periods cover 2018, 2019, and 2020. I employ the Doubly-Robust Difference-in-Differences (DRDiD) estimator \cite{sant2020doubly} as implemented in the \texttt{did} package based on \citeA{CALLAWAY2021200}. This estimator incorporates control variables into treatment effect estimation, helping correct potential bias from observable differences between treated and control groups. It replaces the standard unconditional parallel trends assumption with a more flexible and realistic conditional parallel trends assumption. Methodological details about the estimation are specified in Appendix \ref{appendix:methodological_details}\\

The DRDiD estimator combines two complementary approaches: (1) outcome regression of dependent variables on covariates, performed separately for each group and period, analogous to classical linear regression that directly controls for observable individual differences; and (2) propensity score weighting of control group observations, where weights reflect the estimated probability of treatment conditional on covariates. This weighting emphasizes control individuals with characteristics similar to those of the treated individuals, making the counterfactual group as comparable as possible to the treatment group. This dual mechanism provides the estimator with a double robustness property: estimation remains unbiased as long as either the outcome regression model or the propensity score model is correctly specified. This method represents a significant advantage over traditional estimators that rely entirely on the correct specification of a single model. Since most final-year students use APB or Parcoursup only once - I exclude from my sample the applicants who are switching fields or who were not enrolled precisely in their final year of high school -, I employ repeated cross-sectional data.

\paragraph{Control variables}
Control variables include: academic region, school type, school Social Position Index in 2019 based on \citeA{dauphant2023ips}, French nationality, gender on birth certificate, parental socio-professional categories using PCS-2003 from \citeA{insee2003pcs}, and baccalaureate track (e.g., Science, Economics and Social, Literature for general track). I also use parents' revenue and the number of siblings, which are the two variables that determine eligibility for treatment. Unfortunately, this data is self-reported by the students, with many errors in the self-declaration (many students received scholarships even though they declared themselves to be above the threshold), so I was unable to perform RDD. In the main specification, baccalaureate results are not included as controls, given our hypothesis that quota policies could motivate students to improve academic performance (making it a "bad control"). Individuals with missing gender or baccalaureate track category are excluded from the sample. In summary, I don't have the exact eligibility data, but I have a set of variables that are linked and potentially strongly correlated with scholarship eligibility status.

\paragraph{Linear approximation}
For binary variables, I use linear approximations, which provide good proxies when the mean values are distant from the bounds of 0 and 1. For outcomes with pre-treatment mean values below 2\% or above 98\%, no analysis is conducted due to approximation limitations and package constraints that prevent the use of Logit-DiD estimation.

\paragraph{ITT or ATE?} The empirical method poses the following question: What constitutes treatment? Is it the fact of having a scholarship, or is it the fact of having changed final enrollment thanks to quotas? For the hypotheses concerning application, I assume that all scholarship applicants are "treated" as all of them were aware of the quotas on the program pages. The hypotheses accordingly relate to the ATE effect. The same reasoning applies to academic results. For final enrollment, if I assume that there is no behavioral effect and that scholarship students did not change programs due to the mechanical effect of quotas, I cannot say that they were marginally treated. However, I cannot determine who changed programs due to quotas, so I can only identify an ITT effect.

\subsection{Assumptions}

This section discusses the various assumptions necessary for estimation and inference to be valid.

\label{section:assumptions}

\paragraph{Conditional Parallel Trends}
The DRDiD estimator relies on the conditional parallel trends assumption: in the absence of treatment, the difference in outcomes between treated and control groups would have evolved similarly over time, conditional on observed covariates. This assumption is more plausible than the unconditional parallel trend, as it allows for systematic differences between groups that remain constant over time after controlling for observable characteristics. Verifying the existence of pre-trends over the three pre-treatment periods (excluding the reference period) will enable us to confirm the plausibility of the hypothesis. To summarize, the validity assumptions are the same as \citeA{sant2020doubly}. Furthermore, the HonestDiD robustness tests of \citeA{rambachan2023more} allow this assumption to be partially bypassed, if the bias remains within a certain range of the estimated effect during pre-treatment years.

\paragraph{No Compositional Change} 
Using repeated cross-sectional data rather than panel data requires the additional assumption of no compositional change: scholarship recipients in 2019 must represent the same population as scholarship recipients in 2017. This assumption is questionable: Scholarship recipient numbers grew substantially (8\% annually vs. 2\% for non-recipients in general tracks), with 11.9\% growth between 2017-2018 versus 4.8\% for non-recipients (see \autoref{principal:evolution_scolarship})\footnote{Note that the scholarship allocation campaigns in 2018 were conducted before the quotas were announced, which suggests that the increase in 2018 is not linked to parents' anticipation of the quotas' effect}. Using Haussman tests, I show in Appendix \ref{section:compositionnal_change} that, while there is indeed a compositional change, it is very negligible for students in all categories of high school students, especially for students in the general track. It would still be advisable to use the correction proposed by \citeA{santanna2025differenceindifferencescompositionalchanges} when it becomes available. 

\begin{table}[h!]
\centering
\caption{Evolution of scholarship and non-scholarship student numbers from 2014 to 2020}

\begin{tabular}{lccccccc}
\hline
\textbf{Session} & \textbf{2014} & \textbf{2015} & \textbf{2016} & \textbf{2017} & \textbf{2018} & \textbf{2019} & \textbf{2020} \\
\hline
\textbf{Scholarship} & & & & & & & \\
Value N & 46,626 & 49,622 & 52,212 & 57,419 & 64,676 & 67,977 & 73,883 \\
Growth (\%) &  & 6.23 & 5.09 & 9.51 & 11.90 & 4.98 & 8.33 \\
\hline
\textbf{Non-scholarship} & & & & & & & \\
Value N & 271,424 & 279,947 & 289,446 & 300,725 & 315,238 & 307,973 & 305,493 \\
Growth (\%) & & 3.14 & 3.39 & 3.90 & 4.83 & -2.30 & -0.81 \\
\hline
\end{tabular}

\noindent \tablereading{In 2016, there were 52,212 scholarship recipients in their final year of high school applying on Parcoursup. In 2017, there were 57,419 scholarship recipients, representing a 9.51\% increase.}
\label{principal:evolution_scolarship}
\end{table}
\paragraph{No Selective Attrition}
I assume no selective attrition in the sample inclusion. Specifically, I suppose that treatment doesn't influence the probability of registering on Parcoursup (rather than just applying once registered). I am unable to observe this effect, as our data only includes registered users. However, I think the risk of selective attrition is limited, at least for general and technological tracks students: using public aggregated data, I found out that for general track and technological track, the rate of being on Parcoursup when you're taking the final exam at the end of high school is respectively up to 97\% and 92\%, and that these rate are stable over time between 2016 and 2019 (no sudden increase in 2018 when Parcoursup arrives). Calculations are detailed in Appendix \ref{appendix:selective_attrition}.

\paragraph{Individual Assignment}
This assumption concerns inference rather than estimation. To justify not clustering standard errors, treatment must be assumed to be assigned at the individual level \cite{rambachan2025design}. By reweighting the control group based on the probability of receiving treatment, we ensure that the counterfactual group consists of individuals most likely to be eligible for the scholarship. Consequently, receipt of the scholarship — conditional on eligibility — is mainly driven by the individual's decision to apply, rather than by belonging to a specific social or demographic group. It is therefore crucial to determine whether this decision is truly idiosyncratic — i.e., varying at the individual level — or whether institutional factors, such as high school-level practices, largely shape it. If the latter holds, the relevant unit of assignment would be the school, and standard errors should be clustered at the high school level, accordingly to \citeA{rambachan2025design}. Several elements support the individual-level assignment assumption. Scholarship awareness campaigns are disseminated through diverse channels (e.g., ministry announcements, online platforms, school outreach) and at various educational stages — often during middle school, at a separate institution. While these campaigns are officially standardized across public high schools, variation in actual practices remains plausible. To empirically assess the assumption, I estimate mixed-effects models and find that, conditional on eligibility, the high school attended has no statistically significant effect on the probability of receiving the scholarship. This suggests that, among likely-eligible individuals, treatment assignment is not determined by school. Full details of this analysis are available in Appendix \ref{section:treatment_assignement}. The variation in scholarship application behavior among eligible students thus likely reflects genuine individual-level treatment assignment within a population of similarly eligible candidates. Each individual $i$ in the treatment and control groups has a probability $p_i$ of receiving the scholarship, based on personal characteristics. Although $p_i$ is a complex, non-random, and unknown function, \citeA{rambachan2025design} shows that this does not hinder inference, provided assignment is indeed at the individual level.

\paragraph{Weak Monotonicity}
Suppose I observe a strong effect on the extensive margin of being admitted to a program on Parcoursup. In this case, I can no longer determine whether observed effects on the prestige of obtained admissions (assessed only among admitted students) or on program type truly reflect the intensive margin, as the treatment population now includes "marginally admitted" students compared to the counterfactual. Under weak monotonicity, I assume treatment cannot reduce the probability of scholarship recipients being admitted to at least one program. Formally, for any scholarship recipient $i$:

$$\tau_i = Y_i(1) - Y_i(0)\geq 0$$

This assumption, similar to a "no defier" assumption in encouragement design, ensures that if I observe no extensive margin effects, the population on which I measure intensive margin effects remains identical to the counterfactual. If we measure no extensive margin effect, the assumption would then be highly credible, since quotas can only improve the positions of scholarship recipients, never worsen them. A violation could occur through highly heterogeneous behavioral effects where treatment simultaneously encourages some students to apply while discouraging others to such an extent that aggregate admission rates remain identical despite different underlying populations. However, such massive, perfectly offsetting compositional changes are implausible in this context, notably because there are no theoretical reasons that quotas discourage people from applying\footnote{And as I would see in the Results section, I measure no behavioral effect.}.

\paragraph{No Spillover Effects}
Our difference-in-differences strategy relies on the SUTVA assumption — that is, no spillover effects on non-recipients. In the case of application behaviors, if quotas incentivize recipients while discouraging non-recipients, estimated effects may be overstated. The results on application outcomes should therefore be interpreted as upper bounds. For admissions, spillovers are expected: scholarship recipients admitted to selective programs displace others. However, given their low share in general tracks, negative ITT effects on non-recipients are likely much smaller than the positive effects on recipients. The resulting estimate — a mix of both — may thus be slightly upward-biased, but the distortion is limited. Still, the extent of spillovers among the counterfactual group is uncertain. Since control individuals are reweighted under the DRDiD framework, those with the highest weights (i.e., those most similar to treated individuals) may also be more exposed to spillover effects. A cautious interpretation is therefore that the estimated effect represents an upper bound of the average treatment effect (ATE). To assess the potential increased bias due to reweighting, I replicate a DiD analysis using the same controls but without reweighting, focusing only on robust results identified during my estimation. This approach assigns equal weight to all non-recipients, thereby limiting exposure to negative spillovers in the ITT estimation since the low proportion of scholarship recipients compared to the other students. Results are shown in \autoref{g_reg:event_study}. Compared to the main findings in Section \ref{results}, the estimates are slightly lower but of similar magnitude, suggesting that DRDiD reweighting only modestly inflates the estimated effect.

\subsection{Robustness checks}

\paragraph{Multiple Hypothesis Testing}
Since some hypotheses are tested using multiple outcomes, I apply corrections within predetermined outcome families based on logical and conceptual criteria. For each family of outcomes, I apply Holm's sequential Bonferroni correction \cite{abdi2010holm} due to its simplicity in implementation. I group outcomes into five families, following \citeA{rubin2021adjust} guidelines : 

\begin{enumerate}
    \item Aggregate admission outcomes ("Admitted in any program" and "Prestige of final enrollment"), measuring quota effects on final admission ; 
    \item Exploratory admission outcomes (all "Admitted in ... X") addressing hypothesis \textbf{H1c} regarding admission program types ;
    \item Aggregate application outcomes ("Submitted an application", "Average prestige of applications", and "Maximum prestige of applications") measuring application behavior effects;
    \item Exploratory application outcomes (all "Submitted an application in ...X") addressing hypothesis \textbf{H2c} regarding application program types;
    \item Academic performance (baccalaureate grade). 
\end{enumerate}

Note that, for families 2 and 4, it's clear that I have a \textit{disjunction testing}:  rejection of the joint null hypothesis only requires one significant test result. For example, if only one of the estimates is significant in Family 2 (i.e., the probability of applying to a CPGE-type program increases), then I could confirm the joint hypothesis \textbf{H2c} (i.e., the types of programs where people apply differ). Thus, following the guidelines of \citeA{rubin2021adjust}, I must use a correction for each of these families. For families 1 and 3, the outcomes measure different hypotheses (except for the maximum and mean prestige of applications, which measure the same construct), and I am grouping hypotheses on the extensive and intensive margins. Nevertheless, I have grouped them because the concepts measured remain closely related: Family 1 indicates whether there has been an aggregate effect on enrollment (on the intensive \textbf{or} extensive margin). Family 3 indicates whether there has been an aggregate effect on behavior (intensive or extensive margin), and I prefer to be overly conservative rather than not conservative enough.\\

Corrections are applied separately for each post-treatment year and each population, yielding $5\times 3$ outcome families per population. Why don't I consider that the same outcome tested on two different populations should be grouped in the same family of hypotheses? Again, it all comes down to how the tests are interpreted. I do not make a joint interpretation: if I find an effect on enrollment in a specific population in one particular year, I will not conclude that, in general, the treatment affects enrollment prestige, regardless of the year or population. I will only say that it affects that specific population and period. We are therefore in the case of \textit{individual testing}: each statistical test corresponds to a separate hypothesis, and there is no joint hypothesis that groups together populations or periods. In this situation, \citeA{rubin2021adjust} recommends not making any corrections. 

\paragraph{Parallel Trends Robustness}
To assess robustness to potential parallel trends violations, I employ the Honest DiD method \cite{rambachan2023more}. This approach calculates the "breakdown value" $M^*$ for each significant result, which represents the maximum magnitude of parallel trends violation that remains tolerable before the effects become statistically insignificant. This breakdown value is computed by leveraging information contained in the observed pre-treatment trend violations.

\paragraph{Alternative Specifications}
For continuous outcomes (hypotheses \textbf{H1b} and \textbf{H2b}), I employ Change-in-Changes (CiC) as a non-parametric alternative. This estimator, designed by \citeA{athey2006identification} and implemented in R by \citeA{callaway2018qte}, constructs control groups that match the treatment group's outcome distribution, links deciles across groups, and tracks post-treatment deviations. This approach ensures treated individuals are compared to similar controls (in terms of outcome distribution) while maintaining identical assumptions\footnote{I only made the version without adding controls, due to insufficient computing power.}. Methodological details about the CiC estimation are specified in Appendix \ref{appendix:methodological_details}.

\paragraph{Heterogeneity Analysis}
CiC enables heterogeneous effect evaluation by tracking the evolution of each decile, providing Quantile Treatment Effects for outcomes of interest. Additionally, I conduct DRDiD analysis specifically on high-achieving scholarship recipients (in the first quartile) to detect potential effects on CPGE and Engineering School applications. I avoid additional ad hoc heterogeneity specifications to prevent data mining.

\subsection{Decision criteria}

Results are deemed significant when no significant pre-trend violations occur during pre-treatment periods and measured effects exceed 5\% thresholds adjusted for multiple hypothesis testing. For continuous outcomes, the alternative estimator (if there is no significant pre-trend) must also yield significant results. I consider results to be robust when they withstand degree-1 violations at the 5\% significance level (adjusted for multiple hypotheses).

\section{Results\label{results}}

The results of the effects estimated by DRDiD are presented in \autoref{g_sans_notes:event_study} (general track), \autoref{t_sans_notes:event_study} (technological track), and \autoref{p_sans_notes:event_study} (vocational track). To test robustness, I also present the breakdown values for each estimate, and these results are shown in \autoref{g_sans_notes:breaking_point_unified} (general), \autoref{t_sans_notes:breaking_point_unified} (technological), and \autoref{p_sans_notes:breaking_point_unified} (vocational).

\subsection{Effect on final enrollment of scholarship applicants}

\noindent \textbf{H1a. Access to higher education:} For the general track, there is a barely significant positive effect of 1.3pp in 2018. The breakdown value is 0.4, so the result is not robust. No other significant effects are observed for the three tracks in 2018 or 2019. The effects in 2020 are difficult to interpret: I observe an effect of -4.2pp in the vocational track. However, as the COVID shock affected the probability of obtaining a high school diploma, it also massively affected the probability of obtaining a place at the desired institution, so the observed effect may be due to a break in parallel trends in 2020 caused by a shock, and it isn't easy to believe this result. In summary, I find no robust effects on the extensive margin in 2018 and 2019, with confidence intervals included in $\pm$ 3 percentage points (pp), and the results in 2020 are not interpretable.\\

\begin{figure}[htbp]
  \centering
  \caption{Estimated event study coefficients for final enrollment prestige using DRDiD (general track, ITT, Cohen’s d)}

  \includegraphics[width=1\textwidth, trim=0 0 0 0, clip]{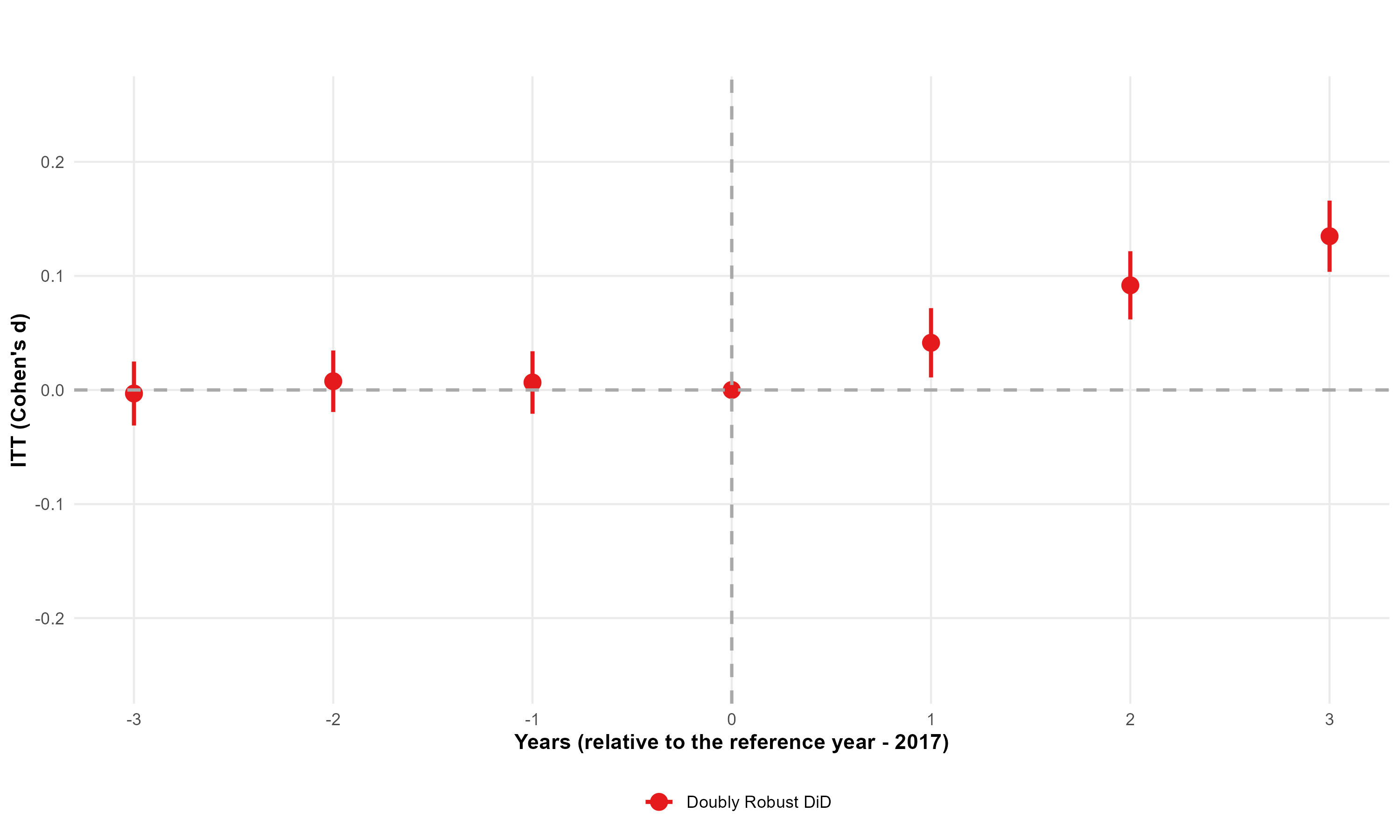}
  \tablereading{According to the DRDiD estimator, the ITT on final enrollment prestige for the general track was 0.132 SD in 2020.}

  \label{principal:score_admission_drdid}
\end{figure}

\begin{figure}[htbp]
  \centering
\caption{Sensitivity analysis of DRDiD estimates for final enrollment prestige (\autoref{principal:score_admission_drdid})}

  \includegraphics[width=1\textwidth, trim=0 0 0 25, clip]{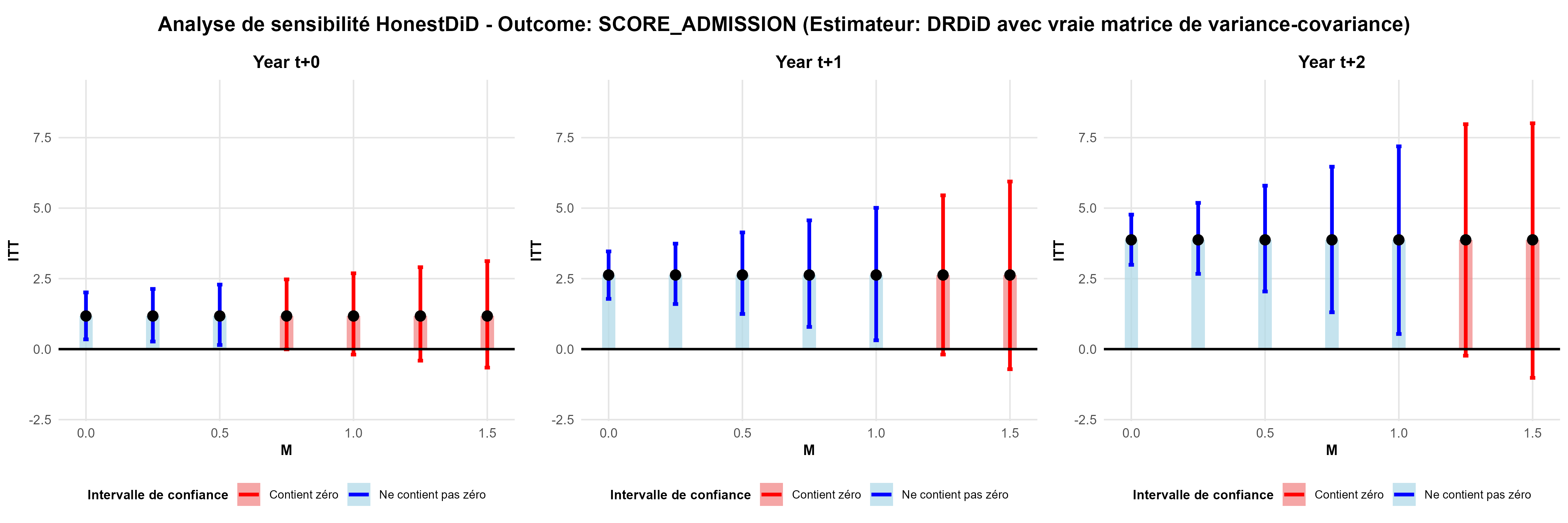}
  \tablereading{In 2019 and 2020, the breakdown value is greater than 1 but inferior to 1.25.}

  \label{principal:sensivity}
\end{figure}

\noindent \textbf{H1b. Prestige of the program in which the individual is finally enrolled:} For the general track, significant and increasing positive effects were observed throughout the post-treatment period with a DRDiD estimation (see \autoref{principal:score_admission_drdid}): 0.041 SD in 2018 ($p<0.01$), 0.092 SD in 2019 ($p<0.001$), and 0.135 SD ($p<0.001$). These estimates are robust to testing, with high breakdown values of 0.8 to 1.2 (\autoref{g_sans_notes:breaking_point_unified}), above the robustness threshold for the last two years of treatment but inferior to 1.25 (see \autoref{principal:sensivity}). The absence of a differential trend before treatment ($p=0.839$) suggests that the parallel trends hypothesis is plausible. For technological and vocational tracks, there is no effect on the prestige of the final program obtained.\footnote{One could be concerned about the population change problem induced by an effect on the extensive margin. However, except for the vocational track in 2020, I have previously observed that the treatment group does not access higher education more following the introduction of quotas, with non-significant effects ranging in $[-0\text{ pp};2\text{ pp}]$ for the general track (on a baseline probability of more than 76\%).
Thus, thanks to the weak monotonicity assumption, I can conclude that the extensive margin effect induces no or very little compositional change, and therefore that I properly identify the intensive margin effect.}

\FloatBarrier

\begin{figure}[htbp]
  \centering
\caption{Estimated event study coefficients for the probability of admission into specific program types, using DRDiD specification (vocational track, ITT)}

  \includegraphics[width=1\textwidth, trim=0 0 0 0, clip]{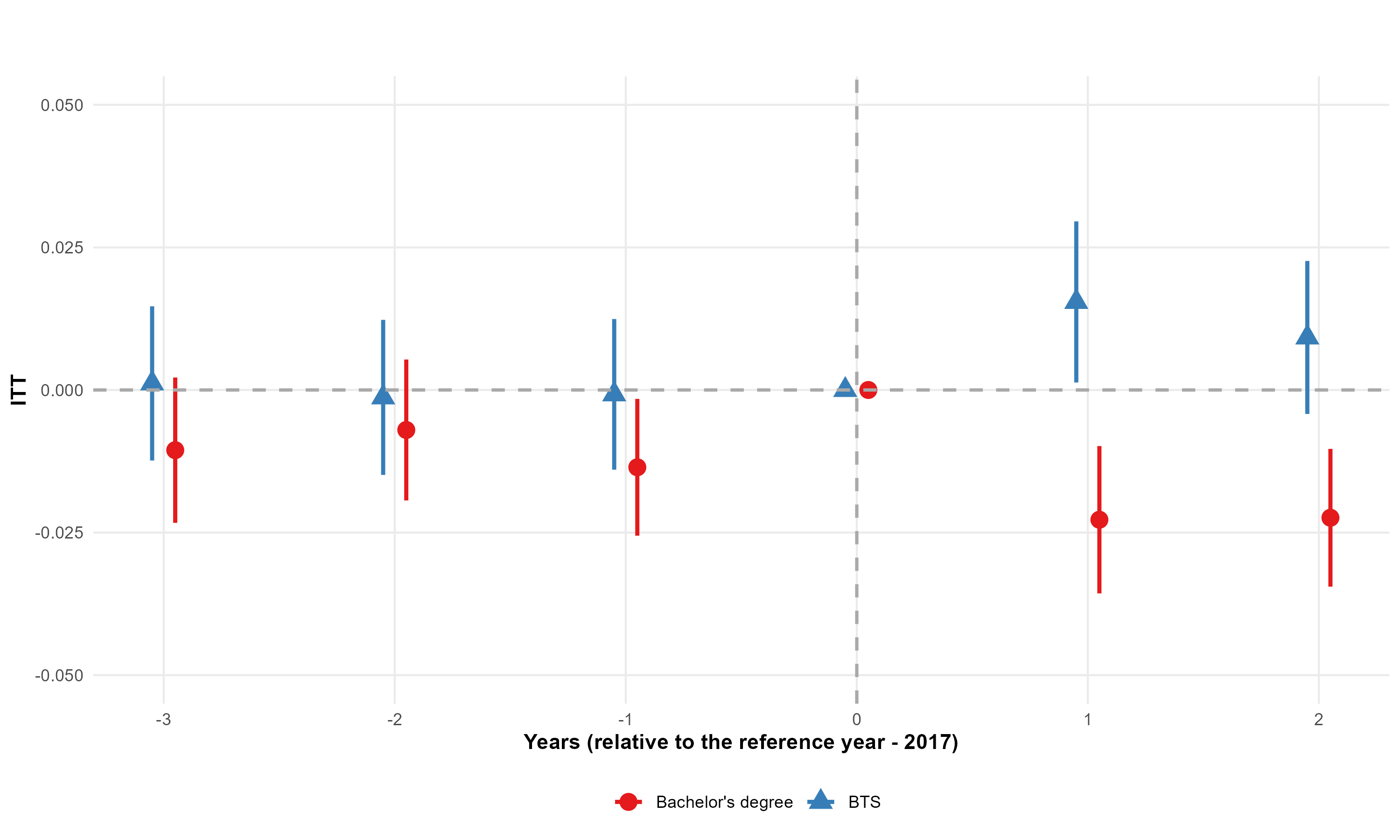}
  \tablereading{According to the DRDiD estimator, in the first post-treatment period, the treatment increased the probability of ultimately enrolling by a 1.5pp increase in the probability of being admitted to a BTS program (non-significant). At the same time, the probability of enrolling in a bachelor's degree program decreased by 2.3pp.}

  \label{principal:type_pro}
\end{figure}

\noindent \textbf{H1c. Type of program in which the individual is finally enrolled} Scholarship recipients in the general track are more likely to be admitted to a DUT program (short-term selective programs): We find significant positive effects of approximately 1pp in all three post-treatment years, with no detectable pre-trend ($p = 0.466$). However, these estimates are not robust, with breakdown values of 0.4, which are below the threshold for robustness. It is challenging to determine whether the effect I am measuring is an extensive margin effect or an intensive margin effect, as I know that an effect on the extensive margin occurred in 2018, which could explain the results for that year. No significant effect is found on other outcomes (CPGE, BTS, etc.) across all years. For technological tracks, no significant effect was observed regardless of the type of program or year. For vocational tracks, the results are shown in \autoref{principal:type_pro}. A significant negative effect is observed: the probability of being admitted to a bachelor's program decreases by 2 points. This result is not robust (breakdown values between 0.5 and 0.9). In comparison, the probability of being admitted to a BTS program (a short-term program targeted for the vocational track) increases by 1 point symmetrically; however, this result is not statistically significant
\footnote{Given the strong possibility of an effect from the COVID shock, which broke the parallel trends in 2020, I am not presenting the results for that year.}
\footnote{I did not test the outcomes for other types of programs, as the baseline values are less than 1\%}. These results could be rather positive for the overall success of scholarship students, as we know that vocational baccalaureate holders enrolled in BTS programs drop out less often than those enrolled in bachelor's courses \cite{NdaoPirus2019}. Still, more in-depth studies will enable us to determine whether being marginally admitted to the BTS, rather than a Bachelor's degree, is beneficial for vocational tracks. 

\paragraph{Summary of the results}: In the general track, students had slightly greater access to higher education ($\approx$ 1 percentage point) via Parcoursup in the first year after treatment. In particular, they were more likely to access a DUT, but these results are not robust. Ultimately, they are admitted to more prestigious programs (robust results). In the technological track, no effect is detected. In the vocational track, the introduction of quotas decreases the probability of pursuing a bachelor's degree by 2-2.5 points; however, this result could be explained by a non-significant increase in admission to BTS of 1-1.5 points. These latter results are not significant or robust. Still, they could be considered positive, as the bachelor's degree is often a fallback option for vocational high school graduates who are unable to be admitted to a BTS program.

\subsection{Effects on the behavior of scholarship applicants}

\noindent \textbf{H2a. Participation in Parcoursup} I don't see any robust effect on participation in Parcoursup. The only significant effect is a -3.2pp decrease in participation in the vocational track in 2020, but this result is not robust (breakdown value of 0.4).\footnote{This hypothesis has not been tested among general baccalaureate holders, as almost 100\% of them make at least one application on Parcoursup.} \\

\begin{figure}[htbp]
  \centering
  \caption{Event study coefficients for the most prestigious (top plot) and average application prestige (bottom plot), using DRDiD specification (general track, ATE, Cohen’s d)}

  \includegraphics[width=1\textwidth, trim=0 0 0 0, clip]{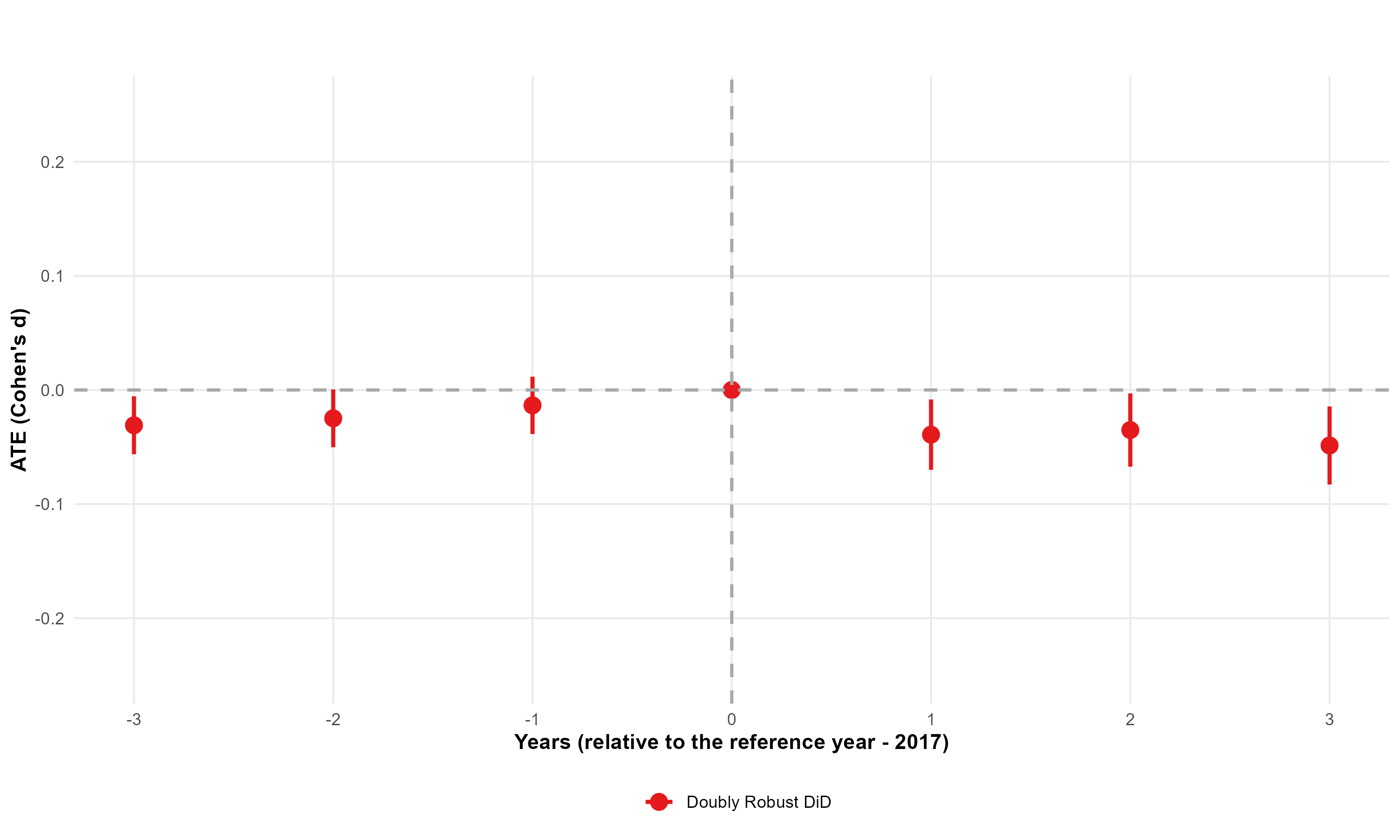}
  \includegraphics[width=1\textwidth, trim=0 0 0 0, clip]{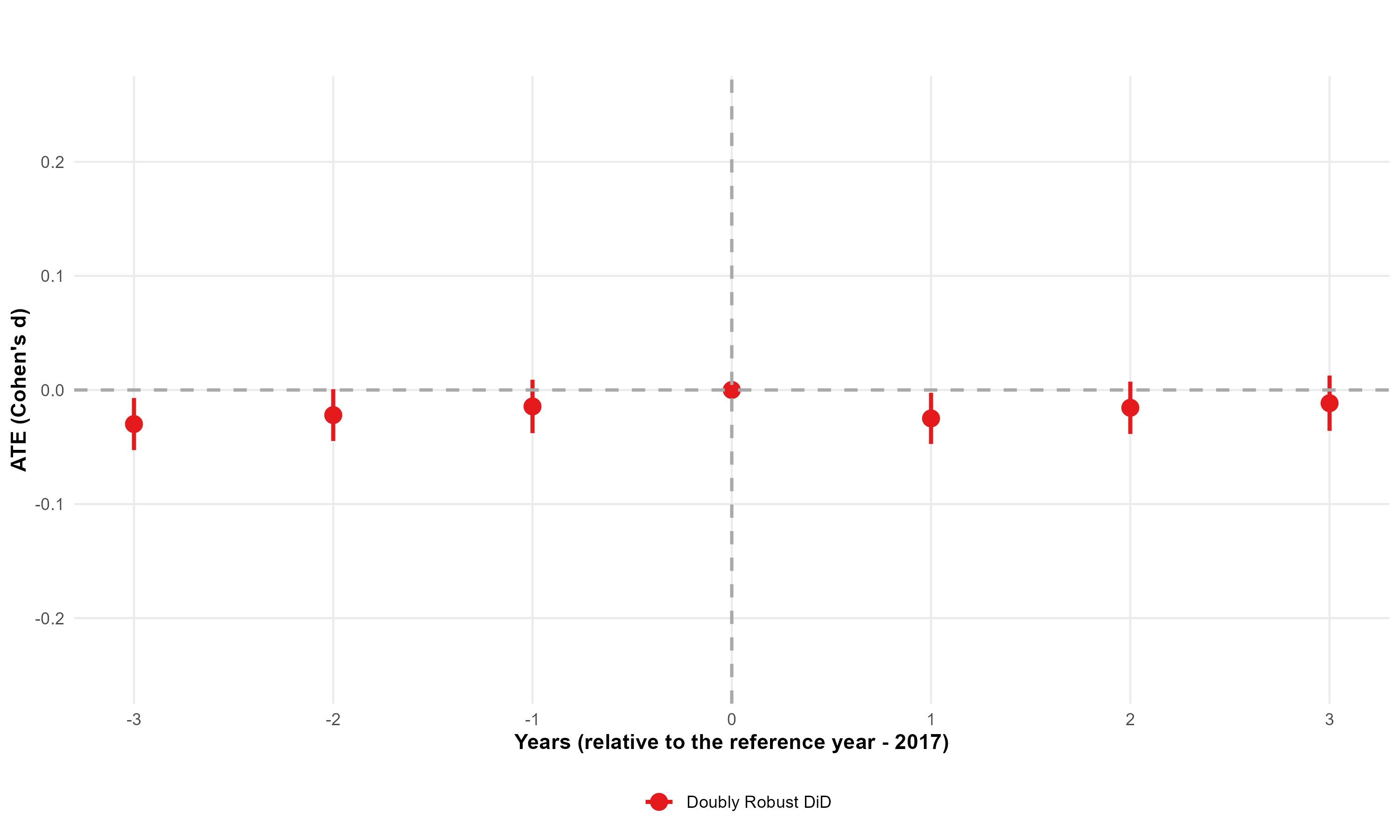}
  \noindent \tablereading{In the general track, according to the DRDiD estimator, in 2020, average prestige in applications decreases by 0.012 SD due to the treatment. According to the DRDiD estimator, the maximum prestige of applications decreases by 0.049 SD in 2020.}

  \label{principal:prestige_application}
\end{figure}

\noindent \textbf{H2b. Prestige of the programs to which the student applies} For all years and all tracks, \emph{no significant positive effect has been measured}. There is a slightly negative effect on the maximum prestige of applications for the general track, but this effect is not very robust (breakdown value between 0.2 and 0.6). Moreover, for the general track, I suspect a possible violation of pre-trend, with $p<0.1$ for both outcomes. Confidence intervals are always between -0.08 SD (for the lower bound) and +0.04 SD (for the upper bound) for all populations and years, meaning precise effects are centered at 0. The estimates for the general track are shown in \autoref{principal:prestige_application}.\\

\noindent \textbf{H2c. Type of program in which the individual applies}: For all the high school tracks, no robust effect was found. In the vast majority of cases, the measured effect is insignificant, or when it is significant, it does not withstand a breakdown value of 0.4 (and most often not even 0.3), representing a very slight violation of the parallel trends assumption. Note that all the confidence intervals are in the range of $\pm$ 3pp (most of the time $\pm$ 1pp), suggesting that even non-robust effects are, at best, very small (\ref{g_sans_notes:confidence_interval_unified},\ref{t_sans_notes:confidence_interval_unified},\ref{p_sans_notes:confidence_interval_unified}).

\paragraph{Summary of the results} I found no robust effect on the behavior of the applicants. In the vast majority of the estimations, I found either no significant effect or a very small \textbf{negative} effect that doesn't withstand a small violation of parallel trend. The general conclusion is that I don't find any robust or strong effect on the prestige or the type of programs recipients apply to, nor on overall participation in Parcoursup.

\subsection{Effects on pre-college academic performance of scholarship applicants}

\noindent \textbf{H3a. Pre-college academic performance} No significant effect on baccalaureate results is observed for the three populations, either in 2018 or 2019. Furthermore, the confidence interval of the estimate remains consistently between $\pm$ 0.03 standard deviations (SD). There is no significant violation of the pre-trend except in vocational baccalaureate programs, but these violations remain very small. In summary, hypothesis \textbf{H3a} is not confirmed for any period or population.

\subsection{Precision}

\paragraph{Type II Error Considerations}
While our robustness tests reduce the risk of Type I errors ("false positives"), they may increase the probability of Type II errors ("false negatives"), leading to the rejection of true effects due to excessive caution. For instance, insufficient statistical power could lead to the rejection of a hypothesis even when confidence intervals contain substantial positive or negative values, suggesting potentially strong effects. To ensure our null estimates represent "true nulls", I compute confidence intervals for each outcome, both in original units and standardized form (Cohen's d). For this table, I apply no correction or robustness check, not even multiple hypothesis testing. Results demonstrate that all confidence intervals (lower \emph{and} upper bounds) for applications final enrollment (except for prestige of final enrollment for the general track) range within $\pm$ 0.08 SD\footnote{except for the probability of participating in Parcoursup, which drops to -0.11 SD for technological and vocational baccalaureate}. In the vast majority of cases, confidence intervals range within $\pm$ 0.04 SD. These narrow confidence intervals centered around zero indicate that non-significant conclusions do not result solely from insufficient power. Any effects potentially rejected due to Type II errors are likely of very small magnitude. The confidence intervals (in Cohen's d) are presented in \autoref{g_sans_notes:confidence_interval_unified} (general track), \autoref{t_sans_notes:confidence_interval_unified} (technological track), and \autoref{p_sans_notes:confidence_interval_unified} (vocational track).

\subsection{Alternative Estimator and Heterogeneity}

The results of the CiC estimators are detailed in \autoref{g_sans_notes:event_study_cic} (general), \autoref{t_sans_notes:event_study_cic}  (technological), and \autoref{p_sans_notes:event_study_cic} (vocational). These tables contain both the ATE/ITT estimators for the three prestige scores (admission, maximum applications, average applications), including Cohen's d, as well as the heterogeneity (by decile) for admission prestige.

\paragraph{CiC for prestige in applications} For general and technological tracks, the estimates on prestige in applications (average and maximum prestige) range in a $\pm 0.06 \text{ SD}$ interval, meaning a very small effect. The only significant estimates in the general and technological tracks are weak ($\pm 0.05\text{ SD}$) and preceded by a pre-trend violation of the same order of magnitude, meaning we cannot interpret these results as robust\footnote{Although we cannot strictly implement Honest DiD for this estimator}. For vocational tracks, I observe a positive effect between 0.05 and 0.1 SD; however, I again encounter a pre-trend violation of more than 0.05 SD, which is highly significant ($p < 0.001$). To summarize, I find only weak effects, and significant effects are observed only when there are very strong pre-trend violations (stronger than those obtained when using the DRDiD estimator). I cannot, thus, conclude that there is any behavioral effect on the prestige of the applications made with this alternative estimator, which seems less precise than DRDiD. However, estimates are in the same range when pre-trends are not violated.

\paragraph{CiC for prestige in admission} For the vocational track, I find a null effect on the prestige of the final enrollment. For the technological track, I found a small, positive, and significant effect in 2019 and 2020 (approximately 0.04 SD). Still, I have a pre-trend that increases up to 0.025 SD, so I cannot infer with confidence that the effects are robust. For the general track, I find a positive and robust effect, which is largely superior to pre-trends but inferior to the ITT I measured with DRDiD in 2020. The effects measured are between 0.08 and 0.063 SD in 2019 and 2020.
To summarize, the CiC estimates could potentially identify a small positive effect for the technological track; however, these results are non-robust. They confirm a positive effect on prestige of enrollment for the general track, with no effect whatsoever for the vocational track.

\begin{figure}[htbp]
  \centering
\caption{Coefficients for final enrollment prestige by outcome decile, based on CiC specification (general track, ITT, prestige score out of 100)}
  \includegraphics[width=1\textwidth, trim=0 0 0 0, clip]{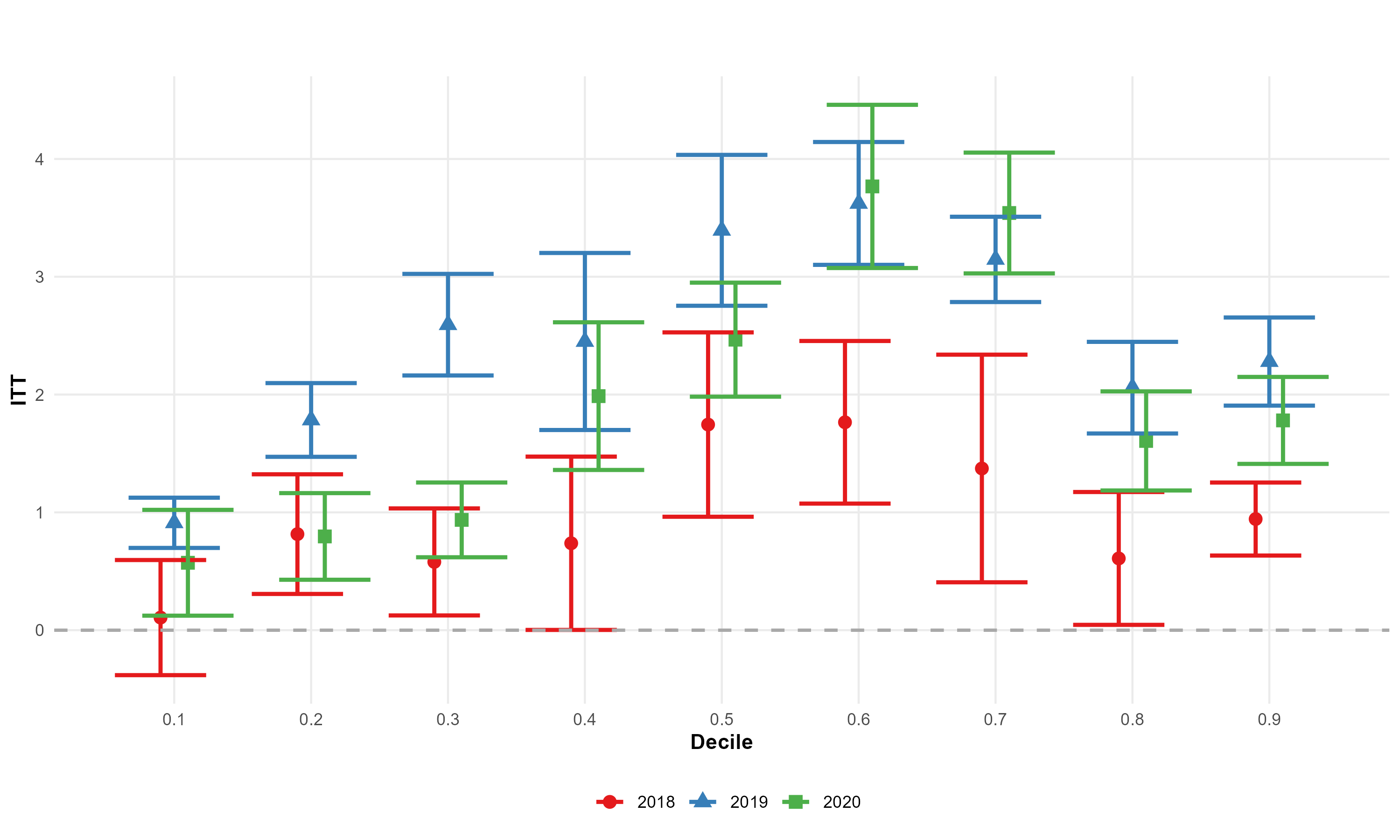}
  \tablereading{For Decile 6, i.e., scholarship recipients who were among the top 30\% to 40\% in terms of prestige in the final enrollment without treatment, the gain was between 3 and 4 points (out of 100) in 2019 and 2020, compared to 2 points in 2018.}
  \label{principal:heterogeneity}
\end{figure}

\paragraph{Heterogeneity of prestige gain for the general track} The CiC estimator enables heterogeneity analysis, examining local treatment effects for each quantile of the treated distribution. I study the effects across deciles, where deciles represent the distribution of scholarship students based on the prestige of programs they have accessed before the introduction of quotas, compared to the counterfactual population. Results are presented in \autoref{principal:heterogeneity}. Decile 0.1 includes the 10\% accessing the least selective programs before quotas, while decile 0.9 represents the 10\% already accessing the most selective programs. The analysis reveals that the strongest effects are concentrated in deciles 0.5-0.7, among scholarship students initially accessing intermediate selectivity programs, with gains reaching 3.5-4 percentile points in the 2019-2020 period. Lower deciles (0.1-0.4), corresponding to students oriented toward the least selective programs, show weak or non-significant effects. Upper deciles (0.8-0.9), grouping those already accessing highly selective programs before quotas, also demonstrate limited effects, suggesting a ceiling effect. The tenth decile (1.0) was excluded because the estimator failed to produce results for this decile, typically generating zero effect with zero standard error for unknown reasons. It should nevertheless be noted that the CiC may be inadequate for the extreme deciles \cite{RePEc:taf:jnlbes:v:42:y:2024:i:2:p:812-824}.

\paragraph{DRDiD on a good general track students} For general track recipients who are part of the first quartile, I do not measure any behavioral effect on the application to CPGE or engineering school. No robust effect on admission or behavior was measured (\autoref{g_avec_tres_bonnes_notes:event_study}). The most robust effect measured was a positive effect on the prestige of admission in 2020, which does not withstand a breakdown value of 0.4 (\autoref{g_avec_tres_bonnes_notes:breaking_point_unified}). The results of the event studies, as well as the breakdown values and confidence intervals, are presented in Appendix \ref{appendix:first_quartile}.

\section{Potential Mechanisms\label{mechanisms}}

The previous section demonstrates that for general track students, quotas had a weak but robust effect on their final admission. In this section, I aim to explain the mechanisms that underlie the results obtained. Since I observed no behavioral effects on applications, I will assume that the effects are solely due to the mechanical effect of quotas on waiting lists, which allowed students to obtain a given and exogenous proposal they preferred. To explore these mechanisms, I conducted three independent statistical analyses addressing three different research questions.

\subsection{Discrimination before and after quotas (Matching)}

\subsubsection{Hypothesis}

The first question is whether quotas effectively increased the probability of being "admissible" for scholarship recipients (i.e., ranked higher than the last person who received an admission offer, meaning the student could have been definitively admitted if desired). This question addresses the effectiveness of treatment, i.e., how much the introduction of quotas changed rankings. I can also examine whether the treatment was "justified", i.e.,  whether scholarship recipients were disadvantaged or discriminated against in admission chances compared to non-scholarship peers. 

\subsubsection{Methodology}

I use two specifications and run regressions for each program type and year to identify heterogeneous effects. Importantly, I compare scholarship and non-scholarship recipients who applied to the same program. I exclude calculations when there are fewer than 10 applicants or no scholarship recipients are available.\\

\noindent In the first specification, I examine the "raw" difference in admissibility chances between scholarship and non-scholarship applicants applying to the same program. Note that these groups may differ substantially, so admission differences could potentially be justified. In the second specification, I employ Coarsened Exact Matching (CEM) to compare scholarship recipients with non-scholarship applicants who have identical characteristics. The characteristics included are: the track (General, Technological, Vocational), the specialization (e.g., for the general track: Scientific, Economics and Social Sciences, Literary), overall average grade during the year (GPA), gender, and geographical area. Methodological details are specified in Appendix \ref{appendix:methodological_details}.\\

\noindent Theoretically, the matching strategy is particularly effective here since I can match applicants on the same observable characteristics that admission committees use for ranking, reducing the risk of self-sorting that could undermine result validity. However, for simplicity, I have included only a limited number of variables, which may not cover all committee evaluation criteria. Notably, I exclude textual analysis of motivation letters and use only the overall GPA rather than subject-specific grades.

\subsubsection{Results}

Results for the BTS programs are presented in \autoref{principal:matching}. Results for all types of selective programs are also available in \autoref{appendix:matching}.\\

\noindent \textbf{Pre-treatment disadvantage:} I observe a clear raw disadvantage for scholarship recipients across nearly all types of higher education programs. In 2017, across all selective programs, scholarship recipients had a lower probability of being "admissible" compared to non-scholarship applicants. For example, for BTS programs, the difference in probability is 3.35 percentage points. However, when using matching methods, this difference decreases to 2.00 percentage points, even without including more complex variables such as motivation letters or detailed grades. This result suggests that observable differences between scholarship and non-scholarship applicants explain at least part of the admissibility gap. For certain program types, such as preparatory classes (CPGE), I find no difference in admission probability before quota introduction when using matched profiles, despite a 4.94 points raw difference in the probability of being admissible when profiles are not matched. While a raw disadvantage undoubtedly exists, I cannot definitively conclude whether this represents unjustified discrimination, as observable characteristics partially explain the gap, and I haven't accounted for all legitimate observables (detailed grades, motivation letters).\\

\noindent \textbf{Treatment effectiveness:} Following quota introduction, scholarship recipients' admissibility probability increased dramatically. By 2020, scholarship recipients in CPGE had a 7.06pp higher probability of being admissible than their counterparts with similar characteristics in the same program. In contrast, there was a non-significant 0.77 percentage point difference in probability in 2017. I observe a progressive increase in admission chances for scholarship recipients relative to non-scholarship recipients across all program types following the 2018 quota implementation. This result strongly supports the hypothesis that quotas substantially increase scholarship applicants' chances of being admitted to their preferred choices, even creating positive discrimination (not merely reducing negative discrimination) compared to non-scholarship recipients.

\begin{figure}[htbp]
  \centering
\caption{Raw and matched-profile differences in admission probabilities between recipients and non-recipients. Only BTS}

  \includegraphics[width=1\textwidth, trim=0 0 0 0, clip]{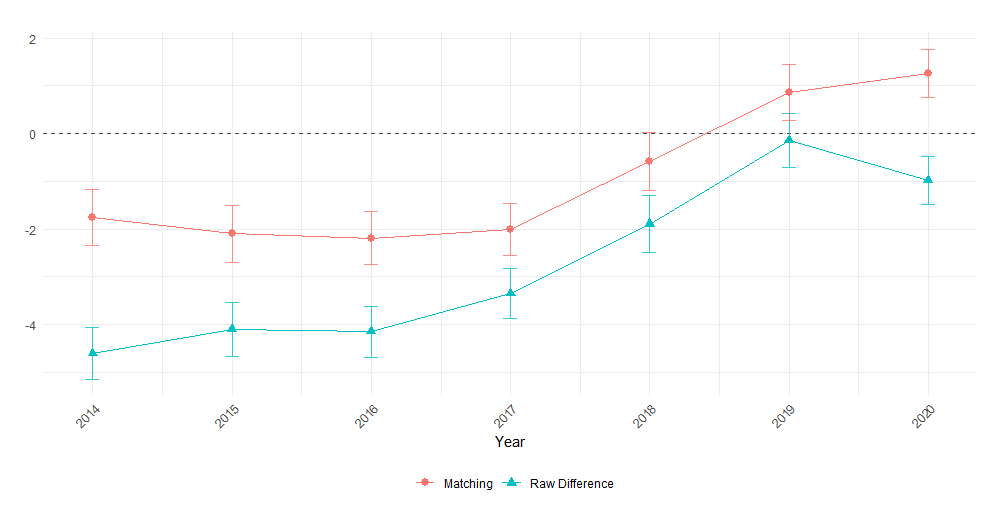}

  \label{principal:matching}
\noindent \tablereading{In 2017, the probability of being eligible (ranked higher than the last person who was finally enrolled) for a BTS program for a scholarship recipient is 3.35pp less than for a non-scholarship recipient who applied for the same program. However, when comparing only to non-scholarship recipients with a similar profile, the difference drops to 2.00pp.}

\end{figure}

\subsection{Explaining the effect on prestige for the general track (Simulations)}

\subsubsection{Hypothesis}

The previous analysis results appear contradictory with the effects measured in the empirical estimation section: despite a substantial increase in admissibility probability, quotas had only weak effects on enrollment selectivity. A possible explanation is that scholarship recipients typically switch to programs of similar selectivity - for example, moving from Bachelor A to Bachelor B, which are equally selective. In summary, while many scholarship recipients might gain admission to their preferred programs through quotas, only a small proportion would "benefit" by accessing significantly more selective programs.\\

I can distinguish two types of "compliers" (scholarship recipients who change programs due to quotas): (1) intra-type compliers who switch programs within the same type of program (would attend Bachelor A without quotas, attends Bachelor B with quotas), and (2) inter-type compliers who transfer between different types of programs (moving from a Bachelor to CPGE). Using the DRDiD, I do not measure a robust effect on the probability of going into one type of program or the other, which does not support the theory of inter-type transfers. However, I could imagine that these transfers still exist, but are too small to be robustly measurable, and that in a second step, these transfers explain at least part of the measured effect on the gain in prestige.

\subsubsection{Methodology}

\paragraph{Simulations} To study the plausibility of these claims, I conducted simulations on APB in 2016. Unlike Parcoursup, APB allows for simulating quota effects, as I have students' ex-ante preferences (ranked choices), enabling the Gale-Shapley algorithm. I run two APB 2016 simulations\footnote{not using 2017 due to dysfunctions that altered preference revelation}: one with quotas (using the $\text{Proportion}_{Recipient} + 2pp$ rule) and one without quotas, then examine which students changed programs and their final enrollment. For simplicity, I consider only the first round of the three-round process, given that the vast majority of final enrollments are made during the first round in APB \cite{CourDesComptes2017_APB}. I conduct a global analysis for all scholarship recipients, then a detailed analysis by track (general, technological, vocational). Based on previous results, I assume that quotas don't influence applicant or program behavior, as scholarship status remains hidden. The simulation is a \textit{static} one, measuring mechanical effects without considering behavioral reactions.

\paragraph{Data Analysis} First, I compare scholarship applicants' situations with and without quotas, counting those who changed programs between simulations. Next, I examine whether these applicants were admitted to more prestigious choices on average (average selectivity gain among compliers) and identify the most frequent transition types (Bachelor → CPGE, Bachelor → BTS, BTS → DUT, etc.). I then determine which transition types explain these gains using decomposition methods. If I measure an average 10pp gain, how much is attributable to Bachelor → BTS transitions? This measure reveals whether gains stem from intra-type transfers (moving from BTS A to more prestigious BTS B). To identify each transfer's contributions, I decompose the average gain as:

$$\text{AverageGain} = \sum_{i,j} p_{ij} \cdot \text{AverageGain}_{ij}$$

\noindent where $\text{AverageGain}_{ij}$ is the average gain for students transitioning from program type $i$ to type $j$ (for intra-type transitions we have $i=j$), and $p_{ij}$ is the proportion of these transitions. The product $p_{ij} \cdot \text{AverageGain}_{ij}$ gives the share of average gain attributable to $i \to j$ transitions. To identify intra-track versus inter-track contribution shares:

$$\text{AverageGain} = \sum_{i=j} p_{ij} \cdot \text{AverageGain}_{ij} + \sum_{i \neq j} p_{ij} \cdot \text{AverageGain}_{ij}$$

\subsubsection{Results}

Results of the simulation are provided in Appendix \ref{appendix:all_2}. I also decline the results for each track: general (\ref{appendix:g_2}), technological (\ref{appendix:t_2}), and vocational (\ref{appendix:p_2}).
\paragraph{Number of Affected Scholarship Recipients} According to simulations, 5249 scholarship recipients received different assignments due to quotas compared to what they would have received without quotas, representing 7.48\% of scholarship students included in the analysis that year. This number breaks down to 2789 (53\%) in general track students (\autoref{appendix:g_2:effectifs}), 1528 (29\%) in technological track students (\autoref{appendix:t_2:effectifs}), and 932 (18\%) in vocational track students (\autoref{appendix:p_2:decomposition}).

\paragraph{Average Gains} On average, treated individuals gain 7.67 points in the selectivity level of their admitted program (\autoref{appendix:all_2:decomposition}). This gain is $\simeq$ 8 points for general and technological tracks (\autoref{appendix:g_2:decomposition},\autoref{appendix:t_2:decomposition}), while only $\simeq$ 4.5 points for the vocational track (\autoref{appendix:p_2:decomposition}). The result corresponds to the ATE, but calculating the ITT by multiplying ATE by compliance yields effects of 0.6 points for general, 0.7 points for technological, and 0.3 points for vocational tracks—smaller effects than those observed empirically for general baccalaureate holders. Note that these are average effects, but many transitions involve prestige losses.

\paragraph{Transition Types} In general tracks, 41\% of transitions are from one bachelor program to another bachelor program, while 21\% are from bachelor to other program types (including 8.5\% to DUT, 4.4\% to BTS, 6.1\% to CPGE) (\autoref{appendix:g_2:effectifs}). Among technological baccalaureate holders, transitions primarily occur from BTS to BTS programs (35\% of transitions) or from bachelor to bachelor programs (16.8\% of transitions) or from bachelor to BTS (13.5\% of transitions) (\autoref{appendix:t_2:effectifs}). For vocational baccalaureate holders, the vast majority of transitions go from bachelor to BTS (10.4\%) or from one BTS to another BTS (71\%) or from bachelor to bachelor (10\%) (\autoref{appendix:p_2:effectifs}). Overall, the total volume of scholarship recipients (compliers and non-compliers) in bachelor programs decreased by -1.08pp for general (\autoref{appendix:g_2:pp}), -0.61pp for vocational (\autoref{appendix:p_2:pp}), and -1.18pp for technological tracks (\autoref{appendix:t_2:pp}). A more surprising result is that there is absolutely no effect on the extensive margin: no scholarship student (who would have obtained nothing without the quotas) received an admission proposal as a result of the quotas. This result must be because all applicants had to apply for a non-selective degree. I also find that for the people who have been admitted to a particular CPGE program thanks to the quotas, more than half would have gone to another CPGE without quotas.

\begin{figure}[htbp]
  \centering  
\caption{Decomposition of ATE gains in prestige by type of transition (general track)}

  \includegraphics[width=1\textwidth, trim=0 50 0 0, clip]{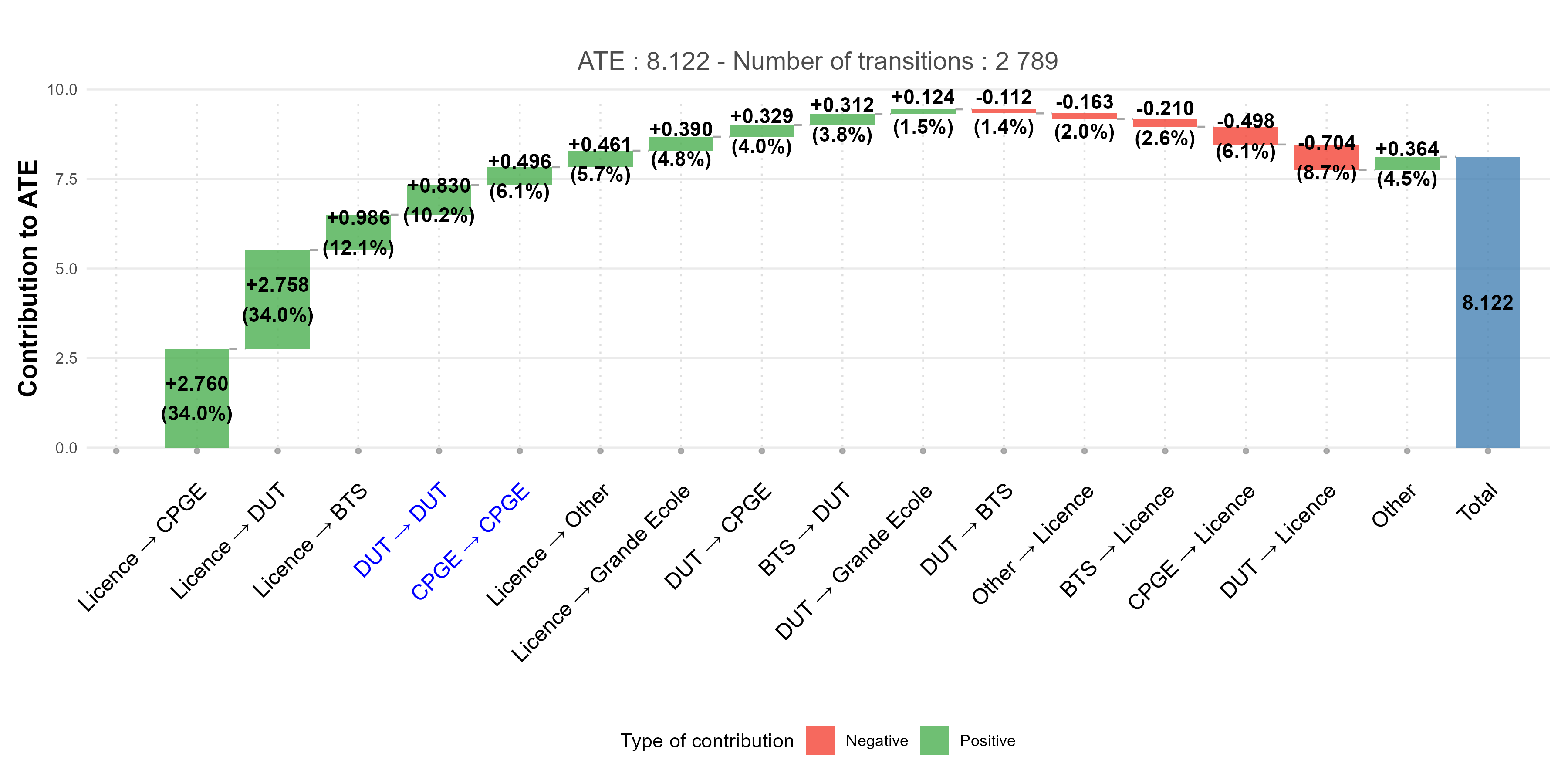}
  \tablenote{The methodology involves simulating the first round of the Gale-Shapley algorithm on the 2016 APB data, and a simulation where all programs have the “proportion of scholarship +2pp” quota rule.}{According to simulations, out of a measured ATE effect of 8.12 points (out of 100) on the prestige of final enrollment, 2.76 points (or 34\% of the effect) can be explained solely by scholarship recipients who would have gone on to a bachelor's degree program but went to CPGE instead thanks to the existence of quotas. Furthermore, 2,79 scholarship general track applicants changed their final enrollment due to quotas.}

  \label{principal:decomposition}
\end{figure}

\paragraph{Decomposition of Gains} This analysis identifies the types of transitions that drove the average prestige gains among scholarship compliers (scholarship holders who change their enrollment due to quotas). For general track students, transitions from bachelor to other program types explain $\simeq$100\% of gains, meaning departing from bachelor programs primarily explains observed gains (see detailed decomposition in \autoref{principal:decomposition}). In contrast, intra-type transitions explain only $\simeq$12\% of gains\footnote{In the end, the sum of these two types of transitions explains more than 100\% of the effect, but it is because there are also transitions which have a negative consequence on prestige, so that when I sum transitions with a positive effect and those with negative effect I arrive at exactly 100\% of the effect explained}. Among the technological tracks, intra-type transitions account for 54\% of the measured effect (see \autoref{appendix:t_2:decomposition}). Finally, for vocational baccalaureate holders, the average gain is primarily explained by BTS-to-BTS transitions (42\% contribution) and Bachelor-to-BTS transitions (35\%) (\autoref{appendix:p_2:decomposition}).

\paragraph{Validity} While the non-behavioral reaction assumptions appear credible (not threatening internal validity), results should be interpreted cautiously due to limited external validity. The simulation evaluates quota effects on APB, not Parcoursup. Moreover, measured ITTs are inconsistent with those from Parcoursup, with substantially lower ITTs for holders of the general baccalaureate. This contradictory result doesn't invalidate the Parcoursup analysis, but suggests that treatment would have had a less pronounced effect on APB than Parcoursup. Multiple factors could explain this: more selective programs on Parcoursup (such as dual bachelor programs), longer waiting lists on Parcoursup, which enable more scholarship recipients to progress due to the quotas algorithm, or different student behaviors on Parcoursup (where all students make more competitive choices). A final explanation could be that bachelor's programs became more selective after Parcoursup's introduction, particularly in certain fields (such as Sports, Psychology, Law, and Medicine). Therefore, transfers to these programs in 2016 weren't associated with increased selectivity (and thus prestige), whereas they were in 2018. However, this hypothesis seems implausible since \citeA{bechichi2021segregation} shows most bachelor programs remain minimally selective on Parcoursup. Note that with random selection in APB 2016, waiting lists weren't based on program preference criteria.

\paragraph{Interpretation}: At minimum, simulation analyses demonstrate that prestige gains among general baccalaureate holders are at least partially explained by inter-type transitions (notably bachelor to other program types) rather than intra-type transitions (or at least a mixture of both), despite relatively small variation in the total proportion of general students attending bachelor programs. This result suggests that using a selectivity score provides statistical power gains over binary variables, such as "admitted in a program of type X". It reveals the cumulative effects of small changes in program types, too small to be statistically significant using binary outcomes in \textbf{H1c} analysis.

\subsection{Effect of a change in the rules for calculating quotas (Simulations)}

\subsubsection{Hypothesis}

Previous analyses show that few scholarship recipients change programs as a result of quotas, which may explain the modest effects observed. One hypothesis is that quotas, calculated endogenously, are too low to have a real influence on the decisions of admissions committees. However, the rule includes an exogenous bonus of +2 percentage points. To test whether the low intensity of the treatment stems from the endogenous part of the calculation, we can examine whether this simple +2 point bonus significantly increases the number of “compliant” scholarship recipients—i.e., those who change programs thanks to quotas. If this is the case, it supports the idea that the endogenous calculation of quotas is indeed the reason for their weak effect, but that adding an exogenous component common to all programs in the quota calculation—as is the practice in other quota systems in India or Brazil—would greatly increase the treatment intensity.

\subsubsection{Methodology}

I reproduce the simulation from the previous section, but I remove the +2pp in the quota calculation rule. Then, I compare the number of scholarship recipients who obtained training thanks to quotas.

\subsubsection{Results}

The average gain in prestige among the "compliers" (individuals who changed courses due to the quotas) remains stable, at approximately 8 points. On the other hand, the number of compliers has fallen sharply: I go from 5,249 students with the +2pp to 3,031 without the bonus (\autoref{appendix:all_0:effectifs}), a decrease of 42\%. The +2pp bonus thus appears to play a crucial role in determining the number of people who benefit from the treatment. We can also see that the ATE is fairly stable compared to the simulation with bonuses, reinforcing the idea that the low treatment intensity indeed explains the low ITT: even a small bonus of 2pp is enough to increase the treatment intensity, considering its current very low level. Nevertheless, the results must be interpreted with caution, particularly in consideration of external validity, because the simulations were carried out on APB and not Parcoursup, which may have affected the number of compliers (see the previous section). Details of the results of this simulation are provided in Appendix \ref{appendix:all_0}.

\section{Discussion\label{discussion}}

\subsection{Interpretation}

\paragraph{Behavior of recipients} Across all student tracks, I observe no positive, strong, or robust effects on application behavior. Interestingly, we see that the effects on prestige are zero regardless of whether we are in the "opaque quotas" condition (the quota amount is not displayed but the existence of quotas applicable to the program is indicated) the first two years, and the "transparent quotas" (the quota amount for the previous year is displayed for each program) the third year. I observe no significant effect on the prestige of applications: the estimates are either non-significant (the vast majority of the estimates) or negative and non-robust to a very small violation of parallel trends (generally <0.4 degrees). Ultimately, I cannot conclude that there is any lasting and robust effect on the behavior of applicants, particularly in a positive direction. 

\paragraph{Prestige and type of final enrollment} The most robust result concerns admission prestige for general track students: general students are admitted to slightly more selective programs, with effects reaching 0.14 SD in 2020 and outlasting violations above 1 degree (considered robust). Note that this is an ITT, as I cannot accurately determine who changed admissions due to. Therefore, a robust effect of 0.14 SD appears to be a relatively large effect, considering that compliance may be quite low.

\paragraph{Potential Mechanisms} Call lists analysis shows that in APB, scholarship recipients had lower eligibility chances than non-scholarship applicants applying to the same programs. Characteristic differences largely explain this difference; however, scholarship recipients still had slightly lower eligibility chances than their comparable non-scholarship peers. Following the introduction of the quota, this trend reversed, with scholarship recipients progressively favored in admission chances relative to their comparable non-scholarship peers. Through this increased intensity, scholarship students gained access to various programs. Many likely made intra-type transitions (moving from program A to program B within the same category). However, gains in prestige among general track students appear to be explained by inter-type rather than intra-type transitions, even though these represent a minority of quota-enabled transitions. Furthermore, the +2pp bonus in the quota calculation rule appears to significantly increase treatment intensity, suggesting that the endogenous quota calculation method is responsible for a potential low treatment intensity.

\paragraph{What this study does not answer} The measured effects don't imply that the APB-to-Parcoursup transition had generally positive effects on scholarship student orientation. They only show that quota policies had positive effects on certain scholarship populations relative to a counterfactual APB-to-Parcoursup transition without quotas. Parcoursup may have had overall "negative" effects on scholarship recipients, with quotas partially mitigating these effects (compared to their counterfactuals). Furthermore, the data analysis cannot quantify individuals who changed programs due to quotas: I only measure the effects on average prestige gains and the types of programs obtained. Then, even without measured effects on technological track students' admissions, scholarship recipients in the technological track likely still had access to preferred choices through quotas. These quota-enabled choices may, on average, be identical to choices without quotas. However, quotas could still significantly affect the satisfaction of students in the technological track with scholarships. Long-term effects are thus also possible: \citeA{insee_apb_tirage2021} shows that obtaining preferred choices has net positive effects on academic success, suggesting potential long-term academic benefits for scholarship recipients who obtained preferred choices, even for populations for whom the quotas didn't initially affect the prestige of the program obtained.

\subsection{Methodological Limitations}

This section examines potential methodological criticisms of my empirical approach and discusses their validity.

\paragraph{Control Variable Selection and Outcome Specification}
Control variables were selected ex ante based on theoretical relevance to both outcomes and scholarship probability. While methods such as Double Machine Learning or LASSO would be preferable for variable selection, the data's high dimensionality - due to the use of raw administrative data - precluded their use. Importantly, controls were not chosen through data-driven trial and error, but rather based on \textit{a priori} theoretical justification, to prevent risks of \textit{p-hacking}. Moreover, some outcomes — such as academic “prestige” — are complex constructs that are difficult to measure. In addition, the calculation of prestige differs slightly in 2020 due to the COVID-19 pandemic: continuous assessment grades have partially replaced final exam grades. This methodological change could theoretically create a small measurement error if certain programs had a very different level of prestige under the 2020 rules compared to what they would have been with traditional final exams. However, this possibility remains low given that the prestige of programs remains very stable from one year to the next (correlation of approximately 0.85), and this stability does not change between the 2018-2019 and 2019-2020 periods\footnote{Even assuming that there is indeed a systematic measurement bias linked to this change, it would not be enough to bias the results. For our estimates to be biased, students in the treatment group (compared to the counterfactual control group) would also have to enroll disproportionately in programs whose prestige was poorly evaluated in 2020. This additional condition makes bias unlikely and provides an additional methodological safeguard.}. To ensure complete methodological rigor, an ideal approach would utilize prestige measures unaffected by pandemic-related assessment changes. Standardized approaches, such as confirmatory factor analysis (CFA) based on multiple indicators (e.g., selectivity rates, average admission grades, graduate outcomes), could have been employed to capture the underlying construct more rigorously. However, a simpler, more interpretable outcome was preferred to enhance clarity and transparency.

\paragraph{Plausibility of Assumptions} Most of the potential limitations due to assumptions were discussed in Section \ref{section:assumptions}. To summarize, the only robust effect (prestige of final enrollment) is measured on students in the general track. We saw in Section \ref{section:assumptions} that for this population, the compositional change, the risks of selective attrition, and the risks of violations of conditional parallel trends are minimal. We also saw that the treatment is well assigned at the individual level (or at least not at the school level) and that spillover effects induced by reweighting only slightly increase the measured effect. In terms of the effects on candidate behavior, we expected an overestimation due to negative spillovers, but we measured zero effects, so we do not have any particular difficulties. The assumption of weak monotonicity is more difficult to study. Still, simulations on APB 2016 suggest that quotas do not affect the extensive margin: no scholarship student who had not received any offers without quotas receives one thanks to the simulated quotas, and vice versa, which is fully compatible with the weak monotonicity hypothesis. In summary, for the only robust effect measured, the assumptions seem largely plausible and do not call into question the results obtained.

\subsection{Future Research Extensions}

\paragraph{Alternative Outcome Measures}
Testing hypotheses with alternative outcomes would strengthen our understanding of underlying mechanisms. I could use alternative prestige measures to distinguish specific, selective Bachelor's programs (which do not necessarily accept all applicants) from non-selective programs, such as separating medical schools from non-selective Bachelor's programs. I could also try alternative continuous measures of prestige. Such refinements would provide deeper insights into how quotas affect different types of educational pathways and enhance our understanding of the selection mechanisms.

\paragraph{Long-term Academic Success}
Analyzing the effects of quotas on the long-term academic performance of scholarship recipients would provide valuable insights into potential academic mismatch theory. Suppose quotas enable students to access more selective programs than they would otherwise attend. In that case, mismatch theory predicts potential adverse effects on completion rates and academic outcomes due to inadequate preparation for program demands. Conversely, quota benefits could outweigh mismatch costs through improved resources and peer effects. Analysis would examine the progression to Master's programs or admission to Grande École for general track students, alongside completion rates to detect mismatch effects. While I theoretically possess data for tracking students through three years post-baccalaureate for the 2020 Parcoursup cohorts (4 years for 2019, and 5 years for 2018), statistical power limitations may be an obstacle, given the potential low proportion of compliers.

\paragraph{Motivational Mechanisms}
A key finding is the absence of motivational effects among scholarship recipients, particularly on intensive margins. Understanding this behavioral non-response merits further investigation. One explanation may be the complexity of quotas and counterintuitive design: quotas are lower for prestigious programs (due to a baseline lower representation of scholarship recipients), yet they still benefit scholarship applicants to these programs. Alternatively, scholarship recipients may be unaware of the quotas' existence despite their prominence on the platform. Mixed sociological methods (interviews and surveys with scholarship recipients) would help explore these mechanisms.

\section{Conclusion\label{conclusion}}

By introducing quotas into Parcoursup, I was able to evaluate a race-neutral alternative to affirmative action in a European country, based on income and family criteria. This alternative had unique characteristics, as the quotas are calculated endogenously for each program, based on the proportion of scholarship applicants in that program, and because they are "call list quotas" (minimum number of admission offers) rather than "places quotas" (minimum number of filled places). The question of the behavioral effect was particularly interesting, as the quotas were displayed in two different ways: for the first two years in the form of a generic paragraph that did not indicate the quota amount for each program ("opaque" condition), and then by displaying the quota amount (different) on each program page ("transparent" condition).\\

\noindent Contrary to what one might fear from endogenously calculated quotas, the program did not reinforce inequalities. Scholarship students in the general track were able to access slightly more prestigious programs (up to a 0.13 SD effect). In contrast, I found no evidence of a behavioral effect on the prestige of scholarship applicants' applications, nor for academic performance, suggesting little or no behavioral response from students, either in the opaque or transparent conditions. In particular, I found no effect on the probability of applying to highly prestigious programs for scholarship students belonging to the top quartile of students in the general track. Analysis of applicant rankings reveals that the introduction of quotas substantially increased the chances of scholarship students being admitted compared to non-scholarship students. Simulations indicate that the effect on prestige is at least partially explained by changes in the types of programs pursued by scholarship students, and that the quotas calculation method may be responsible for the low intensity of treatment.\\

\noindent These results contradict institutional reports on Parcoursup, one of which evaluated a big effect on the probability of pursuing higher education, as well as a strong behavioral effect on the prestige of applications.  However, these results are consistent with the literature on quotas: \citeA{Backes2012} and \cite{Hinrichs2012} explain that affirmative action has an impact on the quality of the college attended, but that no effect is seen on overall college expectations given the existence of non-selective programs that "absorb" students rejected elsewhere. These studies also show that the observed effects of AA bans are primarily explained by changes in the behavior of the admissions panel \cite{card2005would,antonovics2013were,yagan2016supply}. Overall, revenue-based quotas seem a more viable and efficient race-neutral alternative than x-percent rules, which have potentially null effect \cite{Daugherty2014}\\

\noindent Future analyses would enable us to produce more robust results by correcting for potential compositional change biases, as well as to monitor long-term outcomes on success in higher education, which would allow us to test the hypothesis of the \textit{mismatch}.
\newpage

\bibliographystyle{apacite}
\bibliography{references}

\newpage

\appendix

\section{Appendix}

\setcounter{table}{0}
\setcounter{figure}{0}

\renewcommand{\thetable}{A\arabic{table}}
\renewcommand{\thefigure}{A\arabic{figure}}

\subsection{Methodological and Technical Notes}

\subsubsection{Previous evaluations methodologies}
\label{appendix:methodology}
This section of the appendix aims to review the analyses of \citeA{cesp2021} and \citeA{courdescomptes_ore_2020} to study their methodological validity.

\paragraph{Cour des Comptes analysis} \citeA{courdescomptes_ore_2020} estimated that the effects of quotas were very weak or even non-significant using 2019 Parcoursup data. To make this estimation (presented in Appendix 12 of their report), they used a Spearman correlation between the difference $$\Delta_{i,y} = \frac{\text{Scholarship Recipients Admitted}_{i,y}}{\text{applicants Admitted}_{i,y}} - \frac{\text{Scholarship Recipients applicants}_{i,y}}{\text{applicants}_{i,y}}$$ for each program $i$ at year $y$ on the one hand, and the established quota $q_i$ on the other hand. They conclude that the correlation is null or even non-existent in most fields: for example, in CPGE, programs with high quotas do not have a higher admission ratio for scholarship recipients. However, this analysis seems inappropriate for our purposes and could lead to a wrong conclusion, suggesting an absence of effect even when there are massive effects.\\

\noindent To demonstrate this, let us take an extreme (and fictional) example. Suppose that before the introduction of quotas, scholarship recipients were never called in any program: $$\forall i, \quad 
\Delta_{i,2017} = \frac{\text{Scholarship Recipients Admitted}_{i,2017}}{\text{applicants Admitted}_{i,2017}} = 0$$Let's now suppose that after quotas (let's say in 2019), all scholarship recipients were called while scrupulously respecting the quotas. Due to the formula of quotas, I have $$q_{i,2019} = \frac{\text{Scholarship Recipients applicants}_{i,2019}}{\text{applicants}_{i,2019}} + 0.02$$Suppose also that all applicants, scholarship recipients or not, systematically accept their admission offer (or at least, scholarship recipients and non-recipients accept at the same rate). Then, the share of admitted scholarship recipients should also be $$\frac{\text{Scholarship Recipients Admitted}_{i,2019}}{\text{applicants Admitted}_{i,2019}} = q_{i,2019} = \frac{\text{Scholarship Recipients applicants}_{i,2019}}{\text{applicants}_{i,2019}} + 0.02$$

\noindent Thus, I would have $\Delta_i = 0.02$, which is a constant that, by definition, cannot be correlated with any other variable. I would therefore find a null correlation, even though the quotas had massive effects. Added to this are more classic problems of endogeneity: the courses with the highest quotas tend to have the most scholarship applicants in proportion; therefore, they constitute a very different population from those with low quotas. Consequently, the treatment is not exogenous, and there is a risk of omitted variable bias. In summary, the method employed by the Cour des Comptes appears to be inappropriate for evaluating the effect of quotas.\\

\noindent To assess the behavioral effects, the institution uses the same method, but this time by comparing $\Delta_i$ with the pressure rate ($t_i = 1 - \frac{\text{Rank of last admitted}_i}{\text{Number of applicants}_i}$). This time, they compare the correlations from 2017 and 2019. They conclude that "Parcoursup 2019 could have had a positive 'psychological' effect on the acceptance of scholarship students' applications to the most attractive programs." If I understand correctly, this would mean that scholarship students are more likely to accept applications to which they have been admitted, thanks to quotas. The analysis raises several problems to which I have no answer: what is the relationship between the construct (applicant motivation) and the estimator, and why would there be a motivational effect on accepting a wish in which one has already been admitted anyway? Ultimately, the analysis still suffers from the same endogeneity problems and the same problem that the Spearman correlation using $\Delta_i$ is not relevant for making causal inferences.

\paragraph{CESP analysis on the extensive margin} \citeA{cesp2021} publishes an annual report on the procedure's implementation. Some reports analyze the effect of scholarship recipient quotas. The 2021 report highlights very positive effects, utilizing a before-and-after analysis.

\begin{quote}
The Committee points out the success of quotas that are both realistic and ambitious, which encourage scholarship recipients to apply. They have increased the proportion of scholarship recipients among admitted high school students, which rose from 20\% to 25\%.
\end{quote}
\begin{flushright}
--- CESP, 2021
\end{flushright}

\noindent This passage therefore assumes a 25\% increase (5pp on a base of 20pp) in the number of admitted scholarship recipients, representing a massive effect on the extensive margin. However, the number of scholarship recipients in high school has grown strongly in recent years, making it difficult to determine whether the observed increase is attributable to quota effects or simply to the increase in sample size. The CESP retorts that the proportion of scholarship recipients among high school students admitted to Parcoursup grows faster than the proportion of scholarship recipients in high school.

\begin{table}[h]
\centering

\begin{threeparttable}
\caption{Share of scholarship recipients among high school students and Parcoursup applicants}

\begin{tabular}{|c|p{6cm}|p{6cm}|}
\hline
\textbf{Year} & \textbf{\% of scholarship recipients among high school students} & \textbf{\% of scholarship recipients among Parcoursup applicants} \\
\hline
2018 & 25\% & 19\% \\
\hline
2019 & 26\% & 20\% \\
\hline
2020 & 27\% & 25\% \\
\hline
\end{tabular}

\begin{tablenotes}[para,flushleft]
\small
\textit{Source: RERS and appendix Note Flash SIES n°06 April 2020. Reproduced from the CESP 2021 report.}
\end{tablenotes}
\end{threeparttable}
\end{table}

\noindent However, the analysis poses a methodological problem. We're comparing two different populations, namely high school students (\textit{seconde}, \textit{première}, and \textit{terminale} combined) on one side and high school students in their senior year (only terminal) on the other. Thus, the Parcoursup 2020 cohort was already included in the calculation of scholarship recipients in high school in 2018. One could imagine that the percentage of scholarship recipients suddenly increased in 2018 under the impetus of \textit{seconde} students, and that it took two years for this increase to be felt on Parcoursup (hence in 2020). Then, comparing increases in the number of scholarship recipients without accounting for these 'lag' effects makes the comparison invalid.

\paragraph{CESP analysis on the intensive margin} \citeA{cesp2021} also affirms that nearly 13,000 recipients got admitted to a program they couldn't get accepted to without quotas. The calculation method is not public to my knowledge, but I managed to find the method used (arriving at the exact estimates, regardless of the year compared). The estimator is based on the assumption that, without the quotas, the rank of the last person admitted to a training program would have been the same. For example, in a program $i$, if we observe that we admit applicants up to the 500th position on the waiting list, then we assume that without the quotas, we would also have called up to the 500th position. Then, since we have access to the call lists before and after application of the quota, we look at who the scholarship holders are located behind the 500th in the list before application of the quotas, then who pass in front of the 500th after the application of the quotas, and who definitively accept the proposal.\footnote{This method also supposes strictly no behavioral effect, but only mechanical effect}\\

\noindent This analysis is based on a somewhat plausible (but untestable) hypothesis, but it does not allow for a precise counterfactual analysis: if I assume that an individual was admitted to program A thanks to quotas, to which training would they have been admitted without the quota? Would it have been a training of poorer quality? Of a different type? I cannot know this since the applicants' order of preference is not reported on Parcoursup.

\subsubsection{Risk of Selective Attrition}

\label{appendix:selective_attrition}

The purpose of this section is to determine the extent to which high school seniors who took the baccalaureate exams were registered on Parcoursup. As explained, I don't have an exhaustive list of high school students enrolled in the baccalaureate or their final year, but I do have a list of high school seniors on Parcoursup who also took the baccalaureate exams. I can therefore compare the count of these Parcoursup applicants with the official baccalaureate statistics.\\

\noindent For the official baccalaureate statistics, I base my analysis on the statistical analysis produced each year by the Ministry of Education (e.g., \citeA{thomas2020baccalaureat}), which provides the number of students attending the baccalaureate exam each year, by track, for all final year high-school students in the Ministry of Education data, excluding foreign high-schools and overseas communities. I then count the number of high school seniors registered on Parcoursup who also have a baccalaureate grade recorded on Parcoursup. I find that for the years 2016 to 2019, the ratio of applicants to registered students for the baccalaureate is exceptionally high for both technological and general baccalaureate holders, reaching more than 97\% for general baccalaureate holders and 92\% for technological baccalaureate holders, with excellent stability before and after the introduction of Parcoursup in 2018. It appears that the vast majority of students in the technological or general baccalaureate are registered on Parcoursup, and the rates have not changed with the introduction of Parcoursup, making the risk of attrition particularly low.\\

On the other hand, for vocational baccalaureate holders, the attendance rate on APB/Parcoursup increased from 49.2\% in 2014 to 61.8\% in 2019 on Parcoursup. This increase is very linear and progressive, and cannot be attributed to a "Parcoursup" effect (which would mainly target scholarship holders); however, the risk of selective attrition cannot be excluded.\\

\begin{figure}[htbp]
  \centering
\caption{Ratio of high school graduates with a baccalaureate grade and registered on Parcoursup to the number of students who sat the baccalaureate exam, by track}
    \includegraphics[width=1\textwidth, trim=0 0 0 10, clip]{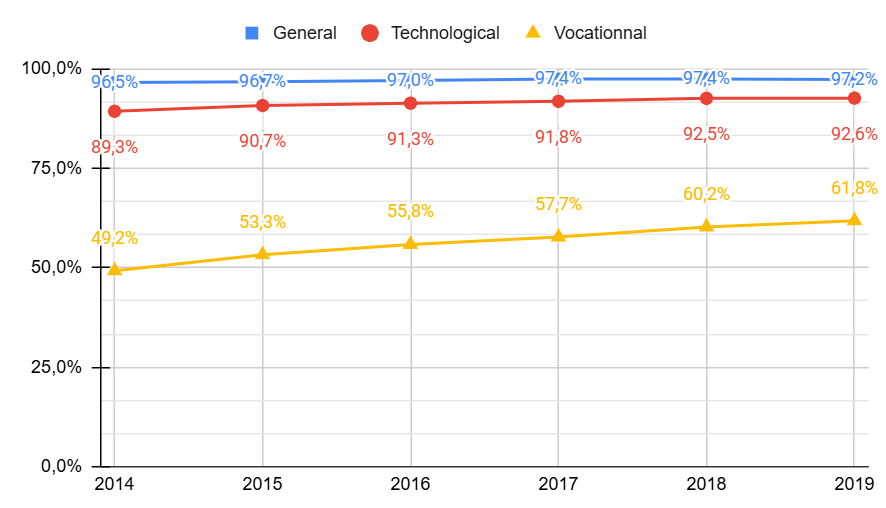}
    \tablereading{By comparing Parcoursup data and aggregated baccalaureate data, I find that 91.3\% of students in the technological track in their final year who took the baccalaureate were also applicants on Parcoursup in 2016. Source: Aggregates from Parcoursup and APB raw data, computed by the author. Inclusion criteria: Any senior high school student included in the baccalaureate data with a grade strictly above 0, in a public or private high school under contract, except military high schools, high schools in an Overseas Collectivity, and high schools abroad. These inclusion criteria are consistent with the aggregated baccalaureate data published by the ministry.}
    \label{appendix:taux_presence_bac}
\end{figure}

\noindent Another concern could arise from the fact that only students who obtained a grade in the baccalaureate are included, which means that they took the exams. This measure was intended to exclude irrelevant applicants from the sample, but it also carries a risk of introducing a similar selection bias. The risk is that the treatment increases the likelihood of attending the baccalaureate exams, which could lead to attrition selection, causing a change in composition that would bias all other estimates. First, I can be reassured that there was no effect on baccalaureate scores: this is very plausible in the case of no effect on the baccalaureate, but it also means that if there were a motivational effect, the compositional change effect created by selective attrition (compliers who take the exams solely because of the existence of quotas) should exactly offset the motivational effect on grades (among students who would have been present with or without quotas), which would be an unlikely coincidence. In addition, \autoref{appendix:passe_bac} shows that the vast majority of students, particularly those in general and technological tracks, as well as scholarship recipients, had a grade recorded on Parcoursup. For general scholarship recipients, for example, this represents at least 98\% of the applicants. The proportion is slightly lower among technological and vocational scholarship students, but it follows a curve that is strictly parallel to that of non-scholarship recipients. In this context, the risk of attrition selection also appears to be very low.\footnote{I could also have done a DRDiD on the probability of taking the exam, but I can see that the curves between scholarship and non-scholarship are already parallel for technological and vocational students even without controls, and for technological and general students the probability is so high at baseline that a linear approximation would not have been effective.}\footnote{As a reminder, the 2020 figures should be treated with caution, as continuous assessment grades replaced the baccalaureate exams due to the COVID crisis. Furthermore, it can be assumed that the COVID crisis, which forced students to attend classes remotely for several months, particularly affected vocational students, who are more prone to academic failure, which would explain the decline observed in 2020.}

\begin{table}[h!]
    \centering
    \caption{Probability of having a high school diploma grade (even in the event of failure) listed in the Parcoursup data (\%)}
    \label{appendix:passe_bac}
    \begin{tabular}{lcccccc}
    \toprule
    \textbf{Year} & \multicolumn{2}{c}{\textbf{General}} & \multicolumn{2}{c}{\textbf{Technological}} & \multicolumn{2}{c}{\textbf{Vocational}} \\
    \cmidrule(lr){2-3} \cmidrule(lr){4-5} \cmidrule(lr){6-7}
     & \textbf{Scholarship} & \textbf{Non-Schol.} & \textbf{Scholarship} & \textbf{Non-Schol.} & \textbf{Scholarship} & \textbf{Non-Schol.} \\
    \midrule
    2014 & 99 & 99 & 96 & 95 & 87 & 84 \\
    2015 & 99 & 99 & 97 & 95 & 90 & 86 \\
    2016 & 99 & 99 & 96 & 95 & 90 & 87 \\
    2017 & 99 & 99 & 97 & 95 & 91 & 87 \\
    2018 & 99 & 99 & 97 & 95 & 91 & 87 \\
    2019 & 99 & 99 & 97 & 95 & 90 & 87 \\
    2020 & 98 & 99 & 95 & 94 & 81 & 77 \\
    \bottomrule
    \end{tabular}
    \tablereading{In 2017, 99\% of scholarship students in the general track obtained a grade for the baccalaureate, and 98\% of non-scholarship students obtained a grade.}
\end{table}

\newpage

\subsubsection{Methodology details}

\label{appendix:methodological_details}

\paragraph{Inclusion criteria} Applicants who are not in their final year of high school, high school students who are not enrolled in either a public high school or a private high school under contract with the ministry, are excluded from our sample. I also exclude applicants whose gender is not specified in official records and those for whom a baccalaureate average cannot be found. For prestige scores for applicants, I only included students who had made at least one confirmed choice. For the final admission prestige score, I only included applicants who were ultimately admitted to a program at the end of the APB or Parcoursup procedure. If the status of scholarship recipient or non-recipient is not specified (missing value), the student is considered not to be a scholarship recipient.

\paragraph{High-school characteristics} To describe the social situation of the high school, I use the Social Position Index (all tracks combined) produced by the Ministry for each high school \citeA{dauphant2023ips}, and I use the 2019 index for all cohorts. I therefore exclude students from high schools that did not exist in 2019, as the IPS does not apply in this case. I also provide information on the type of high school, which is a binary variable indicating whether the high school exclusively enrolls students in general and technological tracks, or whether it offers vocational tracks in addition to its general and technological tracks.

\paragraph{Baccalaureate grade} Students may have two baccalaureate grades, as they may have taken the baccalaureate exam twice (in the main session and the resit session). In this case, only the main session is used, which necessarily corresponds to the lowest overall average. It should be noted that the vast majority of students on Parcoursup have a baccalaureate grade (even if they failed), at least in the general and technological tracks, especially among scholarship recipients, and that the difference between scholarship recipients and non-scholarship recipients is stable over time, which limits the risk of selective attrition (see \autoref{appendix:passe_bac}).

\paragraph{Parent 1 and Parent 2} The Parcoursup databases refer to Parent 1 (or legal guardian 1) and Parent 2 (or legal guardian 2). I don't know precisely how the order of parents is decided. For example, the following rule could be applied: if there are two parents of different sexes, then the father is Parent 1 and the mother is Parent 2. In any case, these rules seem to be stable between APB and Parcoursup, as evidenced by the high stability of the sociodemographic characteristics used.

\paragraph{Confirmed applications} In Parcoursup, for the outcomes on applications, I excluded those that were not confirmed. However, on APB, it is unclear whether the applications in the database are confirmed or not. I also created an alternative specification that excluded unconfirmed applications, and this did not alter any of the results.

\paragraph{Social category} Social categorization utilizes the PCS 2003 reference system produced by \cite{insee2003pcs}. However, this categorization is hierarchical: there are different levels of categories, and the level of detail may vary between APB and Parcoursup. In this case, I "move up" the hierarchy for categories that are too detailed until I find the "highest common denominator", which is the most detailed category that is reported for all years. For example, in some years, unemployed people are grouped in category 82, while in other years they are listed in subcategories of 82: 83, 84, 85, 86. In this case, I re-aggregate all these subcategories to find category 82, which is the most detailed category available for all years.
Additionally, during the matching phase of the Doubly Robust Difference-in-Differences method, it can be challenging to include overly detailed categorical variables that are underrepresented in one of the two groups. For example, suppose there are pastors' children among the non-scholarship recipients but none among the scholarship recipients. In that case, a collinearity problem arises, preventing the model from being estimated. For this reason, all categories representing less than 1\% of scholarship or non-scholarship recipients are grouped into a super-category called "Other". 

\paragraph{Controls included} The DRDiD model includes the following variables: social category of Parent 1 and Parent 2, nationality (French or foreign), gender as recorded in civil records (not gender identity), baccalaureate specialization, type of high school (general and technological only, or not), social position index of the high school, and school district (geographic area). For taxable income and the number of siblings, I apply the following method: First, I bin the taxable income into quintiles, as well as the number of siblings (grouping three or more children into the same bin), with a special bin for unknown values in each case. Then, I add all combinations of bins $Bin(Income) \times Bin(Sibling)$ to my controls. For the general stream, due to a collinearity issue, I grouped all Quintiles 4 (regardless of the number of siblings) in a single factor, as well as Quintiles 5, as there are too few scholarship recipients with these incomes to separate them. Baccalaureate grades are excluded to minimize the risk of "bad control". For the vocational track, I do not include the academy, specialization (due to the large number of options), or type of high school (since it cannot be a general or technological high school by definition). For the CiC model, I do not include controls due to computational power limitations. For pre-college investment, due to a technical issue, the taxable income quintile has not been included for the technological and vocational tracks; however, it has been included for the general track.

\paragraph{DRDiD specification} The reference year is 2017, the last pre-treatment period. Standard errors are calculated using bootstrap (1000 iterations). Confidence intervals are based on an alpha threshold of 5\%. I used the package \texttt{did} by \citeA{CALLAWAY2021200}. Standard errors are not clustered, as I use cross-sectional data.

\paragraph{CiC specification} The reference year is 2017, the last pre-treatment period. Standard errors are calculated using bootstrap (10 iterations). Confidence intervals are based on an $\alpha$-threshold of 5\%. I used the package \texttt{qte} by \citeA{callaway2018qte}.
Since the CiC-function cannot be performed in event-study mode, I perform CiC with all pairs of years between 2014 and 2020, with 2017 specified as the pre-treatment year, so that I have an estimate for each comparison with the reference year, but this prevents us from applying the \citeA{rambachan2023more} correction because it requires the complete covariance matrix (and not only the diagonal). Standard errors are not clustered, as I use cross-sectional data.

\paragraph{Matching} An individual is considered eligible for a program if they are ranked better than the last individual who received an offer of admission, even if that individual has since withdrawn. For the raw difference, I compare the probability of being eligible for a scholarship among scholarship recipients versus non-scholarship high school students for each program, without controlling for any other factors. To do this, I use the application database. I match the preferences of scholarship students with those of non-scholarship students applying for the same programs (exact matching based solely on the program ID). Then, I calculate the average difference (applications across all programs) using a linear regression with scholarship status as the only explanatory variable. This method enables a comparison between scholarship recipients and non-scholarship students, without controlling for the differences in characteristics between scholarship recipients and non-recipients who apply to the same programs. For the profile-matched difference, I match on: gender, overall average in all classes during the school year (overall GPA equivalent, which I have put into 10 bins: between 10/20 and 10.99/20, 11 to 11.99/20, etc.), which is a variable observed by the admissions committees, the academy, the track, and the detailed specialization. I also include the program ID to ensure I compare applications within the same program. However, for computational reasons, I do not match on the content of cover letters or detailed grades by subject or by high school ID. In both specifications, a preference corresponds to a statistical individual, so I cluster standard errors by applicant ID (as there may be several preferences in the same type of program for an applicant). Both matching estimations were realised using the \texttt{MatchIt} package \cite{MatchIt}.

\newpage

\subsubsection{Compositionnal change}

\label{section:compositionnal_change}

One way to see if two populations are identical (in terms of observable characteristics) is to perform a Hausman test to see if the set of control variables predicts membership in a category. In RCTs, this method is used to see if the characteristics of the treated group are identical to those of the control group. Instead, I propose comparing the characteristics of each year's cohort of scholarship recipients with those of the reference year 2017. The purpose of this test is to see whether, by taking only the members of the reference year (2017) and another year (say 2018), we can predict whether an individual belongs to one category or another based on observable characteristics. If the $R^2$ of the model is significantly different from 0, then we do indeed have a composition change effect. I show the $R^2$ obtained for each comparison in \autoref{composition_effect}.\\

We can see that for both the general and technological tracks, there are indeed changes in composition, but they are of the same order as the changes in the pre-treatment years, i.e., an $R^2$ of about 0.003 or less in 2018 and 2019, and an $R^2$ of up to 0.02 in 2020. Although the change in composition is significant due to the large sample size, it remains relatively small. Furthermore, there was no break in the pre-trends even though we can see changes in composition of the same order during the pre-treatment periods, which reinforces our confidence in the significant result on the prestige of the final enrolment, which does not appear to be explainable by composition effects. For vocational tracks, we observe a slightly higher composition effect, with $R^2$ values ranging from 0.02 to 0.04. It is substantially higher than that obtained in the pre-treatment period but remains relatively weak.

\begin{table}[htbp]
\centering
\caption{McFadden $R^2$ of Model Predicting 2017 Cohort Membership vs. Other Cohorts Using Control Variables}
\label{composition_effect}
\begin{tabular}{lcccccc}
\toprule
Population & 2014 & 2015 & 2016 & 2018 & 2019 & 2020 \\
\midrule
General        & 0.003*** & 0.001*** & 0.001*** & 0.001*** & 0.003*** & 0.017*** \\
Professional   & 0.004*** & 0.001*** & 0.000*** & 0.001*** & 0.003*** & 0.018*** \\
Technological  & 0.004*** & 0.002*** & 0.001*** & 0.018*** & 0.020*** & 0.038*** \\
\bottomrule
\end{tabular}
\end{table}

In summary, we suspect a very slight compositional change effect, which will need to be corrected when the estimators proposed by \citeA{santanna2025differenceindifferencescompositionalchanges} become available. Still, these effects remain negligible and do not appear to undermine the validity of the observed results.

\newpage

\subsubsection{Treatment assignment level}

\label{section:treatment_assignement}

To be eligible for a scholarship, you must meet the specific eligibility requirements and submit your application. This last condition raises the question of the level of treatment assignment, and therefore, as a corollary, the question of the level of clustering of standard errors. For example, if applying for a scholarship (conditional on being eligible) depends solely on high school affiliation (extreme example: some high schools require eligible students to apply, while others prohibit it), then once the control sample has been reweighted to include only individuals who are likely to be eligible, the standard errors must be clustered at the high school level, because it is membership in one high school category or another that determines assignment to the treatment \cite{rambachan2025design}.\\

To ensure that the treatment is not assigned to the high school, I use mixed models. The idea is as follows: I estimate a first model with only the control variables to determine treatment assignment. In a second step, I add random effects with the high school identifier. Due to limitations in computing power, I have only included high school students from a sample of 10\% of high schools, and this is limited to the 2019 school year. If treatment is assigned at the high school level (conditional on eligibility), adding random effects from the high school ID should bring the $R^2$ close to 1\footnote{Results of a simulation on fictitious data}. Conversely, suppose the $R^2$ does not change. In that case, it means that, conditional on being likely eligible, there is no high school effect, and the assignment of the treatment is based on other (probably individual) factors. The results of the analysis are presented in \autoref{mixed_model}.

\begin{table}[htbp]
\centering
\caption{Comparison of Marginal $R^2$ (Only FE) and Conditional $R^2$ (FE + RE) for Mixed Models}
\label{mixed_model}
\begin{tabular}{lcccc}
\toprule
 & \multicolumn{2}{c}{Theoretical} & \multicolumn{2}{c}{$\Delta$} \\
\cmidrule(lr){2-3} \cmidrule(lr){4-5}
Group & Marginal $R^2$ & Conditionnal $R^2$ & Marginal $R^2$ & Conditionnal $R^2$ \\
\midrule
General       & 0.646 & 0.648 & 0.466 & 0.467 \\
Technological & 0.598 & 0.614 & 0.480 & 0.493 \\
Vocationnal   & 0.511 & 0.558 & 0.430 & 0.471 \\
\bottomrule
\end{tabular}
\end{table}

We can see that, regardless of the population, adding random effects (moving from Marginal $R^2$ to Conditional $R^2$) yields almost no gain in $R^2$, even though we have not corrected for overfitting. It would therefore seem that the assignment of treatment among those eligible is not decided at the high school level, but rather by idiosyncratic factors that are probably individual.
\newpage

\subsection{Additional Material}

\subsubsection{Parcoursup screenshots}

\label{appendix:parcoursup_screenshots}

\begin{figure}[htbp]
  \centering
\caption{Screenshot of a Parcoursup program page displaying quota information (post-2020). Original in French (bottom) and English transcription (top)}

  \includegraphics[width=0.8\textwidth, trim=0 100 0 0, clip]{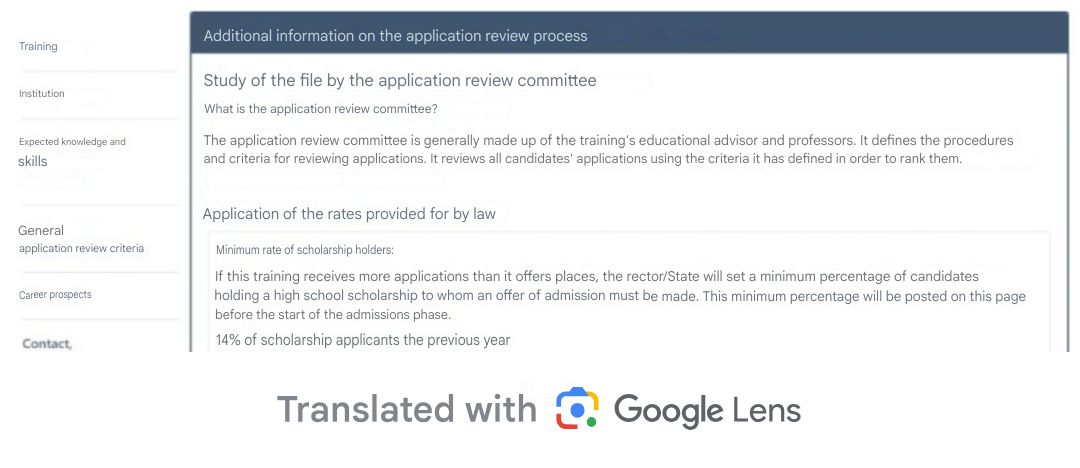}
  \includegraphics[width=0.8\textwidth, trim=0 0 0 0, clip]{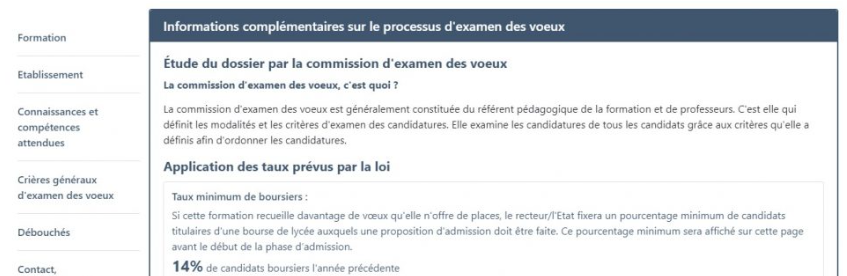}

  \label{appendix:screenshot_apres_2020}
\end{figure}
  \tablenote{The translated image is generated by Google Lens. The original image is taken from an archive of Julien Gossa's EducPro blog.}{On the program's page on Parcoursup, I see that 14\% of the applicants received scholarships the previous year.}

\begin{figure}[htbp]
  \centering
\caption{Screenshot of a Parcoursup program page showing quota information (2018–2019). Original in French (bottom) and English transcription (top)}

  \includegraphics[width=0.8\textwidth, trim=0 100 0 0, clip]{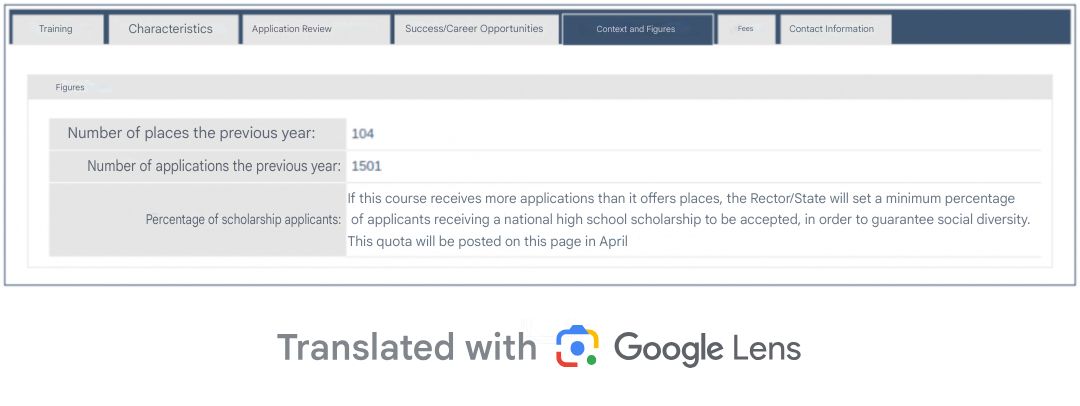}
  \includegraphics[width=0.8\textwidth, trim=0 0 0 0, clip]{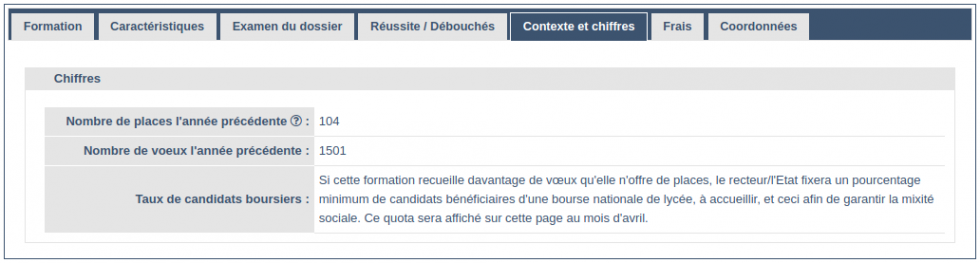}

  \label{appendix:screenshot_avant_2020}
\end{figure}
  \tablenote{The translated image is generated by Google Lens. The original image is taken from an archive of Julien Gossa's EducPro blog.}{On the program page on Parcoursup, there is mention of a scholarship quota, but the value is not displayed.}

\FloatBarrier

\newpage
\subsection{Descriptive statistics}

\subsubsection{General track}

\label{descriptive:general}

\input{productions_drdid/g_sans_notes/latex/latex_stat_descriptives_boursier}
\tablereading{In 2017, 18\% of scholarship students in general education had a first legal guardian who was inactive (but not retired), and 25\% had no second parent or legal guardian.}

\newpage 
\input{productions_drdid/g_sans_notes/latex/latex_stat_descriptives_non_boursier}
\tablereading{In 2017, 2.3\% of non-scholarship students in the general track had a first legal guardian who was inactive (but not retired), and 7,2\% had no second parent or legal guardian.}

\newpage

\input{productions_drdid/g_sans_notes/latex/latex_stat_outcomes_boursier}
\tablereading{In 2017, 15\% of scholarship students in the general track applied to at least one CPGE, but only 5.9\% were ultimately enrolled in a CPGE. The average grade of scholarship students in the general track on the baccalaureate exam was 11.42/20.}

\newpage 
\input{productions_drdid/g_sans_notes/latex/latex_stat_outcomes_non_boursier}
\tablereading{In 2017, 23.4\% of non-scholarship students in the general track applied to at least one CPGE, but only 10.7\% were ultimately enrolled in a CPGE. The average grade of scholarship students in the general track on the baccalaureate exam was 12.54/20.}

\FloatBarrier

\newpage

\subsubsection{Technology track}

\label{descriptive:techno}

\input{productions_drdid/t_sans_notes/latex/latex_stat_descriptives_boursier}

\newpage 
\input{productions_drdid/t_sans_notes/latex/latex_stat_descriptives_non_boursier}

\newpage

\input{productions_drdid/t_sans_notes/latex/latex_stat_outcomes_boursier}

\newpage 
\input{productions_drdid/t_sans_notes/latex/latex_stat_outcomes_non_boursier}

\FloatBarrier

\newpage

\subsubsection{Vocational track}
\label{descriptive: vocational}

\input{productions_drdid/p_sans_notes/latex/latex_stat_descriptives_boursier}
\newpage 
\input{productions_drdid/p_sans_notes/latex/latex_stat_descriptives_non_boursier}
\newpage

\input{productions_drdid/p_sans_notes/latex/latex_stat_outcomes_boursier}
\newpage 
\input{productions_drdid/p_sans_notes/latex/latex_stat_outcomes_non_boursier}
\FloatBarrier

\newpage
\subsection{Detailed results}

\subsubsection{General track}

\input{productions_drdid/g_sans_notes/tables/unified_event_study_table_corrected}
\tablenote{DRDiD estimation.}{In 2020, quotas increased the probability of scholarship students in the general track ultimately enrolling in a DUT by 1.5 percentage points. The effect is significant ($p<0.001$), and there is no significant pre-trend effect ($p=0.466$).}
\newpage

\input{productions_drdid/g_sans_notes/tables/unified_breaking_points_drdid_table}
\tablenote{Breakdown values are tested in increments of 10. A breakdown value of 0 means that the effect was not significant even without Honest-DiD correction.}{The breaking down value for the effect on DUT in 2020 is 0.4, meaning that if I assume a violation of the parallel trend of magnitude 0.4, i.e., the parameter $M=0.4$ in the method of \citeA{rambachan2023more}, the effect is no longer significant.}
\newpage

\input{productions_drdid/g_sans_notes/tables/unified_confidence_intervals_drdid_table}
\tablereading{The estimated effect of DRDiD on admission to DUT is between 1pp and 2pp (between 0.03 SD and 0.07 SD) in 2020.}
\newpage

\input{productions_drdid/g_sans_notes/tables/cic_event_study_table_corrected}
\tablereading{Using a CiC estimator, the ATE of the quota on the maximum prestige of applications is -0.047 SD (Cohen's d), and the effect is highly significant. However, there is a highly significant pre-trend effect: the result is -0.029 SD in 2016 (pre-treatment).}
\FloatBarrier

\newpage

\subsubsection{Technology track}

\input{productions_drdid/t_sans_notes/tables/unified_event_study_table_corrected}
\newpage

\input{productions_drdid/t_sans_notes/tables/unified_breaking_points_drdid_table}
\newpage

\input{productions_drdid/t_sans_notes/tables/unified_confidence_intervals_drdid_table}
\newpage

\input{productions_drdid/t_sans_notes/tables/cic_event_study_table_corrected}
\FloatBarrier

\newpage

\subsubsection{Vocational track}

\input{productions_drdid/p_sans_notes/tables/unified_event_study_table_corrected}
\vspace{0.5ex}%
{\footnotesize \raggedright \textit{Note: Unlike the other specifications, I did not include the specification or the academy because otherwise the estimate did not work.}\par}%
\newpage

\input{productions_drdid/p_sans_notes/tables/unified_breaking_points_drdid_table}
\newpage

\input{productions_drdid/p_sans_notes/tables/unified_confidence_intervals_drdid_table}
\newpage

\input{productions_drdid/p_sans_notes/tables/cic_event_study_table_corrected}

\FloatBarrier

\newpage

\subsubsection{General track, first quartile}

\label{appendix:first_quartile}

\input{productions_drdid/g_avec_tres_bonnes_notes/tables/unified_event_study_table_corrected}
\newpage

\input{productions_drdid/g_avec_tres_bonnes_notes/tables/unified_breaking_points_drdid_table}
\newpage

\input{productions_drdid/g_avec_tres_bonnes_notes/tables/unified_confidence_intervals_drdid_table}

\FloatBarrier

\newpage

\subsubsection{General track, DiD without reweighting}

\input{productions_drdid/g_reg/tables/unified_event_study_table_corrected}
\noindent \tablenoteonly{DiD with controls but without reweighting, using \texttt{reg} method in \texttt{did} package. Standard errors estimated by bootstrap, 200 iterations. Non-clustered standard errors. ***: $p<0.001$}

\input{productions_drdid/g_reg/tables/unified_breaking_points_drdid_table}

\input{productions_drdid/g_reg/tables/unified_confidence_intervals_drdid_table}
\FloatBarrier

\newpage
\subsection{Matching}

\input{productions_matching/tableau_general_2}
\noindent \tablenote{Standard errors are shown in parentheses. + p < 0.1, * p < 0.05, ** p < 0.01, *** p < 0.001. "Applicants" refer to unique applicants (who can do several applications the same year).}{In 2017, the probability of being eligible (ranked higher than the last person who was finally enrolled) for a CPGE for a scholarship recipient is 4.94pp less than for a non-scholarship recipient who applied for the same program. However, when comparing only to non-scholarship recipients with a similar profile, the difference drops to 0.77 percentage points and becomes non-significant.}
\newpage
\subsection{Simulations}

\subsubsection{All track}

\label{appendix:all_2}

\begin{figure}[htbp]
  \centering
\caption{Decomposition of the ATE for average prestige gains in final enrollment (prestige score out of 100). All tracks included}

  \includegraphics[width=1\textwidth, trim=0 0 0 0, clip]{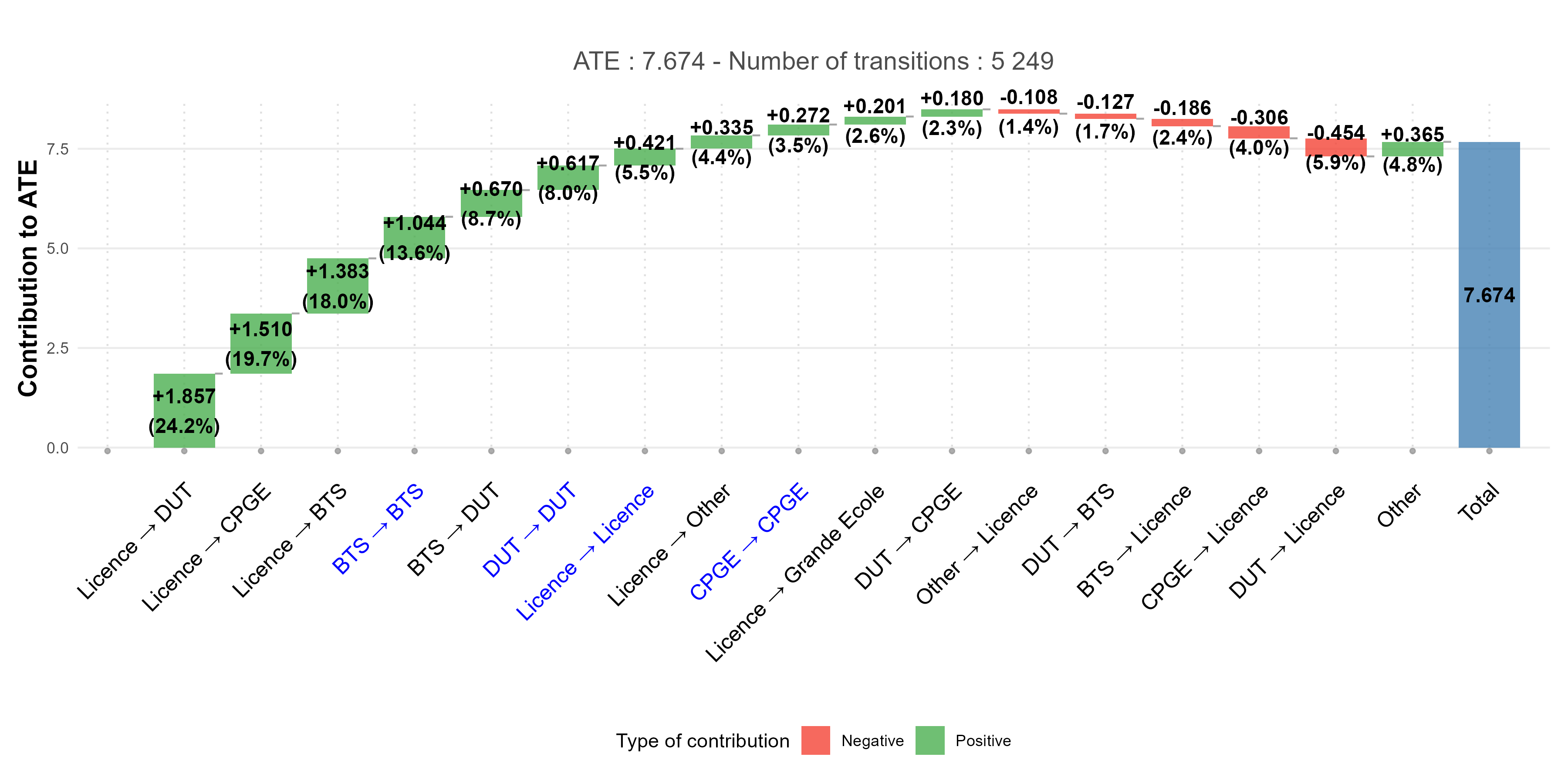}
  \label{appendix:all_2:decomposition}
\end{figure}

\begin{figure}[htbp]
  \centering
\caption{Effects of quotas on the type of programs attended by scholarship students (ITT). All tracks included}

  \includegraphics[width=0.7\textwidth, trim=0 0 0 0, clip]{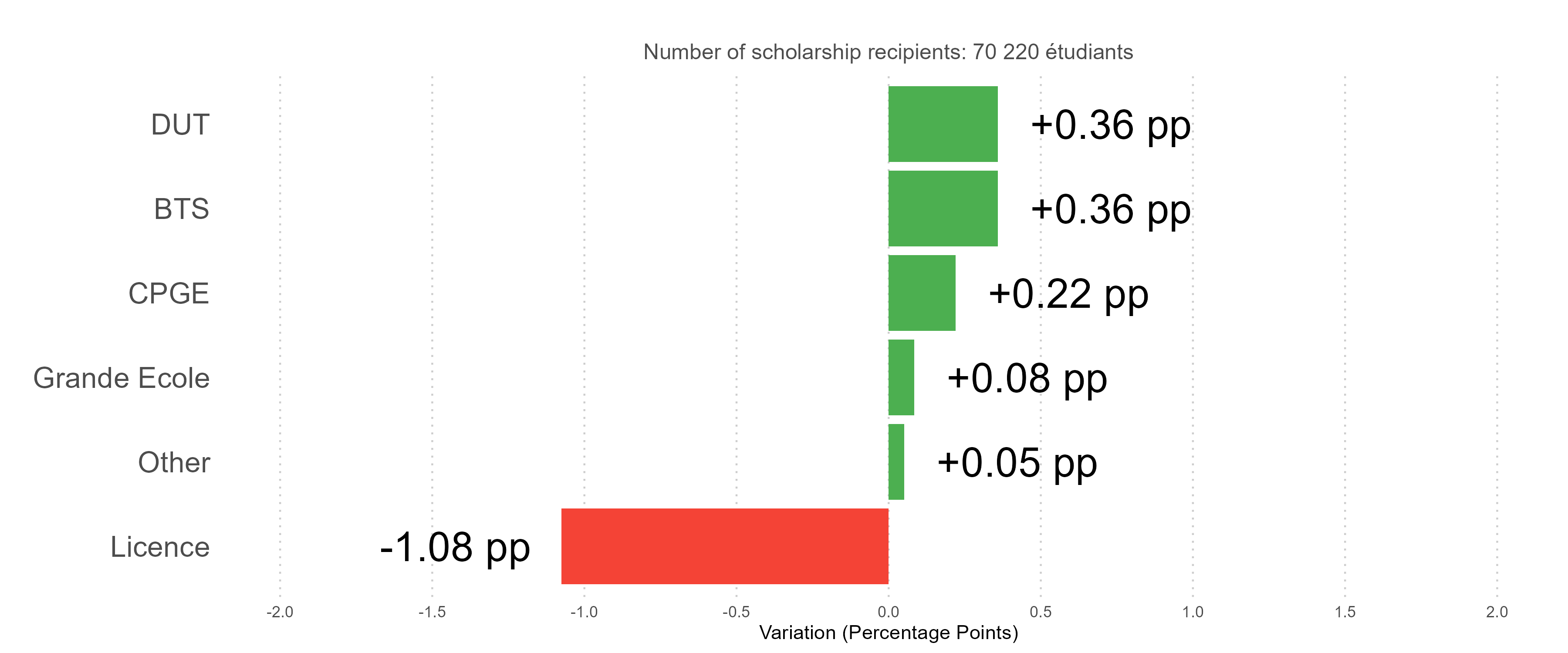}
  \label{appendix:all_2:pp}
\end{figure}
  \tablereading{In ITT (including scholarship recipients who changed programs via quotas and those who did not change), the probability of continuing to a bachelor's degree decreases by 1.08pp overall. Still, the probability of continuing to a DUT increases by 0.36 percentage points with quotas in place.}

\newpage

\subsubsection{General track}

\label{appendix:g_2}
\begin{figure}[htbp]
  \centering
\caption{Decomposition of the ATE for average prestige gains in final enrollment (prestige score out of 100). General tracks only.}

  \includegraphics[width=1\textwidth, trim=0 0 0 0, clip]{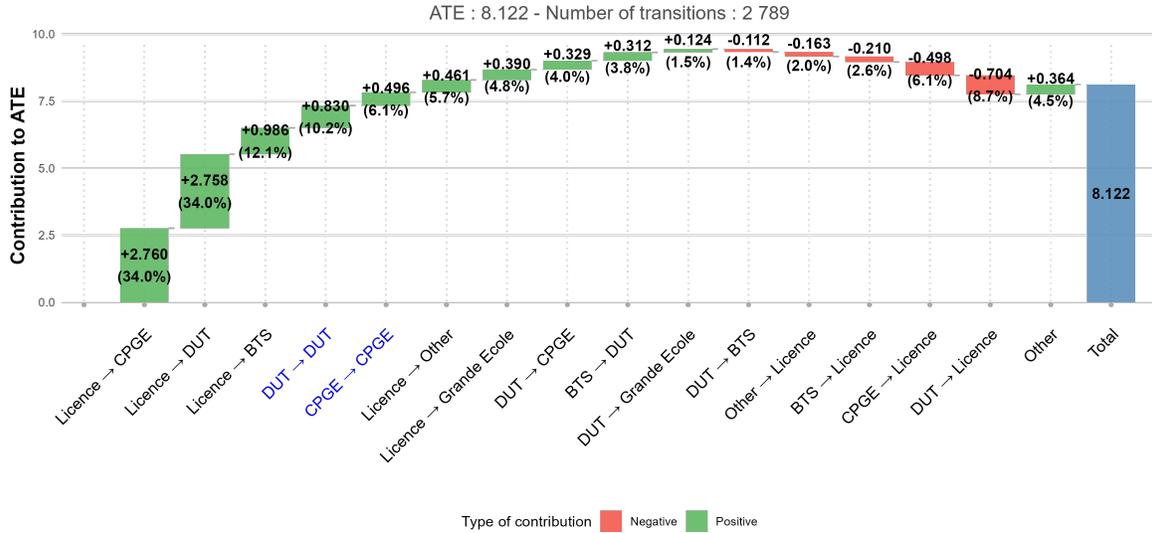}
  \label{appendix:g_2:decomposition}
\end{figure}

\begin{figure}[htbp]
  \centering
\caption{Effects of quotas on the type of programs attended by scholarship students (ITT). General track only.}

  \includegraphics[width=1\textwidth, trim=0 0 0 0, clip]{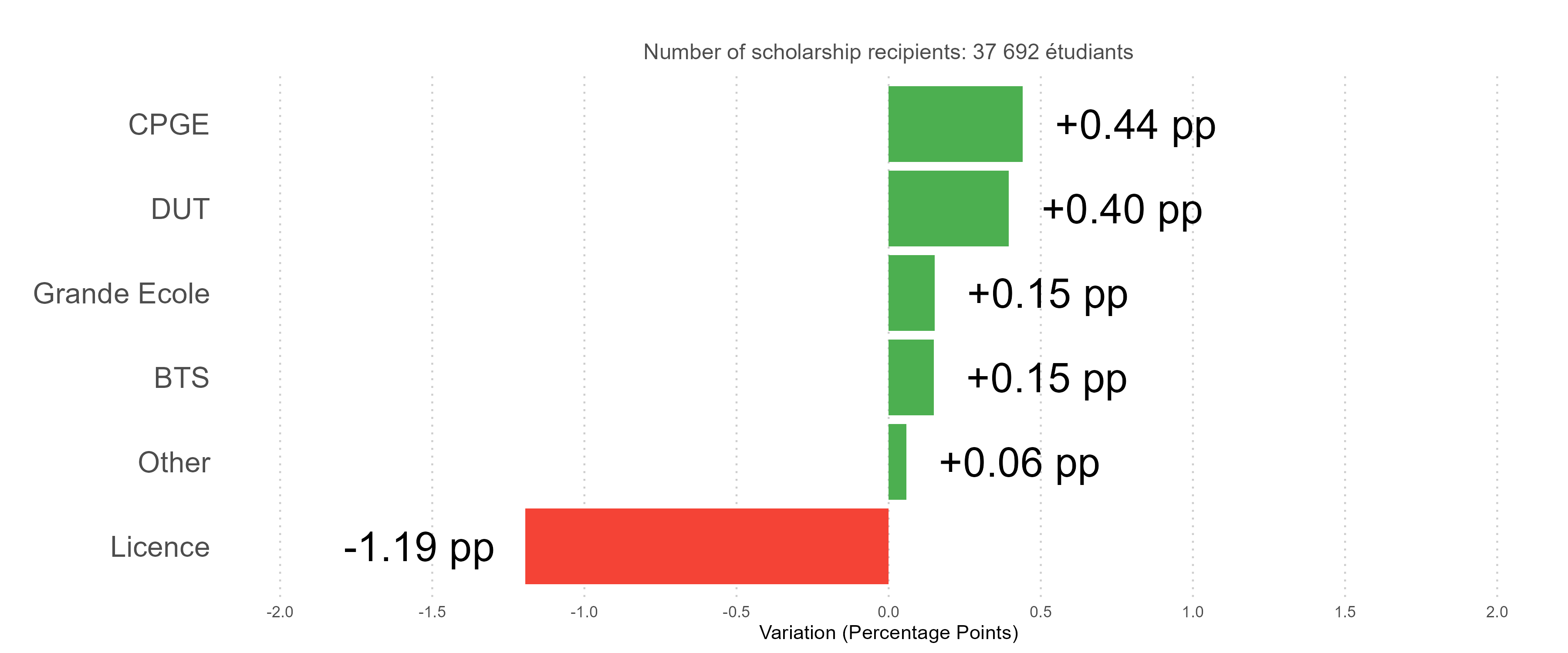}
  \label{appendix:g_2:pp}
\end{figure}

\newpage

\begin{figure}[htbp]
  \centering
\caption{Number of transitions between program types among scholarship students. General track only.}
  \includegraphics[width=0.7\textwidth, trim=0 0 0 0, clip]{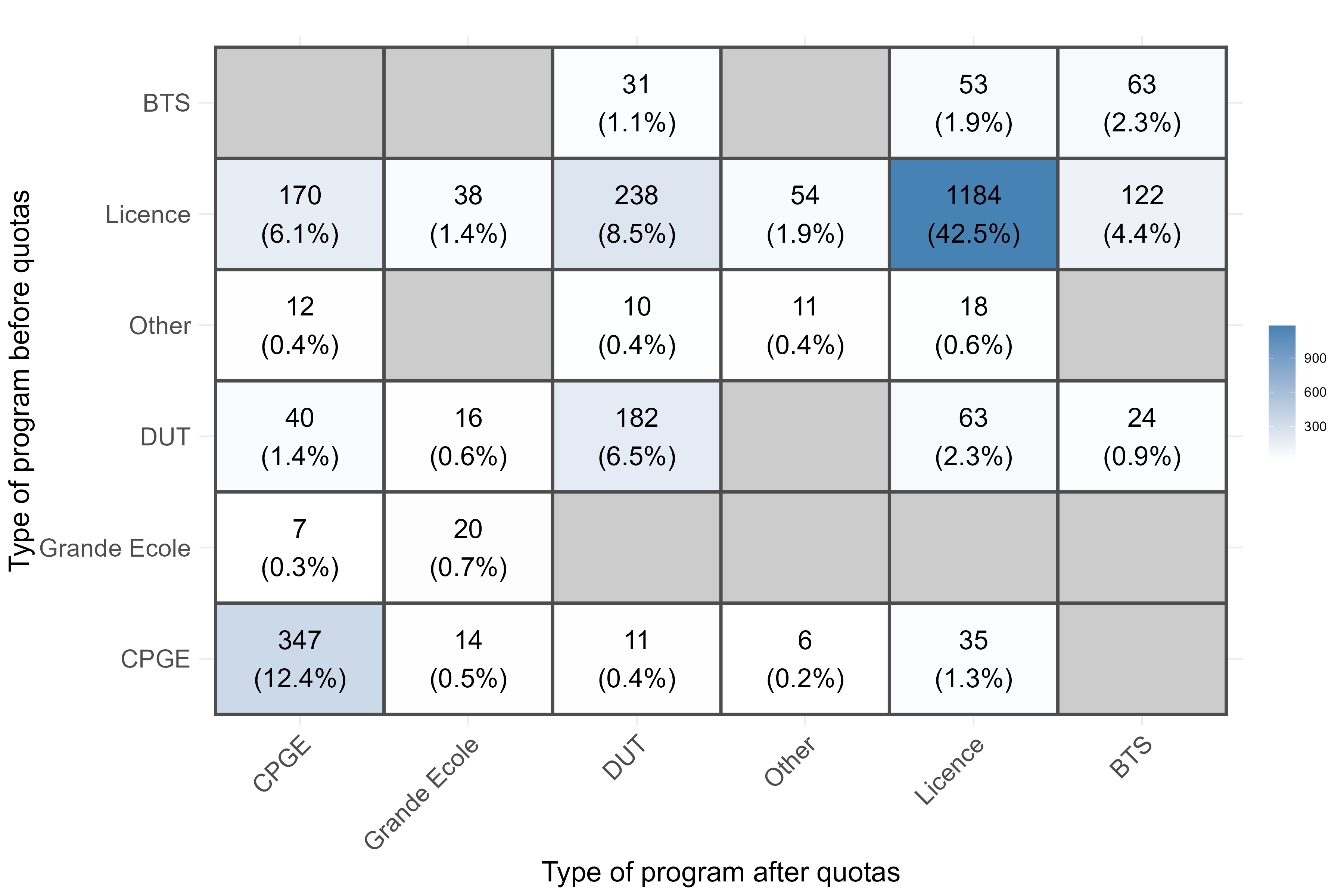}
  \label{appendix:g_2:effectifs}
\end{figure}

\begin{figure}[htbp]
  \centering
\caption{Average prestige gain by type of program transition among scholarship students. General track only.}

  \includegraphics[width=0.7\textwidth, trim=0 0 0 0, clip]{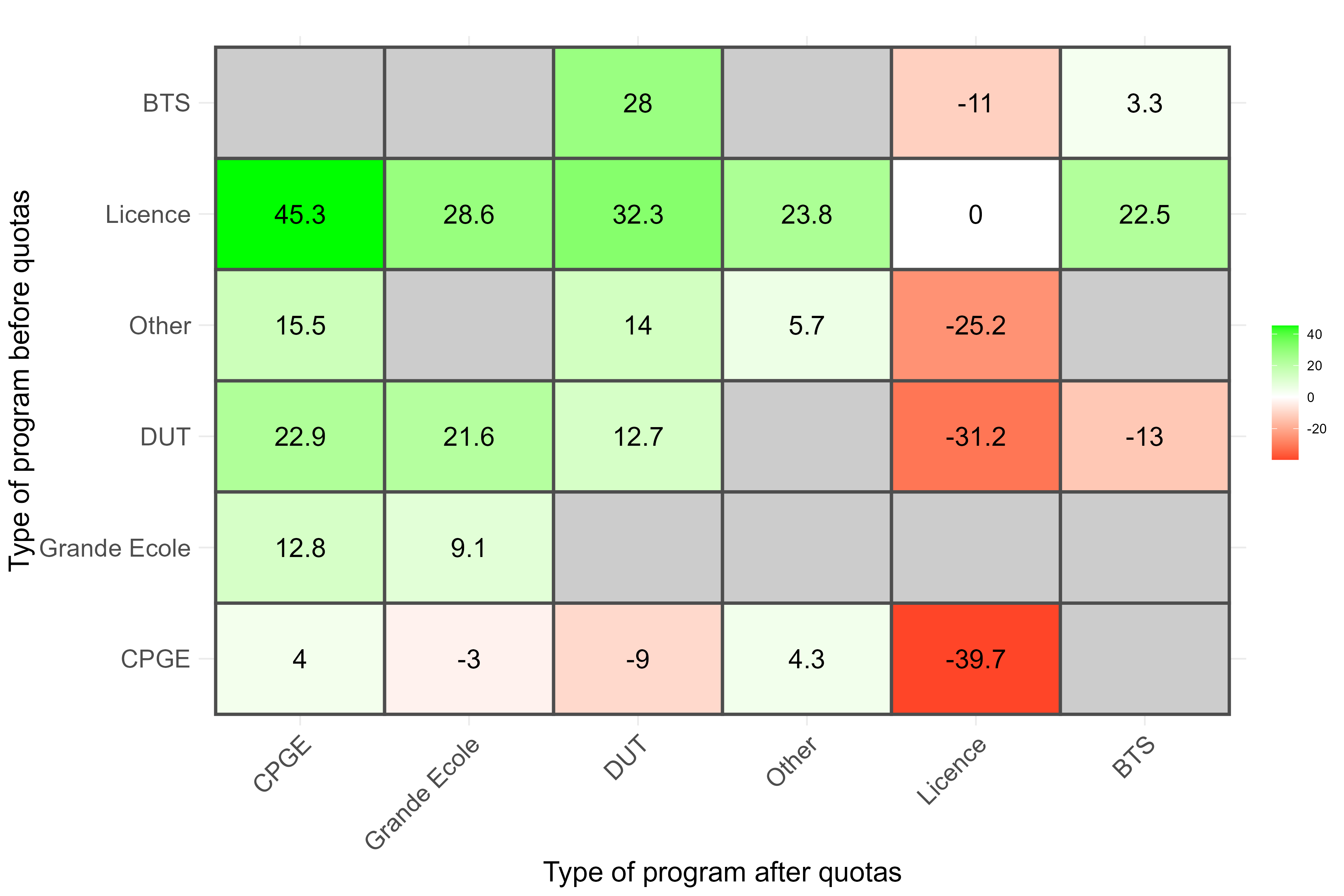}
  \label{appendix:g_2:gains}

\end{figure}

\FloatBarrier

\newpage

\subsubsection{Technology track}

\label{appendix:t_2}

\begin{figure}[htbp]
  \centering
\caption{Decomposition of the ATE for average prestige gains in final enrollment (prestige score out of 100). Technological tracks only.}

  \includegraphics[width=1\textwidth, trim=0 0 0 0, clip]{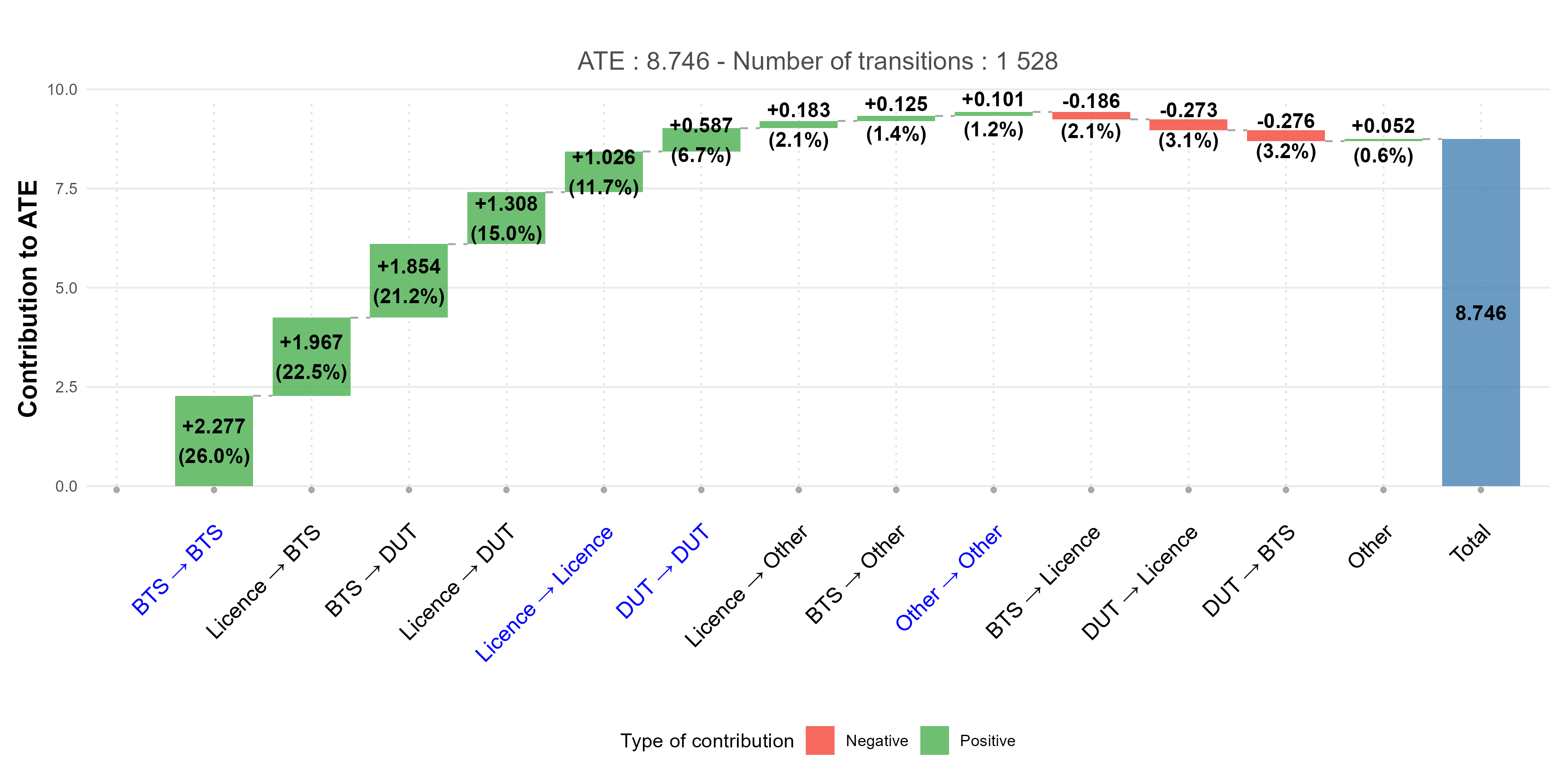}
  \label{appendix:t_2:decomposition}
  
\end{figure}

\begin{figure}[htbp]
  \centering
\caption{Effects of quotas on the type of programs attended by scholarship students (ITT). Technological track only.}

  \includegraphics[width=1\textwidth, trim=0 0 0 0, clip]{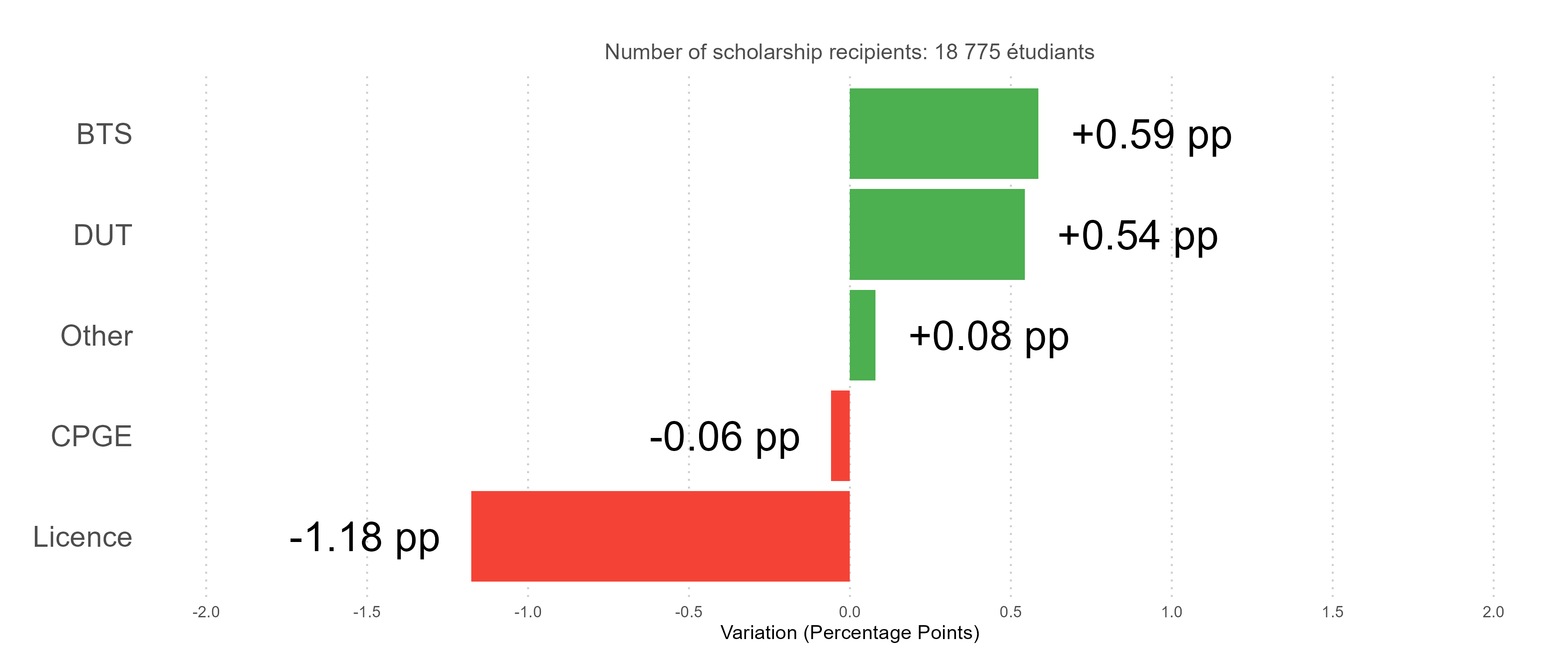}
  \label{appendix:t_2:pp}
  
\end{figure}

\newpage

\begin{figure}[htbp]
  \centering
\caption{Number of transitions between program types among scholarship students. Technological track only.}
  
  \includegraphics[width=0.7\textwidth, trim=0 0 0 0, clip]{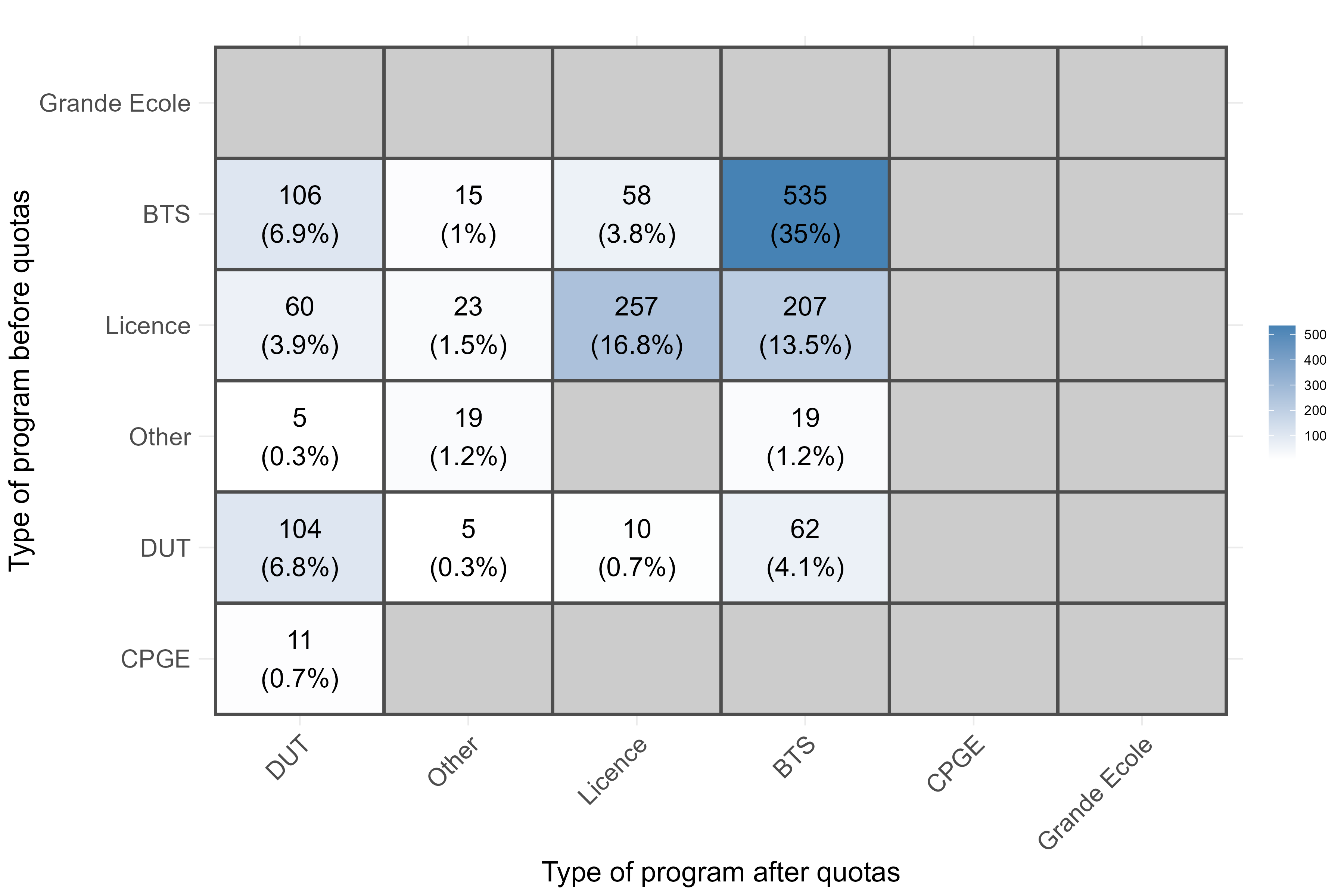}
  \label{appendix:t_2:effectifs}

\end{figure}

\begin{figure}[htbp]
  \centering
\caption{Average prestige gain by type of program transition among scholarship students. Technological track only.}

  \includegraphics[width=0.7\textwidth, trim=0 0 0 0, clip]{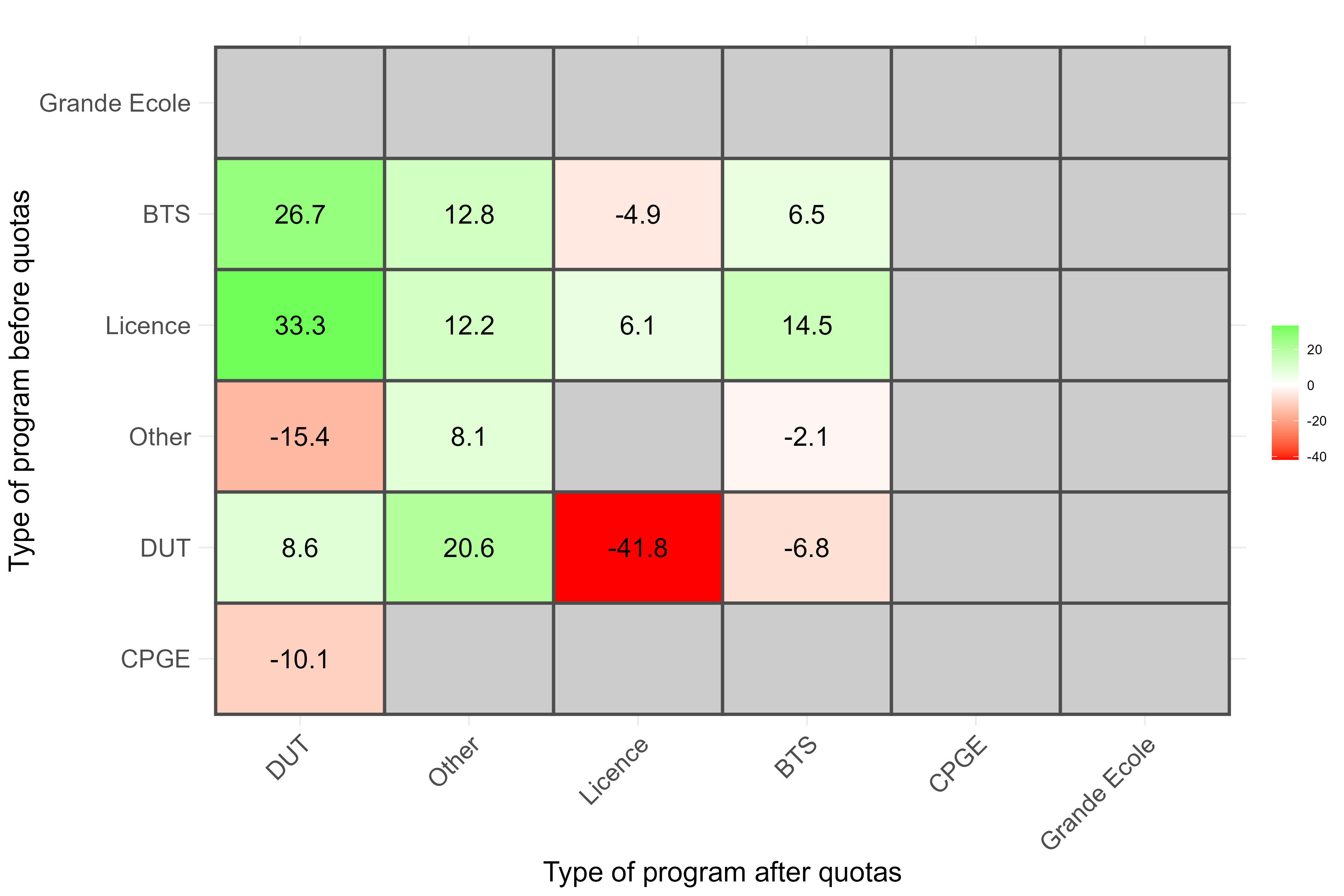}
  \label{appendix:t_2:gains}

\end{figure}

\FloatBarrier

\newpage

\subsubsection{Vocational track}

\label{appendix:p_2}

\begin{figure}[htbp]
  \centering
\caption{Decomposition of the ATE for average prestige gains in final enrollment (prestige score out of 100). Vocational tracks only.}

  \includegraphics[width=1\textwidth, trim=0 0 0 0, clip]{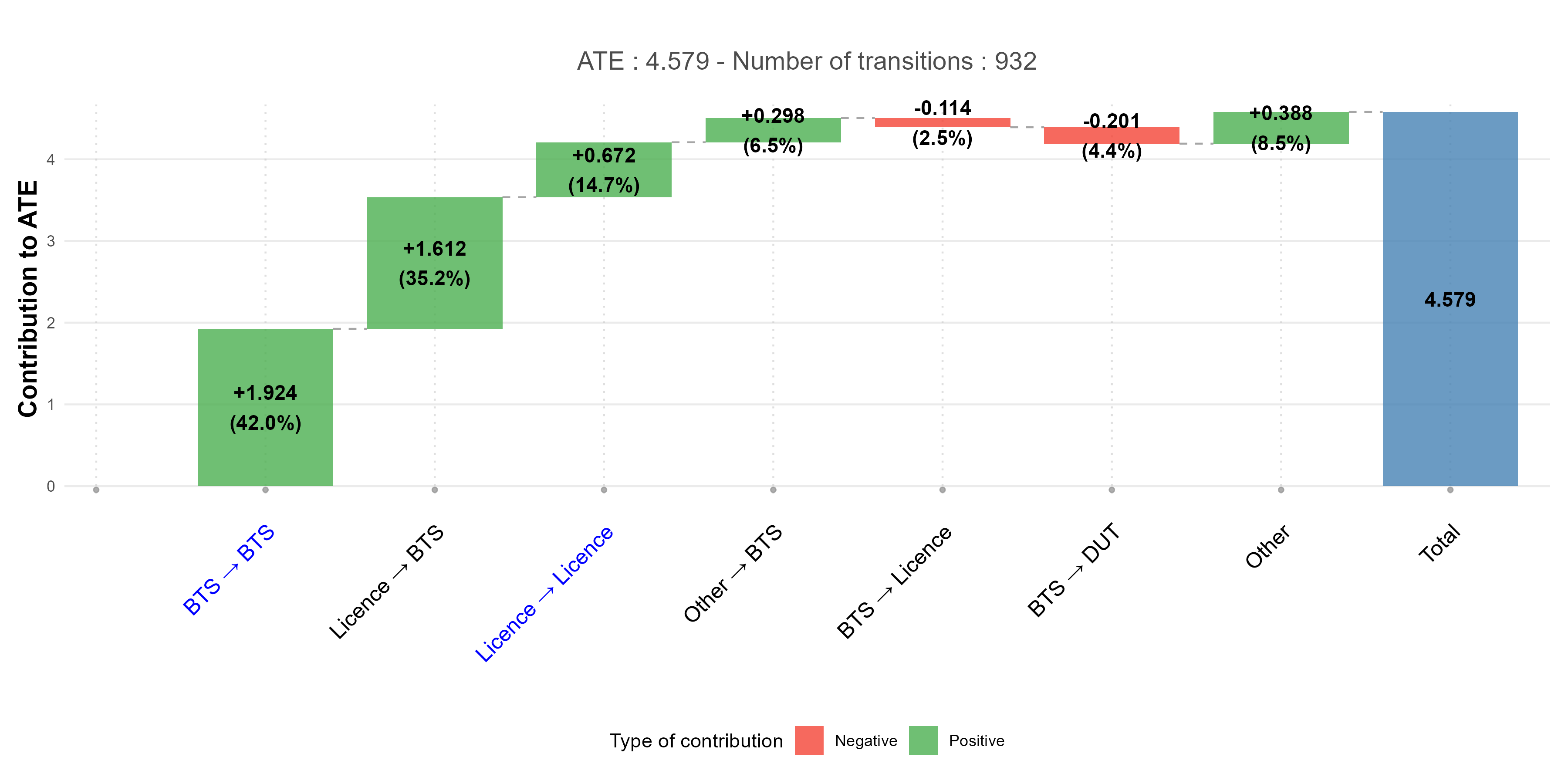}
  \label{appendix:p_2:decomposition}
\end{figure}

\begin{figure}[htbp]
  \centering
\caption{Effects of quotas on the type of programs attended by scholarship students (ITT). Vocational track only.}

  \includegraphics[width=1\textwidth, trim=0 0 0 0, clip]{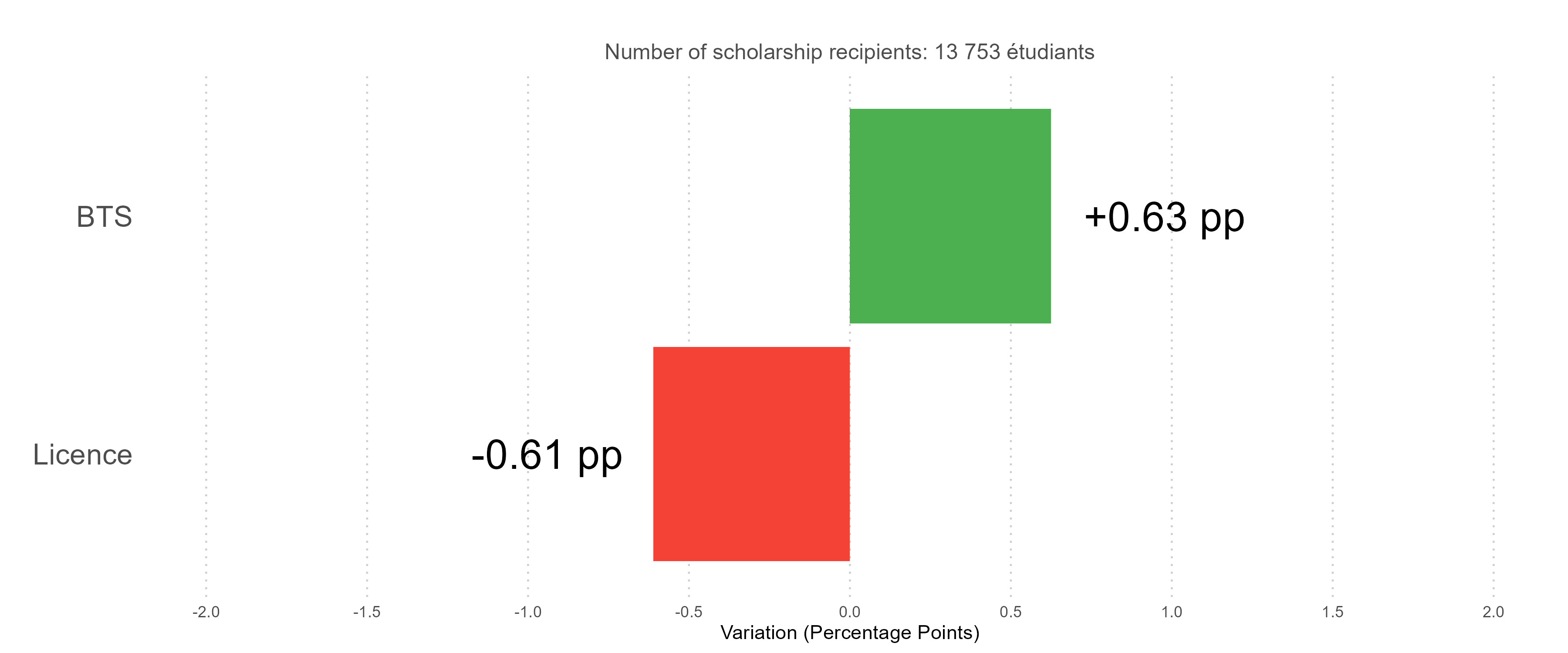}
  \label{appendix:p_2:pp}
\end{figure}

\noindent \tablenoteonly{Changes of less than 0.05pp are not shown.}

\newpage

\begin{figure}[htbp]
  \centering
\caption{Number of transitions between program types among scholarship students. Vocational track only.}

  \includegraphics[width=0.7\textwidth, trim=0 0 0 0, clip]{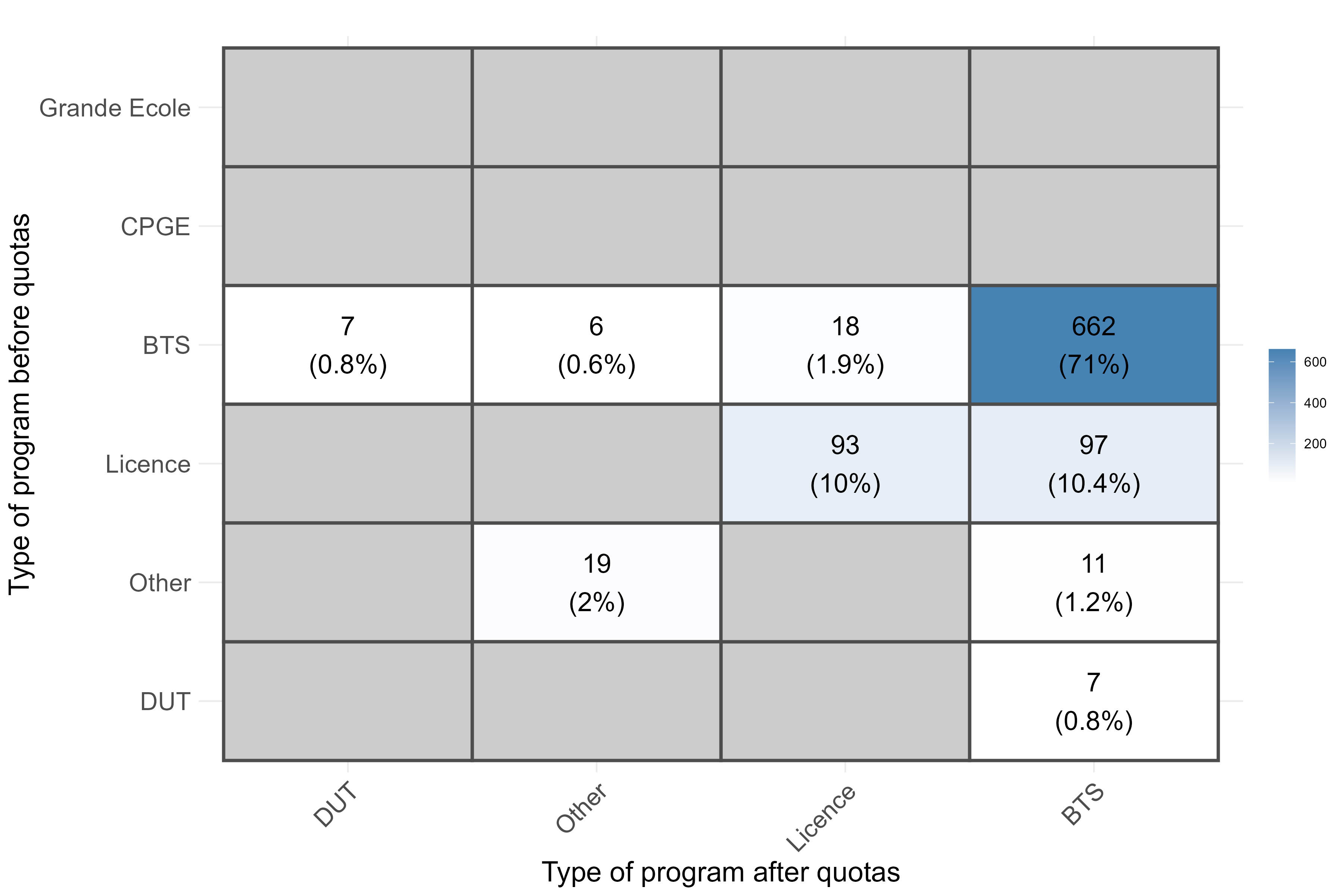}
  \label{appendix:p_2:effectifs}

\end{figure}

\begin{figure}[htbp]
  \centering
\caption{Average prestige gain by type of program transition among scholarship students. Vocational track only.}

  \includegraphics[width=0.7\textwidth, trim=0 0 0 0, clip]{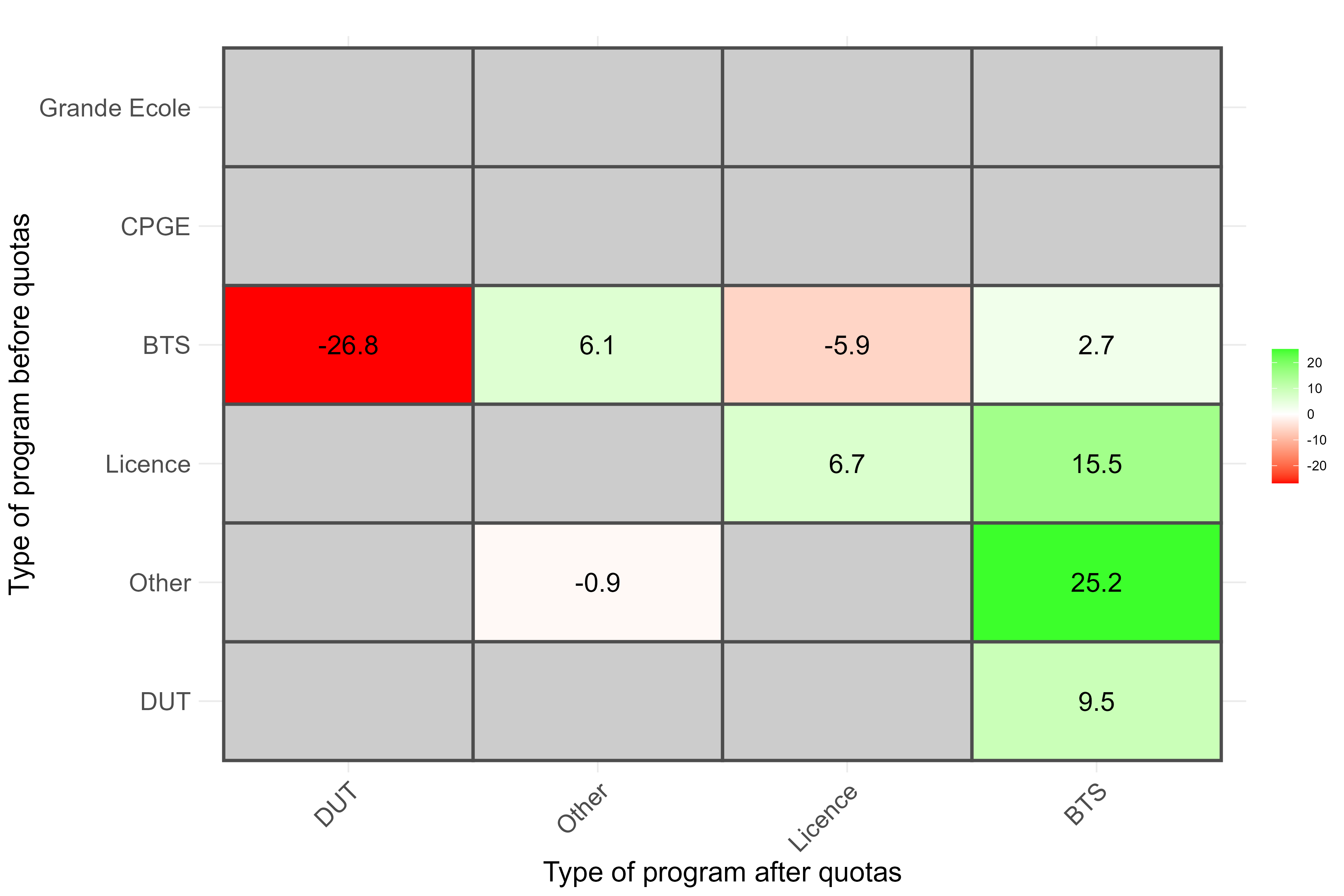}
  \label{appendix:p_2:gains}

\end{figure}

\FloatBarrier

\newpage

\subsubsection{All track, alternative quota rule}

\label{appendix:all_0}

\begin{figure}[htbp]
  \centering 
\caption{Decomposition of the ATE for average prestige gains in final enrollment (prestige score out of 100). All tracks included. Alternative rule of quotas (no +2pp bonus).}
  \includegraphics[width=1\textwidth, trim=0 0 0 0, clip]{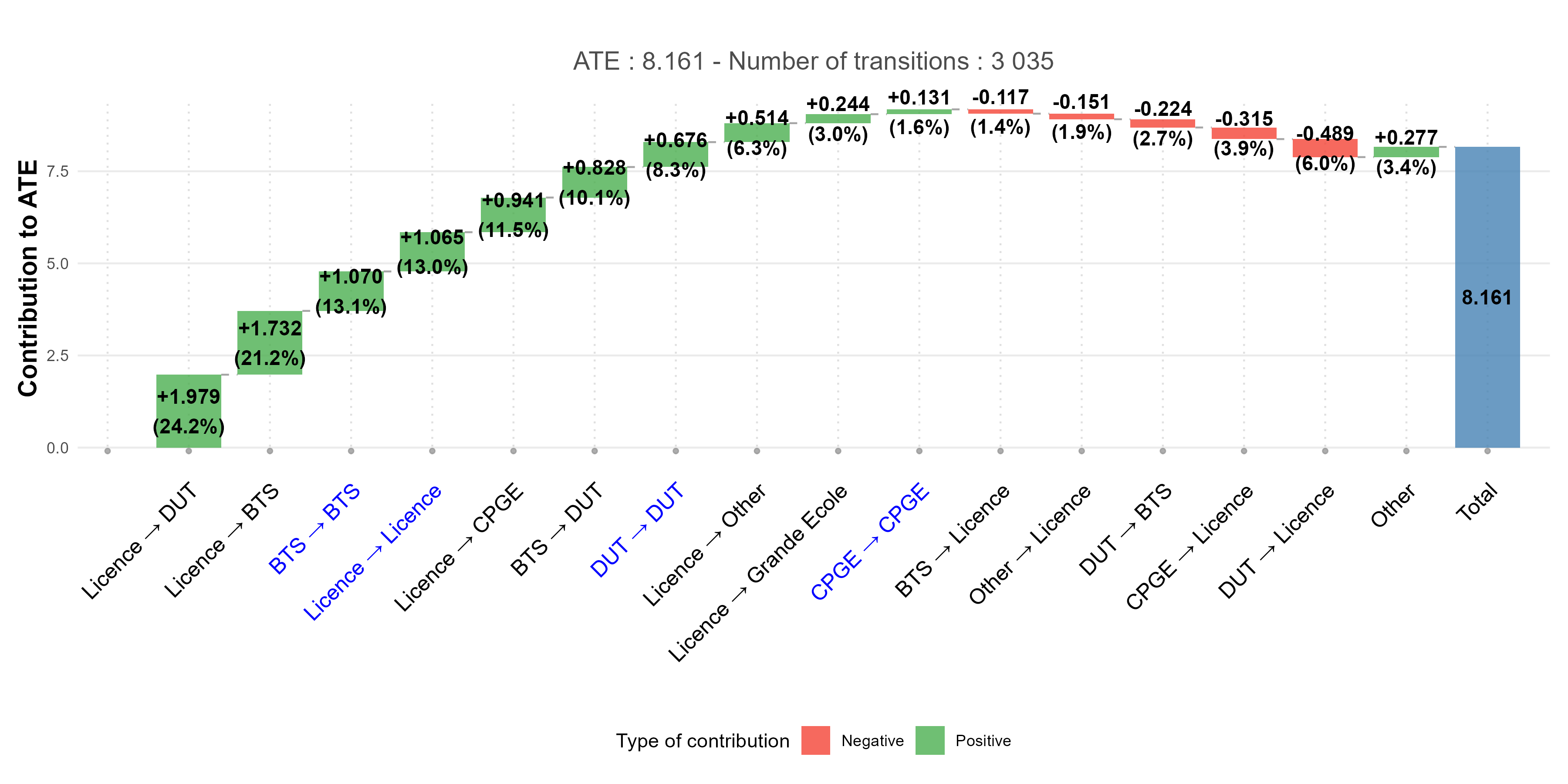}
  \label{appendix:all_0:decomposition}
\end{figure}

\begin{figure}[htbp]
  \centering
\caption{Effects of quotas on the type of programs attended by scholarship students (ITT). All tracks included. Alternative rule of quotas (no +2pp bonus).}

  \includegraphics[width=1\textwidth, trim=0 0 0 0, clip]{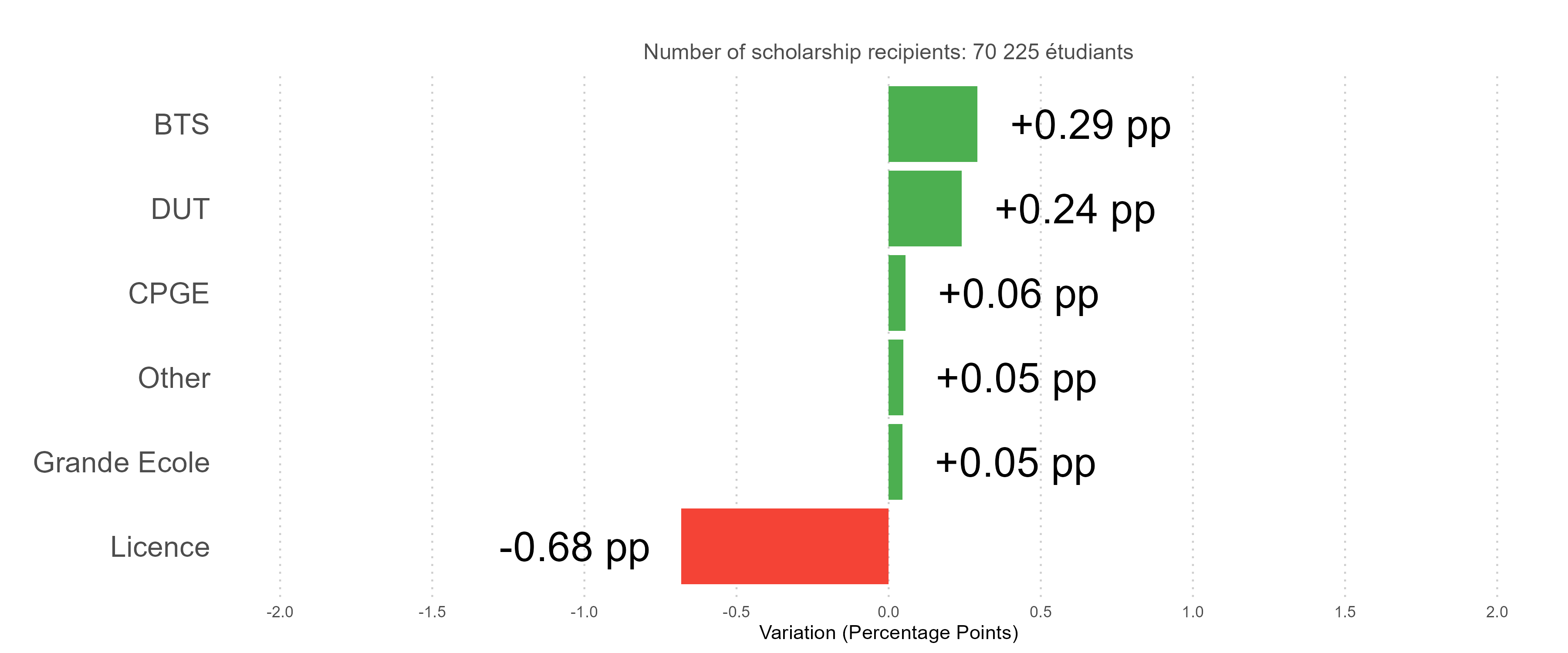}
  \label{appendix:all_0:pp}
\end{figure}

\newpage

\begin{figure}[htbp]
  \centering
\caption{Number of transitions between program types among scholarship students. All tracks included. Alternative rule of quotas (no +2pp bonus).}

  \includegraphics[width=0.7\textwidth, trim=0 0 0 0, clip]{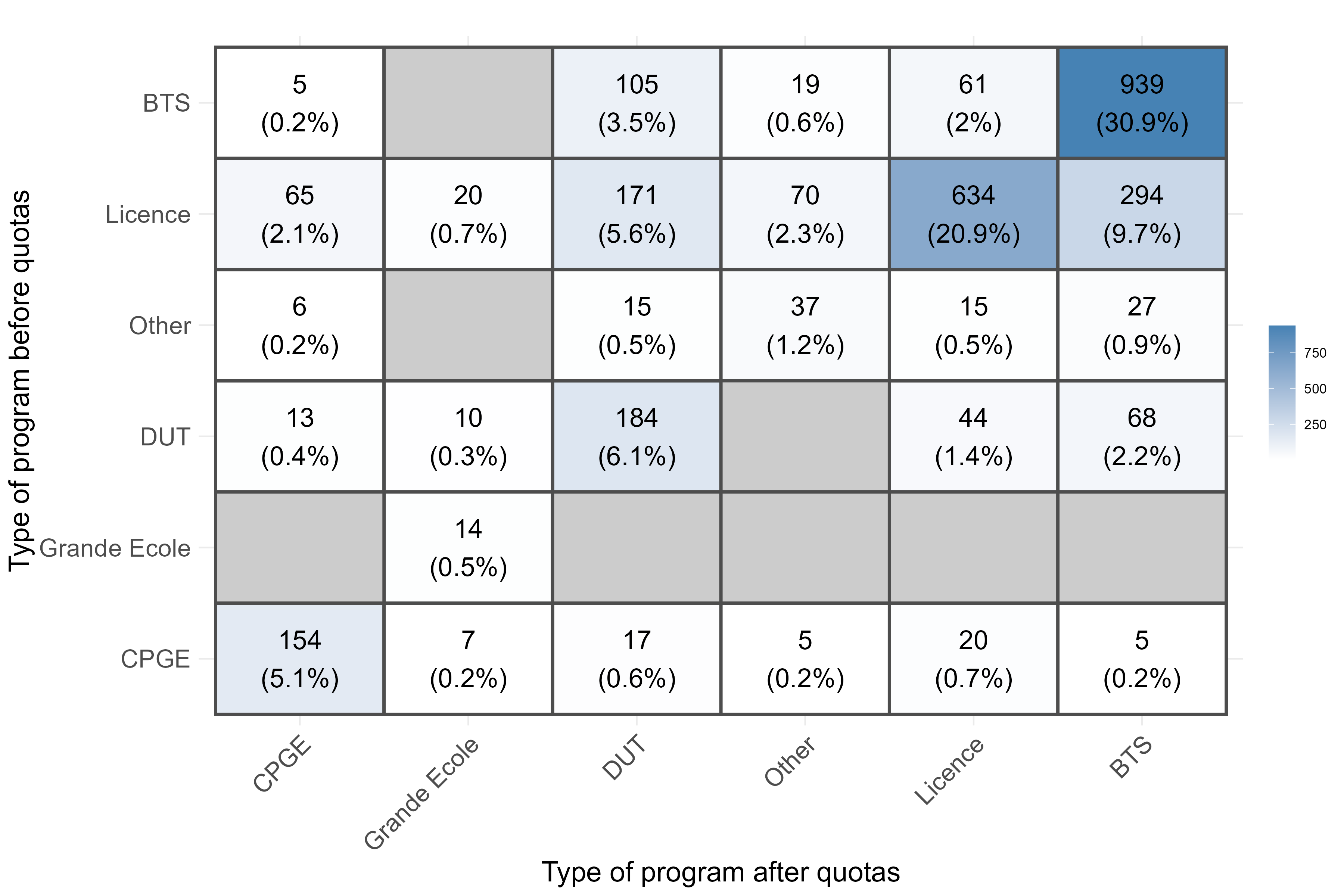}
  \label{appendix:all_0:effectifs}

\end{figure}

\begin{figure}[htbp]
  \centering
\caption{Average prestige gain by type of program transition among scholarship students. All tracks included. Alternative rule of quotas (no +2pp bonus).}

  \includegraphics[width=0.7\textwidth, trim=0 0 0 0, clip]{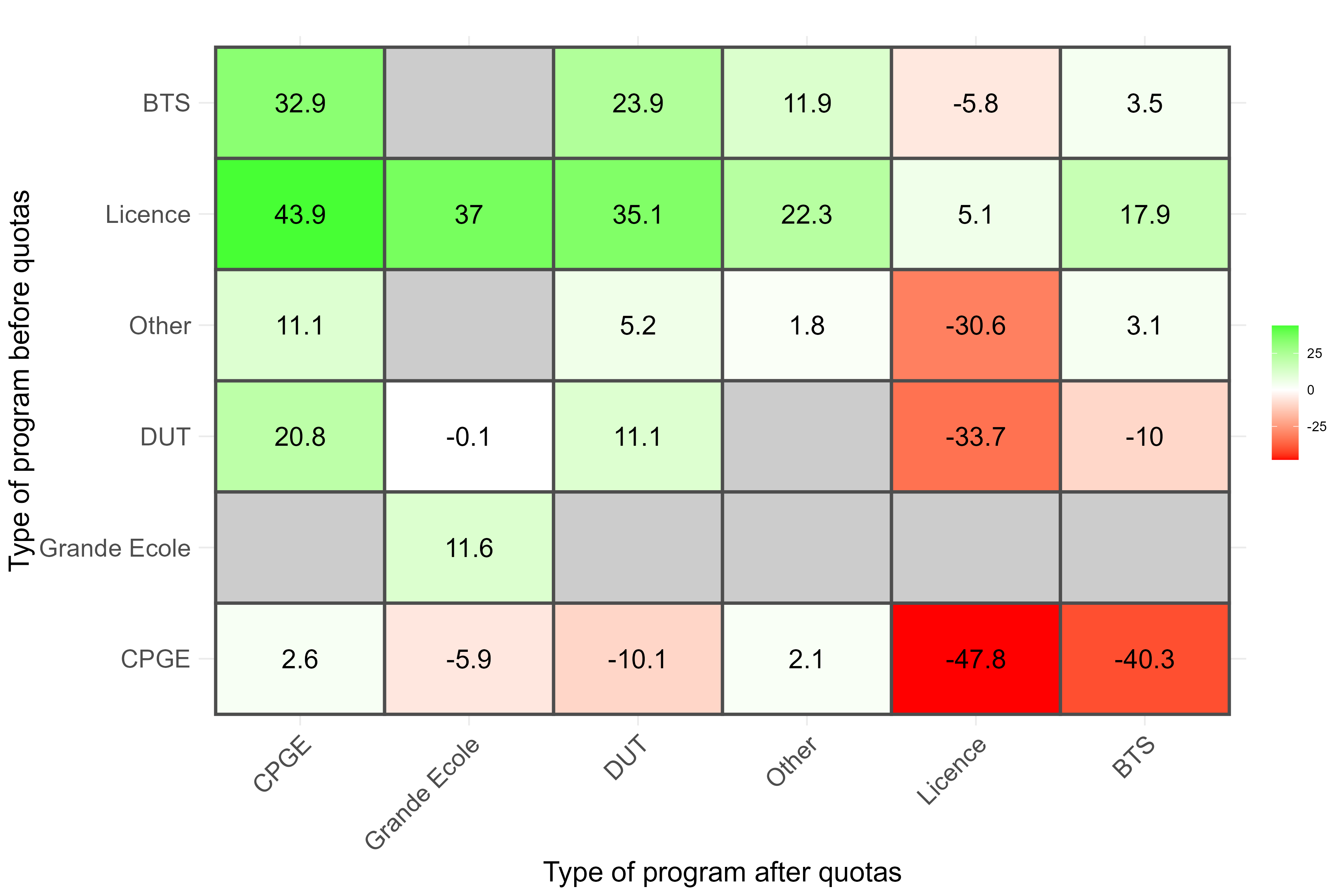}
  \label{appendix:all_0:gains}

\end{figure}

\FloatBarrier

\end{document}

%% file: productions_drdid/g_sans_notes/latex/latex_stat_descriptives_boursier.tex
\footnotesize
\setlength{\tabcolsep}{4pt}
\begin{longtable}[htbp]{@{}l*{7}{c}@{}}
\label{g_sans_notes:desc_boursier} \\
\caption{Socio-demographic characteristics of scholarship students (general track, included in the sample)}\\
\toprule
Outcome & 2014 & 2015 & 2016 & 2017 & 2018 & 2019 & 2020 \\
\midrule
\endfirsthead
\multicolumn{8}{c}{\tablename\ \thetable\ -- \textit{Continued from previous page}} \\
\toprule
Outcome & 2014 & 2015 & 2016 & 2017 & 2018 & 2019 & 2020 \\
\midrule
\endhead
\midrule
\multicolumn{8}{r}{\textit{Continued on next page}} \\
\endfoot
\bottomrule
\endlastfoot
\multicolumn{8}{l}{\textbf{Social category of the first parent}} \\
\midrule
21 - Craftsmen & 0.034 & 0.033 & 0.034 & 0.035 & 0.033 & 0.036 & 0.034 \\
22 - Traders and similar & 0.042 & 0.041 & 0.042 & 0.04 & 0.041 & 0.039 & 0.039 \\
31 - Liberal professions & 0.017 & 0.018 & 0.017 & 0.018 & 0.019 & 0.019 & 0.02 \\
43 - Intermediate health and social work professions & 0.054 & 0.053 & 0.055 & 0.053 & 0.056 & 0.056 & 0.054 \\
45 - Administrative intermediate professions in the civil service & 0.013 & 0.014 & 0.013 & 0.013 & 0.013 & 0.012 & 0.013 \\
47 - Technicians & 0.018 & 0.017 & 0.016 & 0.016 & 0.017 & 0.018 & 0.017 \\
52 - Low-skilled civil servant & 0.073 & 0.075 & 0.079 & 0.077 & 0.08 & 0.082 & 0.083 \\
\end{longtable}

%% file: productions_drdid/g_sans_notes/latex/latex_stat_descriptives_non_boursier.tex
\footnotesize
\setlength{\tabcolsep}{4pt}
\begin{longtable}[htbp]{@{}l*{7}{c}@{}}
\label{g_sans_notes:desc_non_boursier} \\
\caption{Socio-demographic characteristics of non-scholarship students (general track, included in the sample)}\\
\toprule
Outcome & 2014 & 2015 & 2016 & 2017 & 2018 & 2019 & 2020 \\
\midrule
\endfirsthead
\multicolumn{8}{c}{\tablename\ \thetable\ -- \textit{Continued from previous page}} \\
\toprule
Outcome & 2014 & 2015 & 2016 & 2017 & 2018 & 2019 & 2020 \\
\midrule
\endhead
\midrule
\multicolumn{8}{r}{\textit{Continued on next page}} \\
\endfoot
\bottomrule
\endlastfoot
\multicolumn{8}{l}{\textbf{Social category of the first parent}} \\
\midrule
21 - Craftsmen & 0.032 & 0.031 & 0.03 & 0.03 & 0.029 & 0.029 & 0.028 \\
22 - Traders and similar & 0.033 & 0.034 & 0.033 & 0.033 & 0.032 & 0.031 & 0.03 \\
31 - Liberal professions & 0.063 & 0.06 & 0.058 & 0.057 & 0.053 & 0.054 & 0.054 \\
43 - Intermediate health and social work professions & 0.045 & 0.047 & 0.047 & 0.048 & 0.051 & 0.05 & 0.049 \\
45 - Administrative intermediate professions in the civil service & 0.017 & 0.016 & 0.016 & 0.016 & 0.016 & 0.016 & 0.016 \\
47 - Technicians & 0.048 & 0.047 & 0.046 & 0.046 & 0.045 & 0.044 & 0.044 \\
52 - Low-skilled civil servant & 0.053 & 0.054 & 0.055 & 0.059 & 0.059 & 0.06 & 0.059 \\
\end{longtable}

%% file: productions_drdid/g_sans_notes/latex/latex_stat_outcomes_boursier.tex
\footnotesize
\setlength{\tabcolsep}{4pt}
\begin{longtable}[htbp]{@{}l*{7}{c}@{}}
\label{g_sans_notes:outcome_boursier} \\
\caption{Outcomes of scholarship students (general track, included in the sample)}\\
\toprule
Outcome & 2014 & 2015 & 2016 & 2017 & 2018 & 2019 & 2020 \\
\midrule
\endfirsthead
\multicolumn{8}{c}{\tablename\ \thetable\ -- \textit{Continued from previous page}} \\
\toprule
Outcome & 2014 & 2015 & 2016 & 2017 & 2018 & 2019 & 2020 \\
\midrule
\endhead
\midrule
\multicolumn{8}{r}{\textit{Continued on next page}} \\
\endfoot
\bottomrule
\endlastfoot
\multicolumn{8}{l}{\textbf{Pre-college investment}} \\
\midrule
Baccalaureate grade & 11.629 & 11.436 & 11.563 & 11.418 & 11.453 & 11.371 & 12.507 \\
\quad & (2.385) & (2.539) & (2.656) & (2.731) & (2.724) & (2.687) & (2.272) \\
\midrule
\multicolumn{8}{l}{\textbf{Applications}} \\
\midrule
At least one application & 0.98 & 0.983 & 0.993 & 0.996 & 0.974 & 0.984 & 0.989 \\
\quad & (0.139) & (0.131) & (0.084) & (0.066) & (0.16) & (0.124) & (0.102) \\
Max prestige of applications & 70.422 & 71.05 & 70.623 & 70.057 & 77.227 & 80.429 & 83.392 \\
\quad & (24.262) & (24.225) & (23.924) & (24.125) & (20.55) & (19.212) & (17.323) \\
\end{longtable}

%% file: productions_drdid/g_sans_notes/latex/latex_stat_outcomes_non_boursier.tex
\footnotesize
\setlength{\tabcolsep}{4pt}
\begin{longtable}[htbp]{@{}l*{7}{c}@{}}
\label{g_sans_notes:outcome_non_boursier} \\
\caption{Outcomes of non-scholarship students (general track, included in the sample)}\\
\toprule
Outcome & 2014 & 2015 & 2016 & 2017 & 2018 & 2019 & 2020 \\
\midrule
\endfirsthead
\multicolumn{8}{c}{\tablename\ \thetable\ -- \textit{Continued from previous page}} \\
\toprule
Outcome & 2014 & 2015 & 2016 & 2017 & 2018 & 2019 & 2020 \\
\midrule
\endhead
\midrule
\multicolumn{8}{r}{\textit{Continued on next page}} \\
\endfoot
\bottomrule
\endlastfoot
\multicolumn{8}{l}{\textbf{Pre-college investment}} \\
\midrule
Baccalaureate grade & 12.58 & 12.484 & 12.629 & 12.536 & 12.561 & 12.453 & 13.432 \\
\quad & (2.52) & (2.664) & (2.774) & (2.864) & (2.838) & (2.81) & (2.397) \\
\midrule
\multicolumn{8}{l}{\textbf{Applications}} \\
\midrule
At least one application & 0.98 & 0.98 & 0.994 & 0.996 & 0.97 & 0.979 & 0.988 \\
\quad & (0.14) & (0.141) & (0.075) & (0.065) & (0.172) & (0.145) & (0.107) \\
Max prestige of applications & 76.146 & 76.742 & 76.362 & 75.361 & 81.685 & 83.934 & 87.091 \\
\quad & (23.218) & (23.075) & (23.052) & (23.697) & (19.23) & (18.147) & (15.739) \\
\end{longtable}

%% file: productions_drdid/t_sans_notes/latex/latex_stat_descriptives_boursier.tex
\footnotesize
\setlength{\tabcolsep}{4pt}
\begin{longtable}[htbp]{@{}l*{7}{c}@{}}
\label{t_sans_notes:desc_boursier} \\
\caption{Socio-demographic characteristics of scholarship students (technological track, included in the sample)}\\
\toprule
Outcome & 2014 & 2015 & 2016 & 2017 & 2018 & 2019 & 2020 \\
\midrule
\endfirsthead
\multicolumn{8}{c}{\tablename\ \thetable\ -- \textit{Continued from previous page}} \\
\toprule
Outcome & 2014 & 2015 & 2016 & 2017 & 2018 & 2019 & 2020 \\
\midrule
\endhead
\midrule
\multicolumn{8}{r}{\textit{Continued on next page}} \\
\endfoot
\bottomrule
\endlastfoot
\multicolumn{8}{l}{\textbf{Social category of the first parent}} \\
\midrule
21 - Craftsmen & 0.027 & 0.028 & 0.031 & 0.029 & 0.03 & 0.029 & 0.029 \\
22 - Traders and similar & 0.032 & 0.032 & 0.031 & 0.032 & 0.031 & 0.032 & 0.032 \\
31 - Liberal professions & 0.008 & 0.008 & 0.009 & 0.009 & 0.01 & 0.01 & 0.01 \\
43 - Intermediate health and social work professions & 0.043 & 0.044 & 0.049 & 0.049 & 0.049 & 0.048 & 0.048 \\
45 - Administrative intermediate professions in the civil service & 0.01 & 0.01 & 0.012 & 0.011 & 0.01 & 0.01 & 0.01 \\
47 - Technicians & 0.014 & 0.014 & 0.014 & 0.014 & 0.014 & 0.015 & 0.015 \\
52 - Low-skilled civil servant & 0.073 & 0.072 & 0.077 & 0.074 & 0.079 & 0.08 & 0.081 \\
\end{longtable}

%% file: productions_drdid/t_sans_notes/latex/latex_stat_descriptives_non_boursier.tex
\footnotesize
\setlength{\tabcolsep}{4pt}
\begin{longtable}[htbp]{@{}l*{7}{c}@{}}
\label{t_sans_notes:desc_non_boursier} \\
\caption{Socio-demographic characteristics of non-scholarship students (technological track, included in the sample)}\\
\toprule
Outcome & 2014 & 2015 & 2016 & 2017 & 2018 & 2019 & 2020 \\
\midrule
\endfirsthead
\multicolumn{8}{c}{\tablename\ \thetable\ -- \textit{Continued from previous page}} \\
\toprule
Outcome & 2014 & 2015 & 2016 & 2017 & 2018 & 2019 & 2020 \\
\midrule
\endhead
\midrule
\multicolumn{8}{r}{\textit{Continued on next page}} \\
\endfoot
\bottomrule
\endlastfoot
\multicolumn{8}{l}{\textbf{Social category of the first parent}} \\
\midrule
21 - Craftsmen & 0.044 & 0.044 & 0.041 & 0.041 & 0.04 & 0.038 & 0.039 \\
22 - Traders and similar & 0.038 & 0.037 & 0.036 & 0.034 & 0.035 & 0.034 & 0.033 \\
31 - Liberal professions & 0.023 & 0.023 & 0.024 & 0.023 & 0.023 & 0.023 & 0.024 \\
43 - Intermediate health and social work professions & 0.045 & 0.047 & 0.048 & 0.051 & 0.052 & 0.055 & 0.053 \\
45 - Administrative intermediate professions in the civil service & 0.017 & 0.017 & 0.016 & 0.016 & 0.016 & 0.016 & 0.017 \\
47 - Technicians & 0.055 & 0.055 & 0.054 & 0.053 & 0.05 & 0.05 & 0.052 \\
52 - Low-skilled civil servant & 0.077 & 0.077 & 0.079 & 0.082 & 0.083 & 0.085 & 0.086 \\
\end{longtable}

%% file: productions_drdid/t_sans_notes/latex/latex_stat_outcomes_boursier.tex
\footnotesize
\setlength{\tabcolsep}{4pt}
\begin{longtable}[htbp]{@{}l*{7}{c}@{}}
\label{t_sans_notes:outcome_boursier} \\
\caption{Outcomes of scholarship students (technological track, included in the sample)}\\
\toprule
Outcome & 2014 & 2015 & 2016 & 2017 & 2018 & 2019 & 2020 \\
\midrule
\endfirsthead
\multicolumn{8}{c}{\tablename\ \thetable\ -- \textit{Continued from previous page}} \\
\toprule
Outcome & 2014 & 2015 & 2016 & 2017 & 2018 & 2019 & 2020 \\
\midrule
\endhead
\midrule
\multicolumn{8}{r}{\textit{Continued on next page}} \\
\endfoot
\bottomrule
\endlastfoot
\multicolumn{8}{l}{\textbf{Pre-college investment}} \\
\midrule
Baccalaureate grade & 11.495 & 11.299 & 11.264 & 11.295 & 11.127 & 11.046 & 11.935 \\
\quad & (1.887) & (2.106) & (2.084) & (2.158) & (2.22) & (2.293) & (2.021) \\
\midrule
\multicolumn{8}{l}{\textbf{Applications}} \\
\midrule
At least one application & 0.951 & 0.953 & 0.947 & 0.953 & 0.935 & 0.961 & 0.972 \\
\quad & (0.216) & (0.211) & (0.224) & (0.212) & (0.246) & (0.193) & (0.164) \\
Max prestige of applications & 67.511 & 68.559 & 69.353 & 69.641 & 72.718 & 76.811 & 77.295 \\
\quad & (21.865) & (22.057) & (21.213) & (20.214) & (17.964) & (16.83) & (15.37) \\
\end{longtable}

%% file: productions_drdid/t_sans_notes/latex/latex_stat_outcomes_non_boursier.tex
\footnotesize
\setlength{\tabcolsep}{4pt}
\begin{longtable}[htbp]{@{}l*{7}{c}@{}}
\label{t_sans_notes:outcome_non_boursier} \\
\caption{Outcomes of non-scholarship students (technological track, included in the sample)}\\
\toprule
Outcome & 2014 & 2015 & 2016 & 2017 & 2018 & 2019 & 2020 \\
\midrule
\endfirsthead
\multicolumn{8}{c}{\tablename\ \thetable\ -- \textit{Continued from previous page}} \\
\toprule
Outcome & 2014 & 2015 & 2016 & 2017 & 2018 & 2019 & 2020 \\
\midrule
\endhead
\midrule
\multicolumn{8}{r}{\textit{Continued on next page}} \\
\endfoot
\bottomrule
\endlastfoot
\multicolumn{8}{l}{\textbf{Pre-college investment}} \\
\midrule
Baccalaureate grade & 11.853 & 11.751 & 11.697 & 11.747 & 11.589 & 11.536 & 12.35 \\
\quad & (1.821) & (2.064) & (2.043) & (2.109) & (2.182) & (2.257) & (1.995) \\
\midrule
\multicolumn{8}{l}{\textbf{Applications}} \\
\midrule
At least one application & 0.926 & 0.935 & 0.928 & 0.936 & 0.917 & 0.943 & 0.964 \\
\quad & (0.262) & (0.246) & (0.259) & (0.245) & (0.277) & (0.232) & (0.186) \\
Max prestige of applications & 68.886 & 70.571 & 70.83 & 70.644 & 73.466 & 76.879 & 77.891 \\
\quad & (21.843) & (21.68) & (20.914) & (20.177) & (18.136) & (17.198) & (15.493) \\
\end{longtable}

%% file: productions_drdid/p_sans_notes/latex/latex_stat_descriptives_boursier.tex
\footnotesize
\setlength{\tabcolsep}{4pt}
\begin{longtable}[htbp]{@{}l*{7}{c}@{}}
\label{p_sans_notes:desc_boursier} \\
\caption{Socio-demographic characteristics of scholarship students (vocational track, included in the sample)}\\
\toprule
Outcome & 2014 & 2015 & 2016 & 2017 & 2018 & 2019 & 2020 \\
\midrule
\endfirsthead
\multicolumn{8}{c}{\tablename\ \thetable\ -- \textit{Continued from previous page}} \\
\toprule
Outcome & 2014 & 2015 & 2016 & 2017 & 2018 & 2019 & 2020 \\
\midrule
\endhead
\midrule
\multicolumn{8}{r}{\textit{Continued on next page}} \\
\endfoot
\bottomrule
\endlastfoot
\multicolumn{8}{l}{\textbf{Social category of the first parent}} \\
\midrule
21 - Craftsmen & 0.022 & 0.023 & 0.023 & 0.025 & 0.024 & 0.025 & 0.025 \\
22 - Traders and similar & 0.022 & 0.022 & 0.023 & 0.021 & 0.021 & 0.022 & 0.022 \\
31 - Liberal professions & 0.004 & 0.005 & 0.005 & 0.005 & 0.005 & 0.006 & 0.007 \\
43 - Intermediate health and social work professions & 0.034 & 0.033 & 0.036 & 0.037 & 0.038 & 0.039 & 0.04 \\
45 - Administrative intermediate professions in the civil service & 0.008 & 0.007 & 0.008 & 0.008 & 0.007 & 0.007 & 0.008 \\
47 - Technicians & 0.011 & 0.011 & 0.009 & 0.011 & 0.011 & 0.011 & 0.012 \\
52 - Low-skilled civil servant & 0.068 & 0.069 & 0.069 & 0.065 & 0.072 & 0.074 & 0.076 \\
\end{longtable}

%% file: productions_drdid/p_sans_notes/latex/latex_stat_descriptives_non_boursier.tex
\footnotesize
\setlength{\tabcolsep}{4pt}
\begin{longtable}[htbp]{@{}l*{7}{c}@{}}
\label{p_sans_notes:desc_non_boursier} \\
\caption{Socio-demographic characteristics of non-scholarship students (vocational track, included in the sample)}\\
\toprule
Outcome & 2014 & 2015 & 2016 & 2017 & 2018 & 2019 & 2020 \\
\midrule
\endfirsthead
\multicolumn{8}{c}{\tablename\ \thetable\ -- \textit{Continued from previous page}} \\
\toprule
Outcome & 2014 & 2015 & 2016 & 2017 & 2018 & 2019 & 2020 \\
\midrule
\endhead
\midrule
\multicolumn{8}{r}{\textit{Continued on next page}} \\
\endfoot
\bottomrule
\endlastfoot
\multicolumn{8}{l}{\textbf{Social category of the first parent}} \\
\midrule
21 - Craftsmen & 0.046 & 0.044 & 0.043 & 0.043 & 0.041 & 0.041 & 0.041 \\
22 - Traders and similar & 0.033 & 0.033 & 0.031 & 0.029 & 0.03 & 0.03 & 0.029 \\
31 - Liberal professions & 0.012 & 0.012 & 0.012 & 0.012 & 0.011 & 0.013 & 0.013 \\
43 - Intermediate health and social work professions & 0.04 & 0.042 & 0.043 & 0.044 & 0.045 & 0.048 & 0.047 \\
45 - Administrative intermediate professions in the civil service & 0.014 & 0.014 & 0.013 & 0.013 & 0.014 & 0.013 & 0.014 \\
47 - Technicians & 0.046 & 0.044 & 0.043 & 0.04 & 0.041 & 0.04 & 0.043 \\
52 - Low-skilled civil servant & 0.086 & 0.084 & 0.086 & 0.088 & 0.089 & 0.096 & 0.095 \\
\end{longtable}

%% file: productions_drdid/p_sans_notes/latex/latex_stat_outcomes_boursier.tex
\footnotesize
\setlength{\tabcolsep}{4pt}
\begin{longtable}[htbp]{@{}l*{7}{c}@{}}
\label{p_sans_notes:outcome_boursier} \\
\caption{Outcomes of scholarship students (vocational track, included in the sample)}\\
\toprule
Outcome & 2014 & 2015 & 2016 & 2017 & 2018 & 2019 & 2020 \\
\midrule
\endfirsthead
\multicolumn{8}{c}{\tablename\ \thetable\ -- \textit{Continued from previous page}} \\
\toprule
Outcome & 2014 & 2015 & 2016 & 2017 & 2018 & 2019 & 2020 \\
\midrule
\endhead
\midrule
\multicolumn{8}{r}{\textit{Continued on next page}} \\
\endfoot
\bottomrule
\endlastfoot
\multicolumn{8}{l}{\textbf{Pre-college investment}} \\
\midrule
Baccalaureate grade & 11.561 & 11.171 & 11.384 & 11.335 & 11.381 & 11.423 & 12.475 \\
\quad & (2.135) & (2.224) & (2.183) & (2.244) & (2.288) & (2.349) & (2.04) \\
\midrule
\multicolumn{8}{l}{\textbf{Applications}} \\
\midrule
At least one application & 0.78 & 0.781 & 0.752 & 0.778 & 0.773 & 0.807 & 0.815 \\
\quad & (0.414) & (0.413) & (0.432) & (0.416) & (0.419) & (0.395) & (0.388) \\
Max prestige of applications & 61.48 & 61.222 & 61.732 & 61.959 & 62.808 & 66.835 & 67.77 \\
\quad & (22.193) & (22.552) & (22.005) & (20.283) & (19.422) & (19.781) & (19.017) \\
\end{longtable}

%% file: productions_drdid/p_sans_notes/latex/latex_stat_outcomes_non_boursier.tex
\footnotesize
\setlength{\tabcolsep}{4pt}
\begin{longtable}[htbp]{@{}l*{7}{c}@{}}
\label{p_sans_notes:outcome_non_boursier} \\
\caption{Outcomes of non-scholarship students (vocational track, included in the sample)}\\
\toprule
Outcome & 2014 & 2015 & 2016 & 2017 & 2018 & 2019 & 2020 \\
\midrule
\endfirsthead
\multicolumn{8}{c}{\tablename\ \thetable\ -- \textit{Continued from previous page}} \\
\toprule
Outcome & 2014 & 2015 & 2016 & 2017 & 2018 & 2019 & 2020 \\
\midrule
\endhead
\midrule
\multicolumn{8}{r}{\textit{Continued on next page}} \\
\endfoot
\bottomrule
\endlastfoot
\multicolumn{8}{l}{\textbf{Pre-college investment}} \\
\midrule
Baccalaureate grade & 11.894 & 11.544 & 11.701 & 11.603 & 11.633 & 11.708 & 12.816 \\
\quad & (2.094) & (2.199) & (2.18) & (2.258) & (2.325) & (2.365) & (1.993) \\
\midrule
\multicolumn{8}{l}{\textbf{Applications}} \\
\midrule
At least one application & 0.72 & 0.73 & 0.701 & 0.706 & 0.7 & 0.732 & 0.774 \\
\quad & (0.449) & (0.444) & (0.458) & (0.456) & (0.458) & (0.443) & (0.418) \\
Max prestige of applications & 62.448 & 62.744 & 62.689 & 61.984 & 61.724 & 65.98 & 66.426 \\
\quad & (22.365) & (22.511) & (21.998) & (20.555) & (20.34) & (20.413) & (20.133) \\
\end{longtable}

%% file: productions_drdid/g_sans_notes/tables/unified_event_study_table_corrected.tex
\footnotesize
\setlength{\tabcolsep}{4pt} % Réduire l'espacement entre colonnes (défaut: 6pt)
\begin{longtable}[htbp]{@{}l*{6}{c}cc@{}}
\caption{DRDiD estimates of treatment effects on various outcomes (general track)}
 \label{g_sans_notes:event_study} \\
\toprule
 & \multicolumn{3}{c}{Pre-treatment year} & \multicolumn{3}{c}{Post-treatment year} & \\
\cmidrule(lr){2-4} \cmidrule(lr){5-7}
Outcome & 2014 & 2015 & 2016 & 2018 & 2019 & 2020 & N & p-value pretrend \\
\midrule
\endfirsthead
\multicolumn{9}{c}{\tablename\ \thetable\ -- \textit{Continued from previous page}} \\
\toprule
 & \multicolumn{3}{c}{Pre-treatment year} & \multicolumn{3}{c}{Post-treatment year} & \\
\cmidrule(lr){2-4} \cmidrule(lr){5-7}
Outcome & 2014 & 2015 & 2016 & 2018 & 2019 & 2020 & N & p-value pretrend \\
\midrule
\endhead
\midrule
\multicolumn{9}{r}{\textit{Continued on next page}} \\
\endfoot
\bottomrule
\endlastfoot
\multicolumn{9}{l}{\textbf{Pre-college investment}} \\
\midrule
Baccalaureate grade & -0.111*** & -0.064 & -0.030 & -0.028 & 0.014 &  & 2.5M & 0.004 \\
 & (0.033) & (0.033) & (0.034) & (0.033) & (0.032) &  &  &  \\
\midrule
\multicolumn{9}{l}{\textbf{Applications}} \\
\midrule
Max prestige of applications & -0.751* & -0.602 & -0.322 & -0.805** & -0.675* & -0.843** & 2.4M & 0.083 \\
 & (0.314) & (0.313) & (0.307) & (0.283) & (0.275) & (0.264) &  &  \\
Mean prestige of applications & -0.627* & -0.459 & -0.277 & -0.499* & -0.299 & -0.209 & 2.4M & 0.062 \\
 & (0.244) & (0.242) & (0.230) & (0.229) & (0.224) & (0.223) &  &  \\
\end{longtable}

%% file: productions_drdid/g_sans_notes/tables/unified_breaking_points_drdid_table.tex
\footnotesize
\setlength{\tabcolsep}{6pt} % Espacement entre colonnes
\begin{longtable}[htbp]{@{}l*{3}{c}@{}}
\caption{Breakdown value of magnitude of parallel trend violation for a null value (general track)} \label{g_sans_notes:breaking_point_unified}\\
\toprule
Outcome & 2018 & 2019 & 2020 \\
\midrule
\endfirsthead
\multicolumn{4}{c}{\tablename\ \thetable\ -- \textit{Continued from previous page}} \\
\toprule
Outcome & 2018 & 2019 & 2020 \\
\midrule
\endhead
\midrule
\multicolumn{4}{r}{\textit{Continued on next page}} \\
\endfoot
\bottomrule
\endlastfoot
\multicolumn{4}{l}{\textbf{Pre-college investment}} \\
\midrule
Baccalaureate grade & 0 & 0 & -- \\
\midrule
\multicolumn{4}{l}{\textbf{Applications}} \\
\midrule
Max prestige of applications & 0.6 & 0.2 & 0.3 \\
Mean prestige of applications & 0.4 & 0 & 0 \\
\midrule
\multicolumn{4}{l}{\textbf{Applied in at least one...}} \\
\midrule
BTS & 0 & 0.3 & 0.4 \\
CPGE & 0 & 0 & 0 \\
DUT & 0 & 0 & 0 \\
Engineering school & 0 & 0.4 & 0 \\
\midrule
\multicolumn{4}{l}{\textbf{Final enrollment}} \\
\midrule
Enrolled in any program & 0.4 & 0 & 0 \\
Prestige of final enrollment & 0.8 & 1.2 & 1.2 \\
\midrule
\multicolumn{4}{l}{\textbf{Was finally enrolled in a...}} \\
\midrule
BTS & 0 & 0 & 0 \\
CPGE & 0 & 0 & 0 \\
DUT & 0.4 & 0.4 & 0.4 \\
Engineering school & 0 & 0 & 0 \\
Bachelor's degree & 0 & 0 & 0 \\
\midrule
\multicolumn{4}{l}{\footnotesize{}} \\
\end{longtable}

%% file: productions_drdid/g_sans_notes/tables/unified_confidence_intervals_drdid_table.tex
\footnotesize
\setlength{\tabcolsep}{4pt} % Réduire l'espacement entre colonnes (défaut: 6pt)
\begin{longtable}[htbp]{@{}l*{6}{c}@{}}
\caption{Confidence intervals, including standardized effect sizes (Cohen’s d) (general track)}
\label{g_sans_notes:confidence_interval_unified}\\
\toprule
& \multicolumn{3}{c}{\textbf{Pre-treatment}} & \multicolumn{3}{c}{\textbf{Post-treatment}} \\
\cmidrule(lr){2-4} \cmidrule(lr){5-7}
Outcome & 2014 & 2015 & 2016 & 2018 & 2019 & 2020 \\
\midrule
\endfirsthead
\multicolumn{7}{c}{\tablename\ \thetable\ -- \textit{Continued from previous page}} \\
\toprule
Outcome & 2014 & 2015 & 2016 & 2018 & 2019 & 2020 \\
\midrule
\endhead
\midrule
\multicolumn{7}{r}{\textit{Continued on next page}} \\
\endfoot
\bottomrule
\endlastfoot
\midrule
Baccalaureate grade & [-0.18; -0.05] & [-0.13; 0.00] & [-0.10; 0.04] & [-0.09; 0.04] & [-0.05; 0.08] &  \\
 & [-0.07; -0.02] & [-0.05; 0.00] & [-0.04; 0.01] & [-0.03; 0.01] & [-0.02; 0.03] &  \\
\midrule
\multicolumn{7}{l}{\textbf{Applications}} \\
\midrule
Max prestige of applications & [-1.37; -0.13] & [-1.22; 0.01] & [-0.92; 0.28] & [-1.36; -0.25] & [-1.21; -0.14] & [-1.36; -0.32] \\
 & [-0.06; -0.01] & [-0.05; 0.00] & [-0.04; 0.01] & [-0.07; -0.01] & [-0.06; -0.01] & [-0.08; -0.02] \\
Mean prestige of applications & [-1.11; -0.15] & [-0.93; 0.01] & [-0.73; 0.17] & [-0.95; -0.05] & [-0.74; 0.14] & [-0.65; 0.23] \\
 & [-0.05; -0.01] & [-0.04; 0.00] & [-0.04; 0.01] & [-0.05; -0.00] & [-0.04; 0.01] & [-0.04; 0.01] \\
\midrule
\multicolumn{7}{l}{\textbf{Applied in at least one...}} \\
\end{longtable}

%% file: productions_drdid/g_sans_notes/tables/cic_event_study_table_corrected.tex
\footnotesize
\setlength{\tabcolsep}{4pt} % Réduire l'espacement entre colonnes (défaut: 6pt)
\begin{longtable}[htbp]{@{}l*{6}{c}@{}}
\caption{CiC estimates of treatment effects on prestige scores (final enrollment and applications), including heterogeneity by decile for final enrollment (general track)}
 \label{g_sans_notes:event_study_cic} \\
\toprule
& \multicolumn{3}{c}{Pre-treatment year} & \multicolumn{3}{c}{Post-treatment year} \\
\cmidrule(lr){2-4} \cmidrule(lr){5-7}
Outcome & 2014 & 2015 & 2016 & 2018 & 2019 & 2020 \\
\midrule
\endfirsthead
\multicolumn{7}{c}{\tablename\ \thetable\ -- \textit{Continued from previous page}} \\
\toprule
 & \multicolumn{3}{c}{Pre-treatment year} & \multicolumn{3}{c}{Post-treatment year} \\
\cmidrule(lr){2-4} \cmidrule(lr){5-7}
Outcome & 2014 & 2015 & 2016 & 2018 & 2019 & 2020 \\
\midrule
\endhead
\midrule
\multicolumn{7}{r}{\textit{Continued on next page}} \\
\endfoot
\bottomrule
\endlastfoot
\multicolumn{7}{l}{\textbf{Prestige of final enrollment}} \\
\midrule
 0\% - 10\% & 0.029 & 0.204 & -0.225 & 0.107 & 0.912*** & 0.572* \\
 & (0.265) & (0.254) & (0.239) & (0.249) & (0.109) & (0.229) \\
10\% - 20\% & 0.153 & 0.108 & 0.527 & 0.815** & 1.785*** & 0.796*** \\
 & (0.328) & (0.358) & (0.314) & (0.259) & (0.160) & (0.188) \\
20\% -30\% & -0.048 & -0.352 & 0.000 & 0.579* & 2.593*** & 0.936*** \\
 & (0.403) & (0.400) & (0.309) & (0.232) & (0.220) & (0.162) \\
30\% - 40\% & 0.154 & 0.000 & -0.044 & 0.738* & 2.451*** & 1.987*** \\
\end{longtable}

%% file: productions_drdid/t_sans_notes/tables/unified_event_study_table_corrected.tex
\footnotesize
\setlength{\tabcolsep}{4pt} % Réduire l'espacement entre colonnes (défaut: 6pt)
\begin{longtable}[htbp]{@{}l*{6}{c}cc@{}}
\caption{DRDiD estimates of treatment effects on various outcomes (technological track)} \label{t_sans_notes:event_study} \\
\toprule
 & \multicolumn{3}{c}{Pre-treatment year} & \multicolumn{3}{c}{Post-treatment year} & \\
\cmidrule(lr){2-4} \cmidrule(lr){5-7}
Outcome & 2014 & 2015 & 2016 & 2018 & 2019 & 2020 & N & p-value pretrend \\
\midrule
\endfirsthead
\multicolumn{9}{c}{\tablename\ \thetable\ -- \textit{Continued from previous page}} \\
\toprule
 & \multicolumn{3}{c}{Pre-treatment year} & \multicolumn{3}{c}{Post-treatment year} & \\
\cmidrule(lr){2-4} \cmidrule(lr){5-7}
Outcome & 2014 & 2015 & 2016 & 2018 & 2019 & 2020 & N & p-value pretrend \\
\midrule
\endhead
\midrule
\multicolumn{9}{r}{\textit{Continued on next page}} \\
\endfoot
\bottomrule
\endlastfoot
\multicolumn{9}{l}{\textbf{Pre-college investment}} \\
\midrule
Baccalaureate grade & -0.016 & -0.030 & -0.013 & -0.010 & -0.008 &  & 936K & 0.756 \\
 & (0.028) & (0.027) & (0.027) & (0.026) & (0.027) &  &  &  \\
\midrule
\multicolumn{9}{l}{\textbf{Applications}} \\
\midrule
At least one application & 0.010* & 0.001 & 0.007 & 0.004 & 0.002 & -0.009 & 939K & 0.124 \\
 & (0.005) & (0.004) & (0.005) & (0.005) & (0.004) & (0.004) &  &  \\
Max prestige of applications & -0.297 & -0.774 & -0.538 & -0.391 & -0.473 & -0.554 & 883K & 0.242 \\
 & (0.425) & (0.402) & (0.394) & (0.376) & (0.349) & (0.336) &  &  \\
\end{longtable}

%% file: productions_drdid/t_sans_notes/tables/unified_breaking_points_drdid_table.tex
\footnotesize
\setlength{\tabcolsep}{6pt} % Espacement entre colonnes
\begin{longtable}[htbp]{@{}l*{3}{c}@{}}
\caption{Breakdown value of magnitude of parallel trend violation for a null value (technological track)}
\label{t_sans_notes:breaking_point_unified}\\
\toprule
Outcome & 2018 & 2019 & 2020 \\
\midrule
\endfirsthead
\multicolumn{4}{c}{\tablename\ \thetable\ -- \textit{Continued from previous page}} \\
\toprule
Outcome & 2018 & 2019 & 2020 \\
\midrule
\endhead
\midrule
\multicolumn{4}{r}{\textit{Continued on next page}} \\
\endfoot
\bottomrule
\endlastfoot
\multicolumn{4}{l}{\textbf{Pre-college investment}} \\
\midrule
Baccalaureate grade & 0 & 0 & -- \\
\midrule
\multicolumn{4}{l}{\textbf{Applications}} \\
\midrule
At least one application & 0 & 0 & 0 \\
Max prestige of applications & 0 & 0 & 0 \\
Mean prestige of applications & 0 & 0 & 0 \\
\midrule
\multicolumn{4}{l}{\textbf{Applied in at least one...}} \\
\midrule
BTS & 0 & 0 & 0 \\
DUT & 0 & 0 & 0 \\
Bachelor's degree & 0 & 0 & 0 \\
\midrule
\multicolumn{4}{l}{\textbf{Final enrollment}} \\
\midrule
Enrolled in any program & 0 & 0 & 0 \\
Prestige of final enrollment & 0 & 0 & 0 \\
\midrule
\multicolumn{4}{l}{\textbf{Was finally enrolled in a...}} \\
\midrule
BTS & 0 & 0 & 0 \\
DUT & 0 & 0 & 0 \\
Bachelor's degree & 0 & 0 & 0 \\
\midrule
\multicolumn{4}{l}{\footnotesize{}} \\
\end{longtable}

%% file: productions_drdid/t_sans_notes/tables/unified_confidence_intervals_drdid_table.tex
\footnotesize
\setlength{\tabcolsep}{4pt} % Réduire l'espacement entre colonnes (défaut: 6pt)
\begin{longtable}[htbp]{@{}l*{6}{c}@{}}
\caption{Confidence intervals, including standardized effect sizes (Cohen’s d) (technological track)}
\label{t_sans_notes:confidence_interval_unified}\\
\toprule
& \multicolumn{3}{c}{\textbf{Pre-treatment}} & \multicolumn{3}{c}{\textbf{Post-treatment}} \\
\cmidrule(lr){2-4} \cmidrule(lr){5-7}
Outcome & 2014 & 2015 & 2016 & 2018 & 2019 & 2020 \\
\midrule
\endfirsthead
\multicolumn{7}{c}{\tablename\ \thetable\ -- \textit{Continued from previous page}} \\
\toprule
Outcome & 2014 & 2015 & 2016 & 2018 & 2019 & 2020 \\
\midrule
\endhead
\midrule
\multicolumn{7}{r}{\textit{Continued on next page}} \\
\endfoot
\bottomrule
\endlastfoot
\multicolumn{7}{l}{\textbf{Pre-college investment}} \\
\midrule
Baccalaureate grade & [-0.07; 0.04] & [-0.08; 0.02] & [-0.07; 0.04] & [-0.06; 0.04] & [-0.06; 0.04] &  \\
 & [-0.04; 0.02] & [-0.04; 0.01] & [-0.03; 0.02] & [-0.03; 0.02] & [-0.03; 0.02] &  \\
\midrule
\multicolumn{7}{l}{\textbf{Applications}} \\
\midrule
At least one application & [0.00; 0.02] & [-0.01; 0.01] & [-0.00; 0.02] & [-0.00; 0.01] & [-0.01; 0.01] & [-0.02; -0.00] \\
 & [0.00; 0.09] & [-0.03; 0.05] & [-0.01; 0.07] & [-0.02; 0.05] & [-0.03; 0.05] & [-0.10; -0.01] \\
Max prestige of applications & [-1.13; 0.54] & [-1.56; 0.01] & [-1.31; 0.23] & [-1.13; 0.35] & [-1.16; 0.21] & [-1.21; 0.10] \\
 & [-0.05; 0.02] & [-0.07; 0.00] & [-0.06; 0.01] & [-0.06; 0.02] & [-0.07; 0.01] & [-0.08; 0.01] \\
\end{longtable}

%% file: productions_drdid/t_sans_notes/tables/cic_event_study_table_corrected.tex
\footnotesize
\setlength{\tabcolsep}{4pt} % Réduire l'espacement entre colonnes (défaut: 6pt)
\begin{longtable}[htbp]{@{}l*{6}{c}@{}}
\caption{CiC estimates of treatment effects on prestige scores (final enrollment and applications), including heterogeneity by decile for final enrollment (technological track)}
 \label{t_sans_notes:event_study_cic} \\
\toprule
 & \multicolumn{3}{c}{Pre-treatment year} & \multicolumn{3}{c}{Post-treatment year} \\
\cmidrule(lr){2-4} \cmidrule(lr){5-7}
Outcome & 2014 & 2015 & 2016 & 2018 & 2019 & 2020 \\
\midrule
\endfirsthead
\multicolumn{7}{c}{\tablename\ \thetable\ -- \textit{Continued from previous page}} \\
\toprule
 & \multicolumn{3}{c}{Pre-treatment year} & \multicolumn{3}{c}{Post-treatment year} \\
\cmidrule(lr){2-4} \cmidrule(lr){5-7}
Outcome & 2014 & 2015 & 2016 & 2018 & 2019 & 2020 \\
\midrule
\endhead
\midrule
\multicolumn{7}{r}{\textit{Continued on next page}} \\
\endfoot
\bottomrule
\endlastfoot
\multicolumn{7}{l}{\textbf{Prestige of final enrollment}} \\
\midrule
 0\% - 10\% & -0.064 & -0.368 & -0.047 & -0.093 & 0.060 & 0.503 \\
 & (0.288) & (0.214) & (0.181) & (0.135) & (0.223) & (0.281) \\
10\% - 20\% & 0.384 & 0.056 & 0.000 & -0.118 & 0.698* & 0.873*** \\
 & (0.369) & (0.361) & (0.238) & (0.271) & (0.350) & (0.250) \\
20\% -30\% & 0.884* & 0.081 & 0.321 & 0.348 & 0.931 & 1.119*** \\
 & (0.450) & (0.466) & (0.414) & (0.405) & (0.508) & (0.285) \\
30\% - 40\% & 0.346 & -0.073 & 0.000 & -0.159 & 1.308*** & 0.469 \\
\end{longtable}

%% file: productions_drdid/p_sans_notes/tables/unified_event_study_table_corrected.tex
\footnotesize
\setlength{\tabcolsep}{4pt} % Réduire l'espacement entre colonnes (défaut: 6pt)
\begin{longtable}[htbp]{@{}l*{6}{c}cc@{}}
\caption{DRDiD estimates of treatment effects on various outcomes (vocational track)}
 \label{p_sans_notes:event_study} \\
\toprule
 & \multicolumn{3}{c}{Pre-treatment year} & \multicolumn{3}{c}{Post-treatment year} & \\
\cmidrule(lr){2-4} \cmidrule(lr){5-7}
Outcome & 2014 & 2015 & 2016 & 2018 & 2019 & 2020 & N & p-value pretrend \\
\midrule
\endfirsthead
\multicolumn{9}{c}{\tablename\ \thetable\ -- \textit{Continued from previous page}} \\
\toprule
 & \multicolumn{3}{c}{Pre-treatment year} & \multicolumn{3}{c}{Post-treatment year} & \\
\cmidrule(lr){2-4} \cmidrule(lr){5-7}
Outcome & 2014 & 2015 & 2016 & 2018 & 2019 & 2020 & N & p-value pretrend \\
\midrule
\endhead
\midrule
\multicolumn{9}{r}{\textit{Continued on next page}} \\
\endfoot
\bottomrule
\endlastfoot
\multicolumn{9}{l}{\textbf{Pre-college investment}} \\
\midrule
Baccalaureate grade & -0.099*** & -0.090*** & -0.089*** & 0.011 & 0.001 &  & 818K & < 0.001 \\
 & (0.026) & (0.026) & (0.026) & (0.025) & (0.025) &  &  &  \\
\midrule
\multicolumn{9}{l}{\textbf{Applications}} \\
\midrule
At least one application & -0.009 & -0.017* & -0.011 & 0.002 & 0.002 & -0.032*** & 831K & 0.113 \\
 & (0.007) & (0.007) & (0.007) & (0.007) & (0.007) & (0.006) &  &  \\
Max prestige of applications & -0.090 & -0.969* & -0.524 & -0.171 & -0.357 & -0.124 & 618K & 0.089 \\
 & (0.437) & (0.421) & (0.419) & (0.393) & (0.365) & (0.352) &  &  \\
\end{longtable}

%% file: productions_drdid/p_sans_notes/tables/unified_breaking_points_drdid_table.tex
\footnotesize
\setlength{\tabcolsep}{6pt} % Espacement entre colonnes
\begin{longtable}[htbp]{@{}l*{3}{c}@{}}
\caption{Breakdown value of magnitude of parallel trend violation for a null value  (vocational track)}
\label{p_sans_notes:breaking_point_unified}\\
\toprule
Outcome & 2018 & 2019 & 2020 \\
\midrule
\endfirsthead
\multicolumn{4}{c}{\tablename\ \thetable\ -- \textit{Continued from previous page}} \\
\toprule
Outcome & 2018 & 2019 & 2020 \\
\midrule
\endhead
\midrule
\multicolumn{4}{r}{\textit{Continued on next page}} \\
\endfoot
\bottomrule
\endlastfoot
\multicolumn{4}{l}{\textbf{Pre-college investment}} \\
\midrule
Baccalaureate grade & 0 & 0 & -- \\
\midrule
\multicolumn{4}{l}{\textbf{Applications}} \\
\midrule
At least one application & 0 & 0 & 0.4 \\
Max prestige of applications & 0 & 0 & 0 \\
Mean prestige of applications & 0 & 0 & 0 \\
\midrule
\multicolumn{4}{l}{\textbf{Applied in at least one...}} \\
\midrule
BTS & 0 & 0 & 0 \\
DUT & 0 & 0 & 0 \\
Bachelor's degree & 0 & 0 & 0 \\
\midrule
\multicolumn{4}{l}{\textbf{Final enrollment}} \\
\midrule
Enrolled in any program & 0 & 0 & 0.5 \\
Prestige of final enrollment & 0 & 0 & 0 \\
\midrule
\multicolumn{4}{l}{\textbf{Was finally enrolled in a...}} \\
\midrule
BTS & 0.4 & 0 & -- \\
Bachelor's degree & 0.9 & 0.5 & -- \\
\midrule
\multicolumn{4}{l}{\footnotesize{}} \\
\end{longtable}

%% file: productions_drdid/p_sans_notes/tables/unified_confidence_intervals_drdid_table.tex
\footnotesize
\setlength{\tabcolsep}{4pt} % Réduire l'espacement entre colonnes (défaut: 6pt)
\begin{longtable}[htbp]{@{}l*{6}{c}@{}}
\caption{Confidence intervals, including standardized effect sizes (Cohen’s d) (vocational track)}
\label{p_sans_notes:confidence_interval_unified}\\
\toprule
& \multicolumn{3}{c}{\textbf{Pre-treatment}} & \multicolumn{3}{c}{\textbf{Post-treatment}} \\
\cmidrule(lr){2-4} \cmidrule(lr){5-7}
Outcome & 2014 & 2015 & 2016 & 2018 & 2019 & 2020 \\
\midrule
\endfirsthead
\multicolumn{7}{c}{\tablename\ \thetable\ -- \textit{Continued from previous page}} \\
\toprule
Outcome & 2014 & 2015 & 2016 & 2018 & 2019 & 2020 \\
\midrule
\endhead
\midrule
\multicolumn{7}{r}{\textit{Continued on next page}} \\
\endfoot
\bottomrule
\endlastfoot
\multicolumn{7}{l}{\textbf{Pre-college investment}} \\
\midrule
Baccalaureate grade & [-0.15; -0.05] & [-0.14; -0.04] & [-0.14; -0.04] & [-0.04; 0.06] & [-0.05; 0.05] &  \\
 & [-0.07; -0.02] & [-0.06; -0.02] & [-0.06; -0.02] & [-0.02; 0.03] & [-0.02; 0.02] &  \\
\midrule
\multicolumn{7}{l}{\textbf{Applications}} \\
\midrule
At least one application & [-0.02; 0.01] & [-0.03; -0.00] & [-0.02; 0.00] & [-0.01; 0.02] & [-0.01; 0.01] & [-0.04; -0.02] \\
 & [-0.06; 0.01] & [-0.08; -0.01] & [-0.06; 0.01] & [-0.03; 0.04] & [-0.03; 0.04] & [-0.11; -0.05] \\
Max prestige of applications & [-0.95; 0.77] & [-1.79; -0.14] & [-1.35; 0.30] & [-0.94; 0.60] & [-1.07; 0.36] & [-0.81; 0.57] \\
 & [-0.04; 0.03] & [-0.08; -0.01] & [-0.06; 0.01] & [-0.05; 0.03] & [-0.05; 0.02] & [-0.04; 0.03] \\
\end{longtable}

%% file: productions_drdid/p_sans_notes/tables/cic_event_study_table_corrected.tex
\footnotesize
\setlength{\tabcolsep}{4pt} % Réduire l'espacement entre colonnes (défaut: 6pt)
\begin{longtable}[htbp]{@{}l*{6}{c}@{}}
\caption{CiC estimates of treatment effects on prestige scores (final enrollment and applications), including heterogeneity by decile for final enrollment (vocational track)}
 \label{p_sans_notes:event_study_cic} \\
 \toprule
  & \multicolumn{3}{c}{Pre-treatment year} & \multicolumn{3}{c}{Post-treatment year} \\
\cmidrule(lr){2-4} \cmidrule(lr){5-7}
Outcome & 2014 & 2015 & 2016 & 2018 & 2019 & 2020 \\
\midrule
\endfirsthead
\multicolumn{7}{c}{\tablename\ \thetable\ -- \textit{Continued from previous page}} \\
\toprule
 & \multicolumn{3}{c}{Pre-treatment year} & \multicolumn{3}{c}{Post-treatment year} \\
\cmidrule(lr){2-4} \cmidrule(lr){5-7}
Outcome & 2014 & 2015 & 2016 & 2018 & 2019 & 2020 \\
\midrule
\endhead
\midrule
\multicolumn{7}{r}{\textit{Continued on next page}} \\
\endfoot
\bottomrule
\endlastfoot
\multicolumn{7}{l}{\textbf{Prestige of final enrollment}} \\
\midrule
 0\% - 10\% & 0.286 & -0.054 & 0.301* & 0.180 & -0.453* & 0.247 \\
 & (0.181) & (0.302) & (0.137) & (0.145) & (0.227) & (0.207) \\
10\% - 20\% & 0.359 & 0.278 & 0.302 & 1.086*** & 0.109 & 0.743** \\
 & (0.275) & (0.305) & (0.243) & (0.302) & (0.310) & (0.235) \\
20\% -30\% & 0.490 & 0.428 & 1.179** & 2.270*** & 0.935** & 1.492*** \\
 & (0.399) & (0.415) & (0.435) & (0.376) & (0.330) & (0.345) \\
30\% - 40\% & 0.392 & 0.199 & 0.814 & 1.266*** & 0.311 & 0.840 \\
\end{longtable}

%% file: productions_drdid/g_avec_tres_bonnes_notes/tables/unified_event_study_table_corrected.tex
\footnotesize
\setlength{\tabcolsep}{4pt} % Réduire l'espacement entre colonnes (défaut: 6pt)
\begin{longtable}[htbp]{@{}l*{6}{c}cc@{}}
\caption{DRDiD estimates of treatment effects on various outcomes (general track, first quartile)}
 \label{g_avec_tres_bonnes_notes:event_study} \\
\toprule
 & \multicolumn{3}{c}{Pre-treatment year} & \multicolumn{3}{c}{Post-treatment year} & \\
\cmidrule(lr){2-4} \cmidrule(lr){5-7}
Outcome & 2014 & 2015 & 2016 & 2018 & 2019 & 2020 & N & p-value pretrend \\
\midrule
\endfirsthead
\multicolumn{9}{c}{\tablename\ \thetable\ -- \textit{Continued from previous page}} \\
\toprule
 & \multicolumn{3}{c}{Pre-treatment year} & \multicolumn{3}{c}{Post-treatment year} & \\
\cmidrule(lr){2-4} \cmidrule(lr){5-7}
Outcome & 2014 & 2015 & 2016 & 2018 & 2019 & 2020 & N & p-value pretrend \\
\midrule
\endhead
\midrule
\multicolumn{9}{r}{\textit{Continued on next page}} \\
\endfoot
\bottomrule
\endlastfoot
\multicolumn{9}{l}{\textbf{Applications}} \\
\midrule
Max prestige of applications & -0.469 & -0.222 & -0.136 & -0.185 & -0.183 & -0.127 & 524K & 0.928 \\
 & (0.723) & (0.831) & (0.745) & (0.681) & (0.640) & (0.626) &  &  \\
Mean prestige of applications & -0.164 & 0.430 & 1.052 & 0.467 & 0.812 & 1.484 & 524K & 0.292 \\
 & (0.675) & (0.750) & (0.709) & (0.673) & (0.677) & (0.674) &  &  \\
Max prestige of applications (Cohen's d) & -0.023 & -0.011 & -0.007 & -0.011 & -0.012 & -0.009 & -- & -- \\
 & (0.036) & (0.042) & (0.038) & (0.040) & (0.041) & (0.045) &  &  \\
Mean prestige of applications (Cohen's d) & -0.008 & 0.021 & 0.052 & 0.023 & 0.041 & 0.080 & -- & -- \\
\end{longtable}

%% file: productions_drdid/g_avec_tres_bonnes_notes/tables/unified_breaking_points_drdid_table.tex
\footnotesize
\setlength{\tabcolsep}{6pt} % Espacement entre colonnes
\begin{longtable}[htbp]{@{}l*{3}{c}@{}}
\caption{Breakdown value of magnitude of parallel trend violation for a null value (general track, first quartile)} \label{g_avec_tres_bonnes_notes:breaking_point_unified}\\
\toprule
Outcome & 2018 & 2019 & 2020 \\
\midrule
\endfirsthead
\multicolumn{4}{c}{\tablename\ \thetable\ -- \textit{Continued from previous page}} \\
\toprule
Outcome & 2018 & 2019 & 2020 \\
\midrule
\endhead
\midrule
\multicolumn{4}{r}{\textit{Continued on next page}} \\
\endfoot
\bottomrule
\endlastfoot
\multicolumn{4}{l}{\textbf{Applications}} \\
\midrule
Max prestige of applications & 0 & 0 & 0 \\
Mean prestige of applications & 0 & 0 & 0 \\
\midrule
\multicolumn{4}{l}{\textbf{Applied in at least one...}} \\
\midrule
CPGE & 0 & 0 & 0 \\
Engineering school & 0 & 0 & 0 \\
\midrule
\multicolumn{4}{l}{\textbf{Was finally enrolled in a...}} \\
\midrule
CPGE & 0 & 0 & 0 \\
Engineering school & 0 & 0 & 0 \\
\midrule
\multicolumn{4}{l}{\textbf{Final enrollment}} \\
\midrule
Prestige of final enrollment & 0 & 0 & 0.3 \\
\midrule
\multicolumn{4}{l}{\footnotesize{}} \\
\end{longtable}

%% file: productions_drdid/g_avec_tres_bonnes_notes/tables/unified_confidence_intervals_drdid_table.tex
\footnotesize
\setlength{\tabcolsep}{4pt} % Réduire l'espacement entre colonnes (défaut: 6pt)
\begin{longtable}[htbp]{@{}l*{6}{c}@{}}
\caption{Confidence intervals, including standardized effect sizes (Cohen’s d) (general track, first quartile)}
\label{g_avec_tres_bonnes_notes:confidence_interval_unified}\\
\toprule
& \multicolumn{3}{c}{\textbf{Pre-treatment}} & \multicolumn{3}{c}{\textbf{Post-treatment}} \\
\cmidrule(lr){2-4} \cmidrule(lr){5-7}
Outcome & 2014 & 2015 & 2016 & 2018 & 2019 & 2020 \\
\midrule
\endfirsthead
\multicolumn{7}{c}{\tablename\ \thetable\ -- \textit{Continued from previous page}} \\
\toprule
Outcome & 2014 & 2015 & 2016 & 2018 & 2019 & 2020 \\
\midrule
\endhead
\midrule
\multicolumn{7}{r}{\textit{Continued on next page}} \\
\endfoot
\bottomrule
\endlastfoot
\multicolumn{7}{l}{\textbf{Applications}} \\
\midrule
Max prestige of applications & [-1.89; 0.95] & [-1.85; 1.41] & [-1.60; 1.33] & [-1.52; 1.15] & [-1.44; 1.07] & [-1.35; 1.10] \\
 & [-0.09; 0.05] & [-0.09; 0.07] & [-0.08; 0.07] & [-0.09; 0.07] & [-0.09; 0.07] & [-0.10; 0.08] \\
Mean prestige of applications & [-1.49; 1.16] & [-1.04; 1.90] & [-0.34; 2.44] & [-0.85; 1.79] & [-0.51; 2.14] & [0.16; 2.80] \\
 & [-0.07; 0.06] & [-0.05; 0.09] & [-0.02; 0.12] & [-0.04; 0.09] & [-0.03; 0.11] & [0.01; 0.15] \\
\midrule
\multicolumn{7}{l}{\textbf{Applied in at least one...}} \\
\midrule
CPGE & [-0.04; 0.02] & [-0.04; 0.03] & [-0.02; 0.05] & [-0.03; 0.04] & [-0.04; 0.03] & [-0.03; 0.04] \\
 & [-0.09; 0.05] & [-0.08; 0.06] & [-0.04; 0.09] & [-0.06; 0.08] & [-0.07; 0.06] & [-0.06; 0.08] \\
\end{longtable}

%% file: productions_drdid/g_reg/tables/unified_event_study_table_corrected.tex
\footnotesize
\setlength{\tabcolsep}{4pt} % Réduire l'espacement entre colonnes (défaut: 6pt)
\begin{longtable}[htbp]{@{}l*{6}{c}cc@{}}
\caption{DRDiD estimates of treatment effects on various outcomes (general track, DiD without reweighting)}\label{g_reg:event_study} \\
\toprule
 & \multicolumn{3}{c}{Pre-treatment year} & \multicolumn{3}{c}{Post-treatment year} & \\
\cmidrule(lr){2-4} \cmidrule(lr){5-7}
Outcome & 2014 & 2015 & 2016 & 2018 & 2019 & 2020 & N & p-value pretrend \\
\midrule
\endfirsthead
\multicolumn{9}{c}{\tablename\ \thetable\ -- \textit{Continued from previous page}} \\
\toprule
 & \multicolumn{3}{c}{Pre-treatment year} & \multicolumn{3}{c}{Post-treatment year} & \\
\cmidrule(lr){2-4} \cmidrule(lr){5-7}
Outcome & 2014 & 2015 & 2016 & 2018 & 2019 & 2020 & N & p-value pretrend \\
\midrule
\endhead
\midrule
\multicolumn{9}{r}{\textit{Continued on next page}} \\
\endfoot
\bottomrule
\endlastfoot
\multicolumn{9}{l}{\textbf{Final enrollment}} \\
\midrule
Prestige of final enrollment & 0.191 & 0.221 & 0.311 & 0.877** & 2.756*** & 2.522*** & 2M & 0.806 \\
 & (0.333) & (0.325) & (0.324) & (0.322) & (0.322) & (0.312) &  &  \\
Prestige of final enrollment (Cohen's d) & 0.008 & 0.009 & 0.012 & 0.034** & 0.106*** & 0.097*** & -- & -- \\
 & (0.013) & (0.013) & (0.013) & (0.013) & (0.012) & (0.012) &  &  \\
\midrule
\multicolumn{9}{l}{\footnotesize{* p < 0.05, ** p < 0.01, *** p < 0.001 (after BH correction)}} \\
\end{longtable}

%% file: productions_drdid/g_reg/tables/unified_breaking_points_drdid_table.tex
\footnotesize
\setlength{\tabcolsep}{6pt} % Espacement entre colonnes
\begin{longtable}[htbp]{@{}l*{3}{c}@{}}
\caption{Breakdown value of magnitude of parallel trend violation for a null value (general track, DiD without reweighting)} \label{g_reg:breaking_point_unified}\\
\toprule
Outcome & 2018 & 2019 & 2020 \\
\midrule
\endfirsthead
\multicolumn{4}{c}{\tablename\ \thetable\ -- \textit{Continued from previous page}} \\
\toprule
Outcome & 2018 & 2019 & 2020 \\
\midrule
\endhead
\midrule
\multicolumn{4}{r}{\textit{Continued on next page}} \\
\endfoot
\bottomrule
\endlastfoot
\multicolumn{4}{l}{\textbf{Final enrollment}} \\
\midrule
Prestige of final enrollment & 1.1 & 1.8 & 1.1 \\
\midrule
\multicolumn{4}{l}{\footnotesize{}} \\
\end{longtable}

%% file: productions_drdid/g_reg/tables/unified_confidence_intervals_drdid_table.tex
\footnotesize
\setlength{\tabcolsep}{4pt} % Réduire l'espacement entre colonnes (défaut: 6pt)
\begin{longtable}[htbp]{@{}l*{6}{c}@{}}
\caption{Confidence intervals, including standardized effect sizes (Cohen’s d) (general track, DiD without reweighting)}\label{g_reg:confidence_interval_unified}\\
\toprule
& \multicolumn{3}{c}{\textbf{Pre-treatment}} & \multicolumn{3}{c}{\textbf{Post-treatment}} \\
\cmidrule(lr){2-4} \cmidrule(lr){5-7}
Outcome & 2014 & 2015 & 2016 & 2018 & 2019 & 2020 \\
\midrule
\endfirsthead
\multicolumn{7}{c}{\tablename\ \thetable\ -- \textit{Continued from previous page}} \\
\toprule
Outcome & 2014 & 2015 & 2016 & 2018 & 2019 & 2020 \\
\midrule
\endhead
\midrule
\multicolumn{7}{r}{\textit{Continued on next page}} \\
\endfoot
\bottomrule
\endlastfoot
\multicolumn{7}{l}{\textbf{Final enrollment}} \\
\midrule
Prestige of final enrollment & [-0.46; 0.84] & [-0.42; 0.86] & [-0.32; 0.95] & [0.25; 1.51] & [2.13; 3.39] & [1.91; 3.13] \\
 & [-0.02; 0.03] & [-0.02; 0.03] & [-0.01; 0.04] & [0.01; 0.06] & [0.08; 0.13] & [0.07; 0.12] \\
\midrule
\multicolumn{7}{l}{\footnotesize{Line 1: CI with corrected thresholds [LCI; UCI]}} \\
\multicolumn{7}{l}{\footnotesize{Line 2: Cohen's d standardized CI [LCI/SD; UCI/SD]}} \\
\end{longtable}

%% file: productions_matching/tableau_general_2.tex
\begin{table}[htbp]
\centering
\caption{Raw and matched-profile differences in admission probabilities between scolarship recipients and non-recipient, detailed by year and type of program}
\begin{tabular}{lcccccccc}
\label{appendix:matching} \\
\toprule
\multirow{2}{*}{Année} & \multicolumn{2}{c}{BTS} & \multicolumn{2}{c}{BUT} & \multicolumn{2}{c}{CPGE} & \multicolumn{2}{c}{Engineering school} \\
 & Raw & Matching & Raw & Matching & Raw & Matching & Raw & Matching \\
\midrule
2014 & -0.0460*** & -0.0176*** & -0.0690*** & -0.0348*** & -0.0434*** & 0.0034 & -0.1095*** & -0.0864*** \\
 & (0.0028) & (0.0030) & (0.0048) & (0.0050) & (0.0065) & (0.0066) & (0.0127) & (0.0131) \\
[0.5em]
2015 & -0.0409*** & -0.0210*** & -0.0546*** & -0.0327*** & -0.0415*** & -0.0015 & -0.1384*** & -0.1108*** \\
 & (0.0029) & (0.0031) & (0.0051) & (0.0052) & (0.0061) & (0.0062) & (0.0126) & (0.0130) \\
[0.5em]
2016 & -0.0415*** & -0.0219*** & -0.0480*** & -0.0270*** & -0.0380*** & 0.0016 & -0.0871*** & -0.0609*** \\
 & (0.0027) & (0.0028) & (0.0048) & (0.0050) & (0.0062) & (0.0063) & (0.0127) & (0.0129) \\
[0.5em]
2017 & -0.0335*** & -0.0200*** & -0.0470*** & -0.0246*** & -0.0494*** & -0.0077 & -0.1142*** & -0.0939*** \\
 & (0.0026) & (0.0028) & (0.0046) & (0.0047) & (0.0057) & (0.0059) & (0.0115) & (0.0119) \\
[0.5em]
2018 & -0.0189*** & -0.0059+ & -0.0101* & 0.0050 & -0.0137* & 0.0224*** & -0.0594*** & -0.0530*** \\
 & (0.0030) & (0.0031) & (0.0047) & (0.0048) & (0.0055) & (0.0056) & (0.0111) & (0.0115) \\
[0.5em]
2019 & -0.0015 & 0.0087** & 0.0193*** & 0.0428*** & 0.0220*** & 0.0596*** & -0.0552*** & -0.0419*** \\
 & (0.0029) & (0.0030) & (0.0045) & (0.0046) & (0.0052) & (0.0053) & (0.0100) & (0.0102) \\
[0.5em]
2020 & -0.0098*** & 0.0126*** & 0.0255*** & 0.0689*** & 0.0302*** & 0.0706*** & -0.0239** & 0.0023 \\
 & (0.0026) & (0.0026) & (0.0040) & (0.0040) & (0.0051) & (0.0052) & (0.0086) & (0.0088) \\
\midrule
Applicants & 1,455,494 & 1,434,163 & 1,125,519 & 1,105,450 & 611,135 & 586,516 & 227,495 & 218,731 \\
\bottomrule
\end{tabular}
\end{table}